\documentclass[a4paper,11pt]{article}
\pdfoutput=1 

\usepackage{jheppub} 
\usepackage[T1]{fontenc} 
\usepackage{slashed}
\usepackage{arydshln}
\usepackage{amsmath}
\usepackage{amssymb}
\usepackage{amsfonts}
\usepackage{epsfig,multicol,bbm}
\renewcommand{\arraystretch}{1.2}

\newcommand{\be}{\begin{equation}}
\newcommand{\ee}{\end{equation}}
\newcommand{\Eq}[1]{Eq.~\eqref{#1}}

\newcommand{\Refs}[1]{Refs.~\cite{#1}}

\newcommand \tr {{\rm Tr}}

\newcommand{\vev}[1]{\left\langle #1 \right\rangle}
\newcommand{\squared}[1]{\left| #1 \right|^2}

\newcommand{\bs}[1]{\boldsymbol #1}

\def\DR{\Phi}

\DeclareMathOperator{\arctanh}{artanh}
\DeclareMathOperator{\arccoth}{arcoth}

\title{\boldmath Thermal QCD Axions across Thresholds}

\author[a,b]{Francesco D'Eramo}
\author[a,b]{, Fazlollah Hajkarim}
\author[a,b]{, Seokhoon Yun}

\affiliation[a]{Dipartimento di Fisica e Astronomia, Universit\`a degli Studi di Padova, \\ Via Marzolo 8, 35131 Padova, Italy}
\affiliation[b]{Istituto Nazionale di Fisica Nucleare (INFN), Sezione di Padova, \\ Via Marzolo 8, 35131 Padova, Italy}

\emailAdd{francesco.deramo@pd.infn.it}
\emailAdd{fazlollah.hajkarim@pd.infn.it}
\emailAdd{seokhoon.yun@pd.infn.it}

\abstract{Thermal axion production in the early universe goes through several mass thresholds, and the resulting rate may change dramatically across them. Focusing on the KSVZ and DFSZ frameworks for the invisible QCD axion, we perform a systematic analysis of thermal production across thresholds  and provide smooth results for the rate. The QCD phase transition is an obstacle for both classes of models. For the hadronic KSVZ axion, we also deal with production at temperatures around the mass of the heavy-colored fermion charged under the Peccei-Quinn symmetry. Within the DFSZ framework, standard model fermions are charged under this symmetry, and additional thresholds are the heavy Higgs bosons masses and the electroweak phase transition. We investigate the cosmological implications with a specific focus on axion dark radiation quantified by an effective number of neutrino species and explore the discovery reach of future CMB-S4 surveys.
}

\begin{document} 
\maketitle
\flushbottom


\section{Introduction}
\label{sec:intro}

The unexpected invariance of strong interactions under transformations flipping the arrow of time is one of the most challenging puzzles in fundamental physics. This remarkable experimental fact is equivalent to state that Quantum ChromoDynamics (QCD) is invariant, within experimental uncertainties, under the combination of parity and charge conjugation (CP) in agreement with the CPT theorem. Such a CP violation by strong interactions is parameterized by an effective dimensionless parameter $\bar{\theta}$ that is expected to be of order one, but the lack of observation of a fundamental neutron electric dipole moment put the spectacular constraint $\bar{\theta} \lesssim 10^{-10}$~\cite{Baker:2006ts,Pendlebury:2015lrz,nEDM:2020crw}. Anthropic explanations are not viable~\cite{Ubaldi:2008nf,Dine:2018glh}, and understanding this severe inequality is known as the strong CP problem. 

The Peccei-Quinn (PQ) mechanism~\cite{Peccei:1977np,Peccei:1977hh} is one of the most appealing solutions. A new Abelian $U(1)_{\rm PQ}$ symmetry, anomalous under strong interactions and spontaneously broken, plays the role of the main character.  At energies much lower than the PQ breaking scale $f_a$, the only residual degree of freedom is an approximate Nambu-Goldstone boson $a$ known as the axion~\cite{Wilczek:1977pj,Weinberg:1977ma} that acquires the anomalous coupling to gluons
\be
\mathcal{L}_{\rm PQ} \supset \frac{\alpha_s}{8 \pi} \frac{a}{f_a} G^A_{\mu\nu} \widetilde{G}^{A \mu\nu} \ .
\label{eq:LPQaxion}
\ee
Here, we denote the QCD fine structure constant by $\alpha_s = g_s^2 / (4 \pi)$, the gluon field strength by $G_{\mu\nu}^A$ with the index $A = 1, \ldots, 8$ running over the adjoint indices of the color gauge group, and its dual by $\widetilde{G}^{A \mu\nu} \equiv \epsilon^{\mu\nu\rho\sigma} G^A_{\rho\sigma} / 2$. Once strong interactions confine, QCD non-perturbative effects generate a potential that leads to an axion mass~\cite{Bardeen:1978nq,GrillidiCortona:2015jxo}
\be
m_a \simeq 5.7 \, \mu eV \, \left( \frac{10^{12} \, {\rm GeV}}{f_a} \right) \ .
\label{eq:ma}
\ee
The PQ breaking scale $f_a$ suppresses axion couplings as well. Such a scale is constrained by terrestrial and astrophysical axion searches, and accounting only for the axion coupling to gluons in Eq.~\eqref{eq:LPQaxion} leads to the rather conservative bound $f_a \gtrsim 10^8 \, {\rm GeV}$: the axion must be light and weakly-coupled. The field evolution in the early universe goes through two main phases: the axion is initially stuck by Hubble friction, and once its mass becomes comparable to the expansion rate it begins oscillating around the minimum of its potential. The oscillation amplitude gets damped by the Hubble friction, and the axion settles down at its minimum which is CP-conserving as ensured by the Vafa-Witten theorem~\cite{Vafa:1984xg}: QCD dynamics itself solves the strong CP problem. The energy density stored in the field oscillations can account for the observed dark matter abundance~\cite{Preskill:1982cy,Abbott:1982af,Dine:1982ah}. Furthermore, axion interactions with standard model (SM) fields are responsible for a plethora of phenomena in the early universe~\cite{Marsh:2015xka} and the target of an intensive experimental effort~\cite{Graham:2015ouw,Irastorza:2018dyq,Sikivie:2020zpn}.

This work investigates a distinct cosmological signal of PQ theories: the production of relativistic axions from scatterings and/or decays of particles belonging to the primordial thermal bath~\cite{Turner:1986tb}. Given their thermal origin, such hot axions are produced with energies of the size of the bath temperature, and their typical energy will stay of the size of the one for photons as long as they are relativistic. This statement holds regardless of whether they thermalize or not, and the axion mass in Eq.~\eqref{eq:ma} ensures that axions produced thermally are still relativistic as late as at recombination. 

Such hot axions manifest themselves experimentally as an additional contribution to radiation in the early universe. How do we have access to this quantity? Two key events in the expansion history allow us to bound the energy density stored in relativistic particles. Following a chronological order, the first is Big Bang Nucleosynthesis (BBN) when the thermal bath synthesized light nuclei. The successful agreement between predictions and observations gives us information about the expansion rate at BBN and it bounds additional radiation. This effect is parameterized by an effective number of neutrino species on top of the SM contribution, $N_\nu = 3 + \Delta N_\nu$. After the recent measurement of the deuterium burning rate by LUNA~\cite{Mossa:2020gjc}, Ref.~\cite{Yeh:2020mgl} found the constraint $N_\nu = 2.880 \pm 0.144$. 

Another important event is the formation of the Cosmic Microwave Background (CMB) since additional radiation alters the CMB anisotropy spectrum at small angular scales. This effect  is also parameterized in terms of an effective number of additional neutrino species $N_{\rm eff} = N^{\rm SM}_{\rm eff} + \Delta N_{\rm eff}$. The SM naive prediction $N^{\rm SM}_{\rm eff} = 3$ does not hold because neutrino decoupling is not instantaneous~\cite{Mangano:2001iu}, and we have $N^{\rm SM}_{\rm eff} \simeq 3.0440$~\cite{Bennett:2019ewm,Akita:2020szl,Bennett:2020zkv}. The most stringent constraint, $N_{\rm eff} = 2.99 \pm 0.17$, comes from the Planck collaboration~\cite{Aghanim:2018eyx}.

Future CMB surveys forecast an extraordinary improvement in measuring this quantity, and conservative configurations of CMB-S4 can reach $\Delta N_{\rm eff}^{{\rm CMB-S4}}(1 \sigma) \simeq 0.02 - 0.03$~\cite{CMB-S4:2016ple,Abazajian:2019eic}. What does this value imply for fundamental physics? Let us consider a scalar field, such as the axion, and let us assume that it reaches thermal equilibrium in the early universe. Even if we take the most pessimistic hypothesis that decoupling happened well above the weak scale, the resulting contribution would be $\Delta N_{\rm eff} \simeq 0.027$. If decoupling happened at lower temperatures and/or if the field has a larger number of internal degrees of freedom the expected value is larger. Thus future CMB-S4 surveys are sensitive to any relic light particle that was once in equilibrium with the standard model thermal bath~\cite{Brust:2013xpv,Baumann:2016wac}. 

\begin{figure}
\centering
 \includegraphics[width=.95\linewidth]{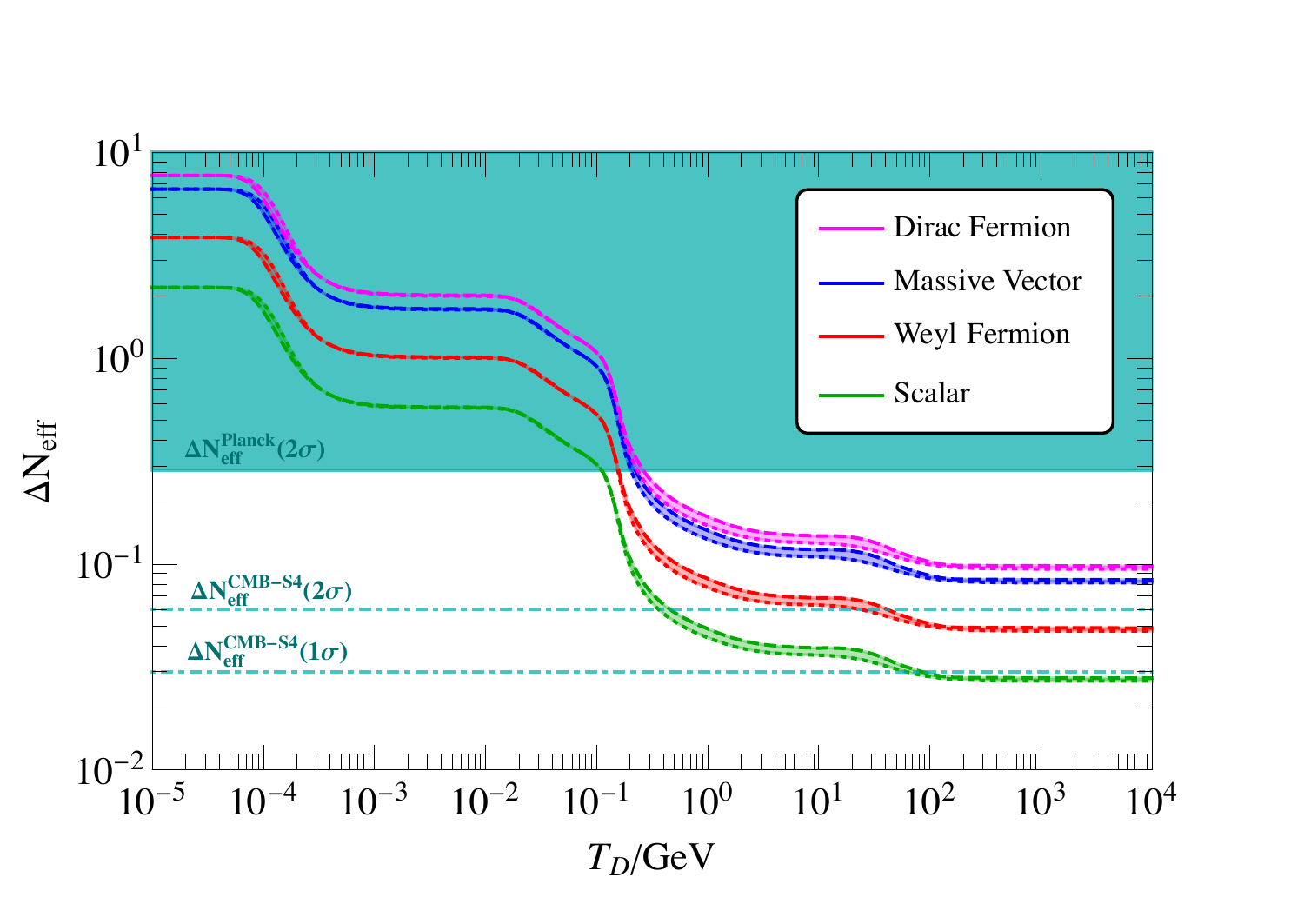}
 \label{fig:neffdec}
\caption{\em Contribution to $\Delta N_{\rm eff}$ from light relics that were once in thermal equilibrium as a function of the decoupling temperature  $T_{\rm D}$. Different colors represent different particle spins, and the width of each line corresponds to the two different treatments of the relativistic degrees of freedom provided in Refs.~\cite{Drees:2015exa} (dashed lines) and \cite{Saikawa:2018rcs} (dashed lines). The shaded region is excluded by Planck~\cite{Aghanim:2018eyx}, the dot-dashed lines show the CMB-S4 discovery reach~\cite{CMB-S4:2016ple,Abazajian:2019eic}.}
\end{figure}

Fig.~\ref{fig:neffdec} shows how future CMB surveys provide a powerful probe of light and elusive physics beyond the SM. Here, we choose the conservative value $\Delta N_{\rm eff}^{{\rm CMB-S4}}(1 \sigma) = 0.03$. For hot relics that thermalized at early times and decoupled when the bath temperature was $T_D$, we take four dark radiation candidates $\DR$: scalar, Weyl fermion, massive vector, and Dirac fermion. The effective relativistic degrees of freedom contributing to the energy density at high temperatures are $g_{* \DR} = \left\{1, 7/4, 3, 7/2 \right\}$. The prediction for $\Delta N_{\rm eff}$ as a function of $T_D$, derived in Eq.~\eqref{eq:DeltaNeffforfig1} of App.~\ref{app:CO}, explicitly reads
\be
\Delta N_{\rm eff} \simeq g_{* \DR} \, \times \, 13.69 \;  g^{\rm SM}_{*s}(T_D)^{-4/3} 
\label{eq:DeltaNeffIntro}
\ee
with $g^{\rm SM}_{*s}(T_D)$ the effective number of SM entropic degrees of freedom at $T_D$. We employ two different sets of data for $g^{\rm SM}_{*s}(T_D)$, Refs.~\cite{Drees:2015exa} (dashed lines) and \cite{Saikawa:2018rcs} (dotted lines), and we notice how the treatment of thermal bath has a tiny effect on the final predictions (see App.~\ref{app:CO} for more discussion). This result is valid for a standard thermal history with no significant releases of entropy that would
dilute the expected amount population of relativistic axions. Exceptions are possible, as for example around the time of the QCD phase transition (QCDPT)~\cite{2010PhRvL.105d1301B} or the electroweak phase transition (EWPT)~\cite{Chaudhuri:2017icn}. 

An important message from Fig.~\ref{fig:neffdec}, which is quantified by the expression in Eq.~\eqref{eq:DeltaNeffIntro}, is that the later the relic decouples the larger is its contribution to $\Delta N_{\rm eff}$. Planck data already exclude dark radiation that decoupled around the time of the QCDPT, and future experiments will probe hot relics that decoupled earlier. Production of dark radiation around or below the QCDPT is particularly timely and promising for future experiments. This is true for any dark radiation candidate, and in particular for the QCD axion which is the subject of our investigation.

These spectacular projections make rigorous calculations a top priority. Recently, several collaborations revised predictions for the hot axion abundance. Production above the weak scale via quark and gluon scattering was investigated by Ref.~\cite{Masso:2002np} with the inclusion of Debye screening effects. Later on, Refs.~\cite{Graf:2010tv,Salvio:2013iaa} provided rigorous treatments of thermal effects, and Ref.~\cite{Salvio:2013iaa} extended the analysis to production via electroweak gauge fields and top quark. These studies considered production above the weak scale. Scatterings of heavy quarks below the electroweak scale were analyzed by Ref.~\cite{Ferreira:2018vjj,Arias-Aragon:2020qtn}, and this analysis is valid only well above the QCDPT. At lower temperatures, hadron scatterings were considered by Refs.~\cite{Berezhiani:1992rk,Chang:1993gm,Hannestad:2005df,DEramo:2014urw,Kawasaki:2015ofa,Ferreira:2020bpb}. The analysis in Ref.~\cite{Giare:2020vzo} bound the axion mass to $m_a < 7.46 \, {\rm eV}$ and $m_a < 0.91 \, {\rm eV}$ for thermalization with gluons and pions, respectively. Ref.~\cite{DEramo:2018vss} investigated production via lepton scatterings and decays, which is immune to QCD complications, and it pointed out a possible connection with the so-called Hubble tension~\cite{Bernal:2016gxb,Verde:2019ivm}.

The presence of mass thresholds across which production rates change dramatically is an issue rarely addressed in the literature.  As a step forward toward the proper treatment of the EWPT, Ref.~\cite{Arias-Aragon:2020shv} provided a smooth connection between rates within an effective field theory framework containing only one Higgs doublet. Although it is not the most general case for PQ theories, it is always a good approximation in the so-called decoupling limit where the heavy Higgs bosons are much more massive than the weak scale. A threshold common to all axion models is the QCDPT where axion interactions with quarks and gluons become non-perturbative, strong interactions confine and one must resort to non-perturbative techniques. Recently, Ref.~\cite{DEramo:2021psx} provided the first smooth treatment of the QCDPT for the axion coupling to gluons in Eq.~\eqref{eq:LPQaxion}.

In this work, we provide predictions for the amount of axion dark radiation within UV complete models. Our methodology features two key steps. First, we evaluate the axion production rate at any temperature with smooth treatments of all mass thresholds. With the rate in hand, we solve the Boltzmann equation for the axion abundance and translate the resulting amount into a correspondent $\Delta N_{\rm eff}$. 

Notwithstanding the broad landscape of PQ theories~\cite{Kim:2008hd,DiLuzio:2020wdo}, we can divide them into two main classes according to the origin of the color anomaly.
\begin{itemize}
\item \textbf{KSVZ axion}~\cite{Kim:1979if,Shifman:1979if}. SM fields are PQ-neutral, and the color anomaly is due to a new heavy-colored fermion $\Psi$ that gets mass from PQ breaking. Following the chronology of the expansion history, the first mass scale that we encounter is the fermion $\Psi$. Binary collisions involving $\Psi$ are the main production channel above such a threshold. Gluon scatterings mediated by the operator in Eq.~\eqref{eq:LPQaxion}, which is generated once we integrate out $\Psi$, control the production rate at temperatures below the $\Psi$ mass. These processes are the main production channel until we reach the scale where strong interactions confine. This is the second and the last threshold that we need to treat for this framework, and we connect the production rate above confinement with the one where the relevant degrees of freedom become hadrons. 
\item \textbf{DFSZ axion}~\cite{Zhitnitsky:1980tq,Dine:1981rt}. SM quarks are responsible for the color anomaly and the Higgs sector is extended with another weak doublet. We work in the decoupling limit where the heavy Higgs bosons have a mass $m_A$ substantially larger than the weak scale. Such a high mass scale is the first threshold: we have a two Higgs doublet model (2HDM) above and the SM below, respectively. The second threshold is the EWPT, not present for the KSVZ case because the axion did not couple to the Higgs field; PQ charges of the Higgs and SM fermions make it relevant for this case. Finally, the QCDPT is also something to account for as we did for the previous case, although the details of the matching are different as a consequence of different axion couplings. 
\end{itemize}

We compute the production rate at any temperature for the KSVZ and DFSZ axion in Secs.~\ref{sec:KSVZ} and \ref{sec:DFSZ}, respectively. We employ these results in Sec.~\ref{sec:results} to quantify how many axions are produced thermally in the early universe and to predict the resulting contribution to $\Delta N_{\rm eff}$. Sec.~\ref{sec:conclusions} contains our conclusions, and we defer all technical details to appendices.


\section{The KSVZ Axion}
\label{sec:KSVZ}

The minimal ingredients for the KSVZ framework are an electroweak singlet complex scalar $\varphi$ and a vector-like colored fermion $\Psi$. Electroweak charges for the fermion are allowed but not mandatory, and we work in the scenario where it is only charged under the fundamental of the $SU(3)_c$ gauge group. The Lagrangian for this case reads
\be
\mathcal{L}_{\rm KSVZ} = \left(\partial^\mu \varphi\right)^\dagger \partial_\mu \varphi +\bar{\Psi} i \slashed{D} \Psi - V_{\rm KSVZ}\left(\varphi\right) - \left(y_\Psi \,  \varphi^\dagger \bar{\Psi}_L \Psi_R + {\rm h.c.} \right) \ .
\label{eq:KSVZUVcomplete}
\ee
A bare fermion mass term is forbidden by some suitable symmetry, and we identify well-defined chiralities $\Psi_{L,R} = P_{L,R} \Psi$ through the action of chiral projectors $P_{L,R} = (1\mp \gamma^5)/2$. 

The theory features a global symmetry that acts on the fields as follows
\be
\varphi \rightarrow e^{i q_\varphi \alpha} \varphi \ , \qquad \qquad \qquad \Psi_L \rightarrow e^{i q_L \alpha}\Psi_L \ , \qquad \qquad \qquad \, \Psi_R \rightarrow e^{i q_R \alpha }\Psi_R \ .
\ee
For any value of the transformation parameter $\alpha$, the Lagrangian in Eq.~\eqref{eq:KSVZUVcomplete} is invariant as long as the scalar potential $V_{\rm KSVZ}(\varphi)$ does not change and the charges satisfy $q_\varphi = q_R - q_L$. Furthermore, two crucial ingredients must be satisfied for this to be a viable PQ symmetry: broken in the vacuum state and anomalous under strong interactions. The potential
\be
V_{\rm KSVZ}(\varphi) = \lambda_{\varphi}\left(\left|\varphi\right|^2 - v_\varphi^2/2\right)^2 \ ,
\ee
where $\lambda_{\varphi}$ is the quartic self-coupling for the field and $v_\varphi$ its vacuum expectation value (vev), satisfies the first requirement. The condition on the anomaly is satisfied as long as the global symmetry is not vector-like, $q_L \neq q_R$ (i.e., $q_\varphi \neq 0$). Thus the phase of the complex scalar $\varphi$, which corresponds to the KSVZ axion, appears in the gluon anomaly operator in Eq.~\eqref{eq:LPQaxion} and it eventually leads to a natural solution of the strong CP problem. 

The KSVZ axion originates solely from the phase of the complex scalar $\varphi$. As long as the quartic coupling is $\lambda_\varphi \sim \mathcal{O}(1)$, the radial mode of $\varphi$ is rather heavy with a mass $\mathcal{O}(v_\varphi)$. We neglect fluctuations along the radial direction, and we identify the axion $a$ as the phase of the complex field, $\varphi \rightarrow \left( v_\varphi/\sqrt{2}\right)e^{i a/v_\varphi}$. The Yukawa coupling is responsible for a fermion mass, $m_\Psi = y_\Psi v_\varphi/\sqrt{2}$, which can be smaller than the symmetry breaking scale $v_\varphi$ if $y_\Psi$ is small, but not in conflict with collider searches for heavy colored states ($m_\Psi \gtrsim {\rm TeV}$).

The effective Lagrangian below the symmetry breaking scale reads
\be
\mathcal{L}^{\rm (linear)}_{\rm KSVZ}  = \frac{1}{2} \partial^\mu a \partial_\mu a + \bar{\Psi} i \slashed{D} \Psi - \left[ m_\Psi e^{-i a / v_\varphi} \bar{\Psi}_L \Psi_R + {\rm h.c.} \right] \ .
\label{eq:KSVZhighESpurionic}
\ee
We find this Lagrangian convenient to compute the axion production rate at temperatures above $m_\Psi$. At lower temperatures, it is preferable to employ a different field basis. Let us describe in detail the difference between these two choices to exploit the interplay between the axion and PQ-charged fields. On one hand, the PQ symmetry can be linearly realized and the axion appears as the phase of the PQ breaking scalar as in Eq.~\eqref{eq:KSVZUVcomplete}. On the other hand, the PQ symmetry can be non-linearly realized and the axion shifts under a PQ transformation, $a\rightarrow a + {\rm const}$. The second option can be reached by performing the axion-dependent chiral rotation $\Psi \rightarrow \exp\left[ i \frac{a}{2 v_\varphi} \gamma^5 \right] \Psi$, and the resulting Lagrangian contains the changes at the classical level as well as the effects of the anomaly through Eq.~\eqref{eq:anomalousrot}
\be
\mathcal{L}^{\rm (non-linear)}_{\rm KSVZ}  =  \frac{1}{2} \partial^\mu a \partial_\mu a + \bar{\Psi} i \slashed{D} \Psi - \frac{\partial_\mu a}{2v_\varphi}\bar{\Psi}\gamma^\mu \gamma^5 \Psi + 
\frac{\alpha_s}{8 \pi} \frac{a}{v_\varphi} G^A_{\mu\nu} \widetilde{G}^{A \mu\nu}  \ .
\label{eq:KSVZhighELagrangian}
\ee
We introduce the axion decay constant $f_a$ and we set it to $f_a = v_\varphi$ so we reproduce the normalization in Eq.~\eqref{eq:LPQaxion}. Although the field basis to describe axion interactions is not unique, the scattering cross sections calculated in App.~\ref{app:XS} are independent on such a choice.

The production of the KSVZ axion goes through three main cosmological phases that are separated by two mass thresholds: (i) the mass of $\Psi$; (ii) the confinement scale. 

\subsection{Matching at the heavy PQ fermion threshold}

Above the heavy fermion mass $m_\Psi$, axion production is driven by the scatterings
\be
\Psi + \bar{\Psi} \rightarrow g + a \ , \qquad \qquad \qquad \Psi/\bar{\Psi} + g \rightarrow \Psi/\bar{\Psi} + a \ ,
\ee
where $g$ is a gluon. Below $m_\Psi$, the rate is controlled by quark ($q$) and gluon scatterings 
\be
g+g \rightarrow g +a \ , \qquad \qquad  q + \bar{q} \rightarrow g +a  \ , \qquad \qquad q/\bar{q}+g \rightarrow q/\bar{q}+a \ .
\ee
The long range nature of gluon interactions leads to IR divergences that need some care. The prescription to regularize such divergences of Ref.~\cite{Braaten:1991dd}, which holds for soft external momenta ($p \simeq g_s T$), works only in the weak-coupling regime for axion production~\cite{Graf:2010tv}. Once the QCD coupling $g_s$ gets stronger one needs to go beyond the hard thermal loop (HTL) approximation. Ref.~\cite{Salvio:2013iaa} parameterized the rate in such a regime as follows~\footnote{Expressions analogous to Eq.~\eqref{eq:F3function} hold for subdominant processes mediated by electroweak gauge bosons with appropriate modifications of $F_3(T)$, group theory factors, and gauge coupling constants~\cite{Salvio:2013iaa}.}
\be
\gamma_{gg} \equiv \frac{d N_a}{dV dt} = \frac{2 \zeta (3) d_g}{\pi^3}\left(\frac{\tilde{c}_{g}^{\,\Psi}(T) \, \alpha_s }{8\pi f_a}\right)^2 F_3\left(T\right) \, T^6  \ .
\label{eq:F3function}
\ee
Here, $\tilde{c}_{g}^{\,\Psi}(T)$ denotes the effective gluon anomaly coefficient discussed in the next paragraph. The numerical factors are $d_g = 8$ and $\zeta(3) \simeq 1.2$ for the dimension of the $SU(3)$ adjoint representation and the Riemann zeta function, respectively. The result for $F_3(T)$ provided by Ref.~\cite{Salvio:2013iaa} for temperatures well above the weak scale allows us to deal with the heavy PQ fermion threshold, but it is not enough to approach the QCDPT. We evaluate $F_3(T)$ in App.~\ref{app:TH} at any temperature in the QCD perturbative regime, and we keep into account the decoupling of heavy quarks. We evaluate the rate with the aid of the `\texttt{RunDec}'~\cite{Chetyrkin:2000yt} code that accounts for the running of the strong coupling constant $\alpha_s$ up to four loops.

The UV origin of gluon scatterings is due to a heavy PQ-charged colored fermion. At temperatures much larger than its mass the effect is negligible, and it becomes relevant only once we integrate out the fermion for physical processes with typical energies smaller than the mass of the fermion itself. We can make this statement quantitative by evaluating the 1PI effective action.\footnote{This important difference between Wilsonian and 1PI effective coupling was pointed out by Ref.~\cite{Bae:2011jb} within the context of axino production for SUSY PQ theories.} For a generic colored fermion $\chi$ charged under PQ, we can perform a chiral rotation to induce the trilinear axion anomalous coupling to gluons
\be
\mathcal{L}_a^{\rm (Wilson)} = c_g^{\, \chi} \times \frac{\alpha_s}{8 \pi} \frac{a}{f_a} G^A_{\mu\nu} \widetilde{G}^{A \mu\nu} \ .
\ee
Here, the Wilson coefficient $c_g^{\, \chi}$ is a constant number that depends on the quantum numbers of the fermion $\chi$. However, if one cares about the axion production rate the relevant quantity to consider is the 1PI effective action that we parameterize as follows
\be
\mathcal{L}_a^{\rm (1PI)} = \tilde{c}_g^{\, \chi}(q^2) \times \frac{\alpha_s}{8 \pi} \frac{a}{f_a} G^A_{\mu\nu} \widetilde{G}^{A \mu\nu} \ .
\ee
Contrarily to the previous case, the effective coupling $\tilde{c}_g^{\, \chi}(q^2)$ depends on the momentum exchanged in the physical process under consideration. The relation between the Wilson coefficient and the 1PI effective coupling reads
\be
\frac{\tilde{c}_g^{\, \chi}(q^2)}{c_{g}^{ \, \chi}} =  \frac{1}{\tau_\chi} \left\{ 
\begin{tabular}{cc}
$\arcsin^2 \sqrt{\tau_\chi}$ & $\tau_\chi \leq 1$ \\
$-\frac{1}{4}\left[\log \frac{1+\sqrt{1-\tau_\chi^{-1}}}{1-\sqrt{1-\tau_\chi^{-1}}}-i\pi\right]^2$ & $\tau_\chi > 1$ 
\end{tabular}
\right. \, ,
\label{eq:1PIgluonAnomaly}
\ee
where we express the momentum dependence in terms of the dimensionless $\tau_\chi = q^2/4m_\chi^2$. 

After this general discussion, we get back to the KSVZ axion ($c_g^{\, \Psi} = 1$) and set $q^2 \approx T^2$ in Eq.~\eqref{eq:1PIgluonAnomaly}. When $\Psi$ is a degree of freedom of the thermal bath, with the universe much hotter than $m_\Psi$, one-loop 1PI corrections effectively cancel out the Wilson coefficient leading to $\tilde{c}_g^{ \, \Psi}(T^2) / c_g^{\, \Psi} \sim (m_\Psi/T)^2 \log^2\left[T/m_\Psi\right]$. The diminished gluon scatterings at $T \gg m_\Psi$ ensures the dominance of $\Psi$ collisions. At temperatures below $m_\Psi$, the effective coefficient $\tilde{c}_g^{\, \Psi}(T \ll m_\Psi)$ becomes nearly unity and this is the radiative remnant of $\Psi$.

\begin{figure}
\centering
\includegraphics[width=0.9\textwidth]{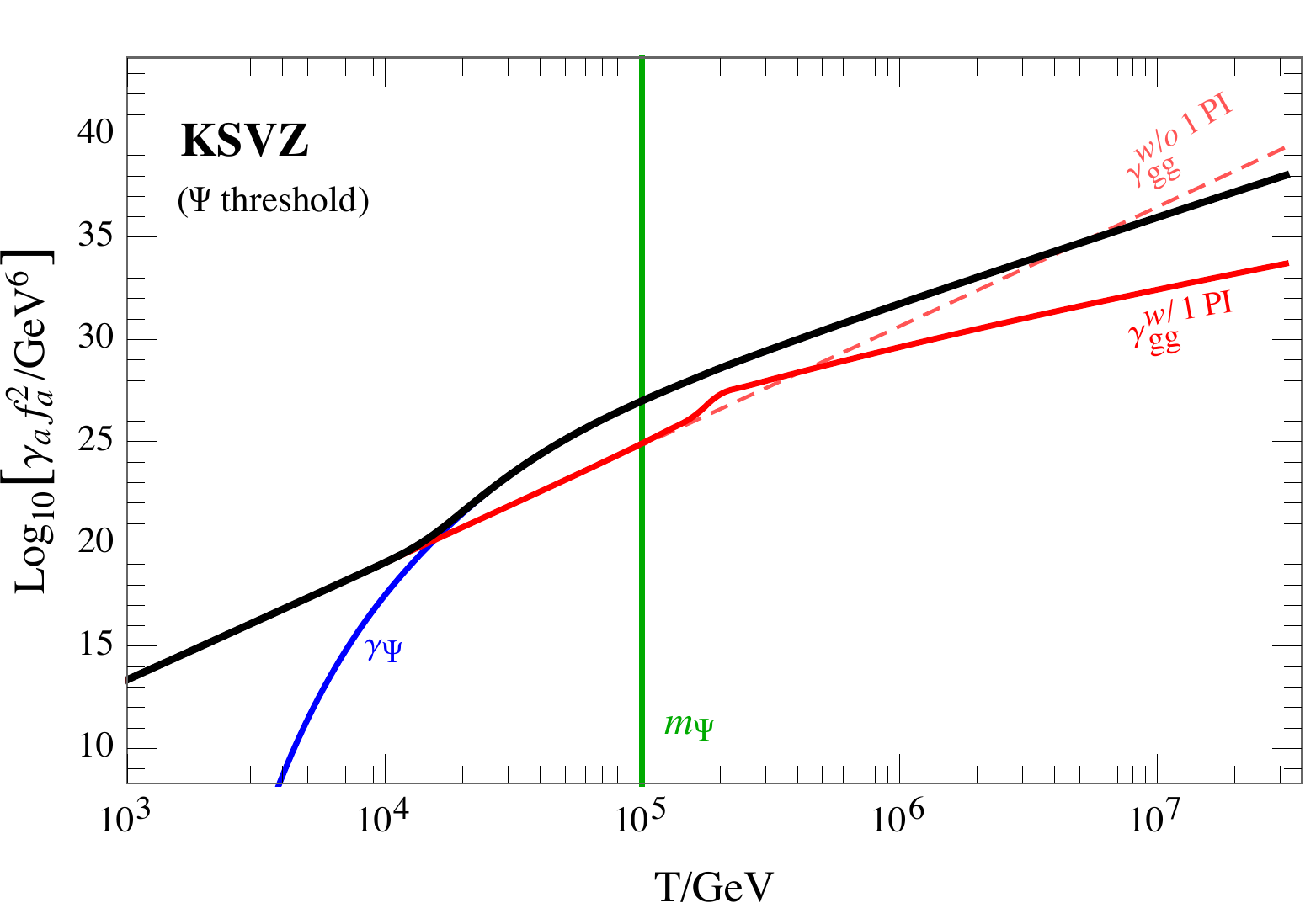}
\caption{\em KSVZ axion production rate across the heavy colored PQ fermion mass $m_{\Psi} = 10^5\,{\rm GeV}$. The black solid line indicates the total rate, and it is the sum of $\Psi$ scatterings (solid blue) and thermal gluon scattering (solid red). We show for comparison the gluon scattering rate obtained with only the Wilsonian action contribution (dashed red).}
\label{fig:KSVZhighE}
\end{figure}

We set $m_{\Psi} =10^5\,{\rm GeV}$ (vertical green line), and we plot the combination $\gamma_a f_a^2$ as a function of the temperature in Fig.~\ref{fig:KSVZhighE}. Axion interactions below $m_\Psi$ mediated by the dimension 5 operator with gluons lead to the scaling $\gamma_a \propto f_a^{-2}$. In the opposite regime, the axion interacts via a renormalizable Yukawa coupling $y_\Psi = \sqrt{2} m_\Psi / f_a$, and the rate scales as $\gamma_a \propto y_\Psi^2 \propto f_a^{-2}$ also at high temperatures once we fix the fermion mass. The solid black line denotes the total rate that is the sum of $\Psi$ (solid blue line) and gluon (solid red line) scatterings. In order to appreciate the difference between Wilsonian and 1PI descriptions, we also show the gluon scattering rate that we would get by accounting for the Wilsonian contribution only (dashed red line); this corresponds to setting $\tilde{c}_{g}^{\, \Psi}(T) = 1$ in Eq.~\eqref{eq:F3function} at any temperature. Consistently with our picture, the solid and dashed red lines are in exact agreement for  $T<m_\Psi$, and they differ significantly for larger temperatures. In particular, $\gamma_{gg}^{\rm w/o \, 1PI} \propto T^6$ dominates even at large temperatures while the correct functional dependence for the production rate in the UV scales as $\gamma_\Psi \propto T^4$. 

\subsection{Matching at the QCD threshold}
\label{sec:KSVZlowthreshold}

The picture where the axion field interacts with quarks and gluons breaks down once we approach the scale $\Lambda_{\rm N} \sim 2\,{\rm GeV}$ and strong interactions become non-perturbative. Quarks are confined within hadrons at lower energies, and one must resort to non-perturbative techniques. Here, we employ the ones of chiral perturbation theory (ChPT) to compute the axion production rate from hadron collisions, and we provide a smooth result across the QCDPT. As we review in App.~\ref{app:PP}, the correct prescription to determine axion couplings to hadrons is to match currents with the same symmetry properties between the UV and IR theories~\cite{Srednicki:1985xd,Georgi:1986df}. Such a procedure is straightforward within the KSVZ framework since the axion interacts with the strong sector only through the anomalous coupling to gluons. We find it convenient to rotate it away due to the large instanton effects at low energy, and this is done via the axion-dependent field redefinition of the light quarks
\be 
q \rightarrow e^{-i c_q \frac{a}{f_a}\gamma^5}q \ ,
\ee
where $c_q = M_q^{-1}/2 \, \tr [M_q^{-1}]$ and $M_q = {\rm diag} (m_u,m_d,m_s)$. As a result, we switch to axion interactions to quark currents via the derivative interactions
\be
\sum_{q=u,d,s} c_q\frac{\partial_\mu a}{f_a} \bar{q}\gamma^\mu\gamma^5 q \,.
\label{eq:KSVZlowEqCoup}
\ee
The matching conditions provided in App.~\ref{app:PP} allow us to find axion couplings to mesons (for coupling to baryons see Ref.~\cite{Chang:1993gm}). 

Up to what UV cutoff $\Lambda_{\rm ChPT}$ can we push ChPT? We treat the primordial plasma within the hadron resonance gas (HRG) approximation~\cite{Hagedorn:1984hz,Huovinen:2009yb,Megias:2012hk} which is inconsistent with lattice QCD results above $T>150\,{\rm MeV}$~\cite{Venumadhav:2015pla}. Furthermore, leading order ChPT for axion production suffers perturbativity problems at $T>62\,{\rm MeV}$~\cite{DiLuzio:2021vjd}. Notwithstanding both values of $\Lambda_{\rm ChPT}$ being smaller than $\Lambda_{\rm N}$, we find it plausible that the axion production rate between $\Lambda_{\rm ChPT}$ and $\Lambda_{\rm N}$ is connected smoothly since the QCDPT is a crossover where thermodynamic variables are continuos~\cite{Aoki:2006we,HotQCD:2014kol}. Such a connection for the anomalous coupling to gluons, and in particular for the KSVZ axion, was provided recently by Ref.~\cite{DEramo:2021psx}.

Contributions to axion production from processes involving baryons (e.g. nucleons) and heavy mesons (e.g. $K$ and $\eta$ mesons) are highly suppressed since $\Lambda_{\rm ChPT} \sim \mathcal{O}(100)\,{\rm MeV}$. The leading contribution, within the region where the ChPT formalism is reliable, comes from scatterings of pions that couple to the axion via the interactions
\be
\mathcal{L}^{(a\pi\pi\pi)}_{\rm KSVZ} = \frac{\partial_\mu a}{f_a} \, \frac{c_{a\pi\pi\pi}^{\rm KSVZ}}{f_\pi} \, \left(\pi^0\pi^+\partial^\mu \pi^- + \pi^0\pi^-\partial^\mu \pi^+ - 2 \pi^+\pi^-\partial^\mu \pi^0\right) \, .
\label{eq:KSVZapi}
\ee
Here, $c_{a\pi\pi\pi}^{\rm KSVZ} = (2/3) \tr [\lambda^3 c_q]$ with $\lambda^3$ a Gell-Mann matrix. We follow Ref.~\cite{GrillidiCortona:2015jxo} and take the average of the values provided by Refs.~\cite{deDivitiis:2013xla,Horsley:2015eaa,MILC:2015ypt}: we find $m_u/m_d = 0.48$ and $m_u/m_s = 0.024$ that leads to $c_{a\pi\pi\pi}^{\rm KSVZ} \simeq 0.12$. The processes producing axions in this regime are
\be
\pi^+ + \pi^-\rightarrow \pi^0 + a  \ , \qquad \pi^+ + \pi^0 \rightarrow \pi^+ + a \ , \qquad \pi^- + \pi^0 \rightarrow \pi^- + a \ .
\ee
We summarize their cross sections in App.~\ref{app:XS}.

\begin{figure}
\centering
\includegraphics[width=0.9\textwidth]{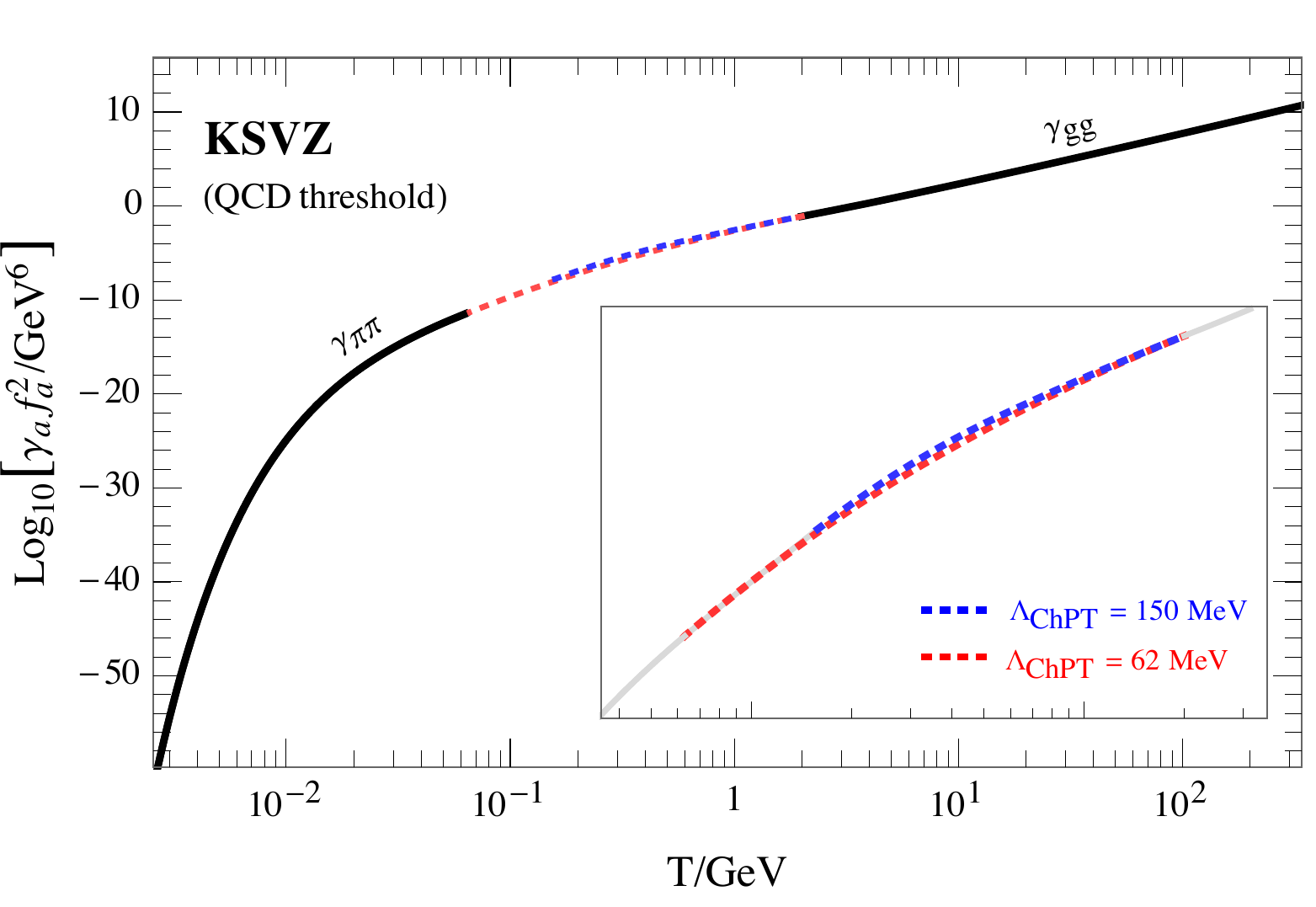}
\caption{\em KSVZ axion production rate across the QCDPT. Pion scatterings dominate at low temperatures ($T \lesssim \Lambda_{\rm ChPT}$) whereas thermal gluon scatterings control the rate where QCD is perturbative ($T \gtrsim \Lambda_{\rm N}$). We interpolate between these two regimes for the two different values $\Lambda_{\rm ChPT} = 62\,{\rm MeV}$ (dashed red) and $150\,{\rm MeV}$ (dashed blue).
}
\label{fig:KSVZlowE}
\end{figure}

Fig.~\ref{fig:KSVZlowE} illustrates the production rate across the QCDPT. Solid black lines show numerical results in the two regions where we have control of our calculations: low temperatures ($T \lesssim \Lambda_{\rm ChPT}$) by pion scatterings ($\gamma_{\pi\pi}$), and high temperatures ($T \gtrsim \Lambda_{\rm N}$) by thermal gluon scatterings ($\gamma_{gg}$). We interpolate them with the cubic `spline' method. The dashed red and dashed blue lines correspond to the best-fit interpolated results with $\Lambda_{\rm ChPT} =  62 \, {\rm MeV}$ and $150 \, {\rm MeV}$, respectively. They closely coincide with each other, and the consistency of the interpolation demonstrates confidence in our inference of a seamlessly connected rate.

\subsection{Summary: production rate for the KSVZ axion}

\begin{figure}
\centering
\includegraphics[width=0.9\textwidth]{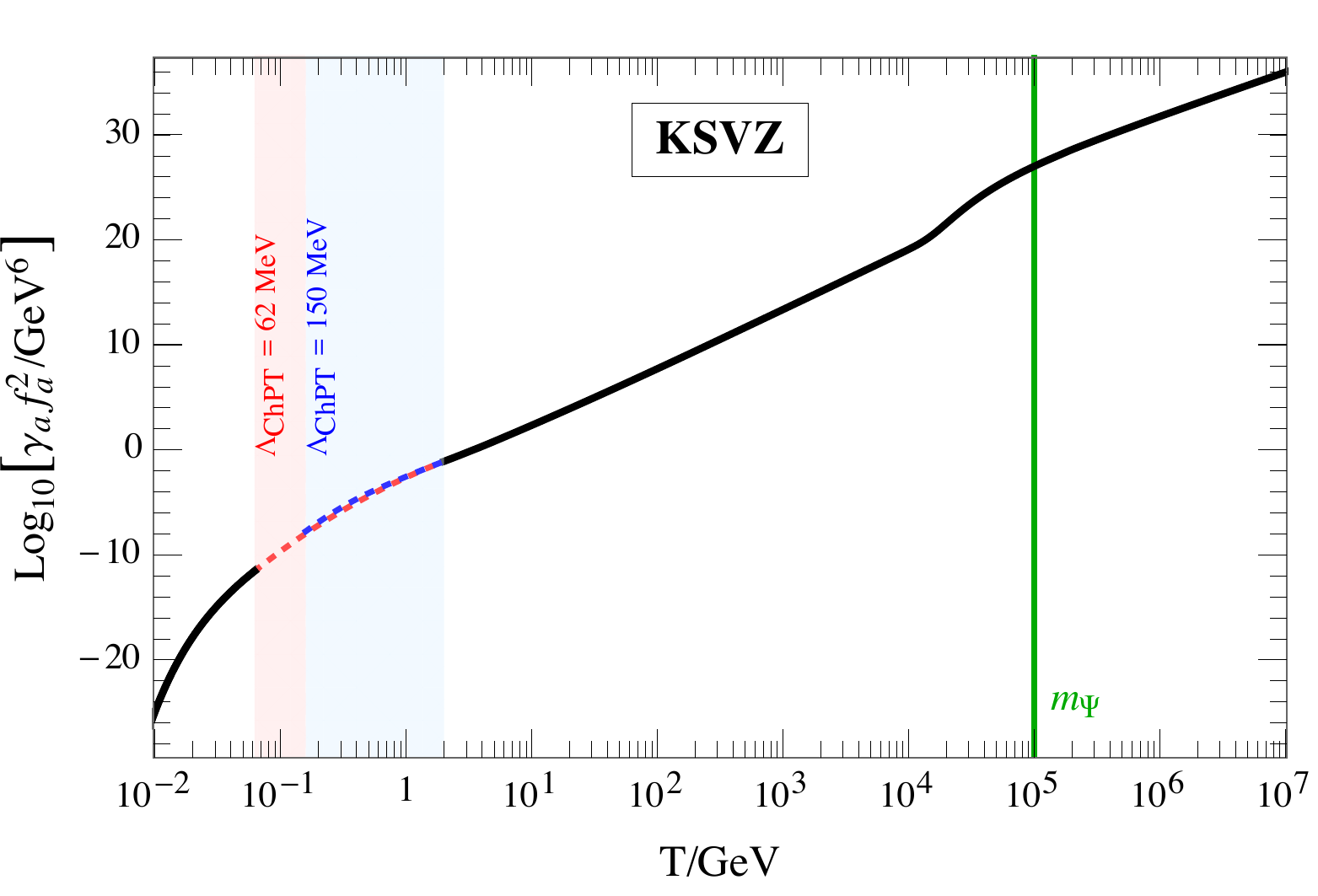}
\caption{\em KSVZ axion production rate for a heavy colored PQ fermion mass $m_{\Psi} = 10^5\,{\rm GeV}$.}
\label{fig:KSVZTot}
\end{figure}

We summarize the KSVZ axion production rate in the whole temperature range in Fig.~\ref{fig:KSVZTot}. Renormalizable interactions at large temperatures give $\gamma_\Psi \propto T^4$, gluon scattering mediated by a dimension 5 operator below $m_\psi$ give $\gamma_{gg} \propto T^6$. The rate drops below the QCDPT because of an exponential Maxwell-Boltzmann suppression for the pion number density.


\section{The DFSZ Axion}
\label{sec:DFSZ}

A complex scalar singlet $\varphi$ is also the starting point for the DFSZ framework. However, instead of introducing a new colored fermion we extend the SM Higgs sector with another weak doublet. We consider a two Higgs doublet model (2HDM) where the Higgs fields $H_u$ and $H_d$ carry opposite hypercharges $\pm 1/2$ and couple to up- and down-quarks, respectively. The essential 2HDM features needed for our discussion are summarized in App.~\ref{app:PP}.

The Lagrangian for scalars within the DFSZ framework takes the schematic form
\be
\mathcal{L}_{\rm DFSZ} = \left(D^\mu H_u \right)^\dagger D_\mu H_u + \left(D^\mu H_d \right)^\dagger D_\mu H_d + \left(\partial^\mu\varphi\right)^\dagger  \partial_\mu \varphi - V_{\rm DFSZ}\left(H_u, H_d, \varphi\right)  \ .
\ee
Unlike the previous case, SM fields are PQ-charged and in particular the combination $H_u H_d$ carries a non-vanishing PQ charge. The specific scalar potential is model-dependent, but it must ensure the spontaneous breaking of two symmetries: PQ at the scale $v_\varphi$, and electroweak at the scale $v$. The latter is due to the vevs of the Higgs field $H_u$ and $H_d$ that we parameterized as $v_u/\sqrt{2}$ and $v_d/\sqrt{2}$, respectively. Following a standard convention in the literature, we parameterize their ratio as $\tan\beta = v_u/ v_d$. Moreover, scalar potential interactions must couple the PQ breaking field $\varphi$ with the two Higgs doublets to have a solution to the strong CP problem. Options for this latter constraint include the renormalizable coupling $\varphi^{\dag 2} H_u^T i\sigma^2 H_d$ and the super-renormalizable coupling $\varphi^\dag H_u^T i\sigma^2 H_d$. We keep our discussion general and we parameterize this coupling as follows
\be
V_{\rm DFSZ}\left(H_u, H_d, \varphi\right) \supset B \left(\frac{\varphi^\dag}{v_\varphi/\sqrt{2}}\right)^{r} H_u^T i\sigma^2 H_d + {\rm h.c.} \ ,
\label{eq:VDFSZ}
\ee
where we introduce the vev of the complex scalar field $\vev{\varphi} = v_\varphi / \sqrt{2}$. Upon appropriate field redefinitions, it is always possible to set the $B$ parameter to be real and positive. The exponent $r$ is connected to the so-called domain wall number via the relation $N_{\rm DW} = 3 r$, and $N_{\rm DW}$ corresponds to the number of degenerate vacua of the axion potential. Finally, SM fermions have the following Yukawa interactions with the Higgs fields
\be
\mathcal{L}_{\rm Yukawa}^{\rm 2HDM-II} = 
 - \bar{Q}_L \, \widetilde{H}_u \, y^{(u)} u_R + \bar{Q}_L \, \widetilde{H}_d \, y^{(d)} d_R + \bar{L}_L \, \widetilde{H}_d \, y^{(e)} e_R + {\rm h.c.} \ .
\label{eq:Yukawa2HDM-II}
\ee
where $\tilde{H}_\alpha =i \sigma^2 H_\alpha^{*}$ and $y^{(\alpha)}$, with $\alpha = u, d, e$, are Yukawa matrices for the type-II 2HDM. 

At energies below PQ breaking, the phase of the scalar field $\varphi$ corresponds to the axion. As in the KSVZ scenario, there are the two typical ways to delineate the effective axion interactions to other fields. When the PQ symmetry is linearly-realized, $\varphi \rightarrow \left(v_\varphi / \sqrt{2}\right) e^{-i a/v_\varphi}$, the effective axion Lagrangian is given by
\be
\mathcal{L}_{\rm DFSZ}^{\rm linear} = \frac{1}{2}\partial^\mu a  \partial_\mu a - B \left[e^{ - i \frac{N_{\rm DW}}{3} \frac{a}{v_\varphi}} H_u^T i\sigma^2 H_d + {\rm h.c.}\right] \, .
\label{eq:DFSZlinear}
\ee

On the contrary, we can realize the PQ symmetry non-linearly via axion-dependent field redefinitions $\xi \rightarrow \exp\left[i q_{\xi} \frac{a}{v_\varphi}\right]\xi$, where we rotate all fields $\xi$ carrying a non-vanishing PQ charge $q_\xi$. The PQ invariance of the scalar potential in Eq.~\eqref{eq:VDFSZ} imposes the constraint $q_{H_u} + q_{H_d} = r = N_{\rm DW} / 3$. Likewise, the invariance of the Yukawa couplings in Eq.~\eqref{eq:Yukawa2HDM-II} imposes the relations among global charges: $q_{H_u} = -q_{Q_L}+q_{u_R}$, $q_{H_d} = - q_{Q_L} + q_{d_R}$, and $q_{H_d} = - q_{L_L} + q_{e_R}$. After these rotations, we find the Lagrangian
\be
\begin{split}
\mathcal{L}_{\rm DFSZ}^{\rm non-linear} = &\frac{1}{2}\partial^\mu a  \partial_\mu a - \frac{\partial_\mu a}{v_\varphi}\left[\sum_f q_f \bar{f}\gamma^\mu f + \sum_{\alpha} q_{H_\alpha} H_\alpha^\dagger i \overleftrightarrow{D}^\mu H_\alpha\right] \\
& + \frac{a}{v_\varphi}\left[N_{\rm DW}\frac{g_s^2}{32\pi^2}G_{\mu\nu}^A\widetilde{G}^{A \mu\nu} + c_W \frac{g^2}{32\pi^2} W_{\mu\nu}^I\widetilde{W}^{I\mu\nu} + c_Y \frac{g^{\prime 2}}{32\pi^2}B_{\mu\nu}\widetilde{B}^{\mu\nu}\right] \, ,
\end{split}
\label{eq:DFSZnonlinear}
\ee
where $f$ denotes SM fermions and we introduce the spin-one Higgs currents $H_\alpha^\dagger i \overleftrightarrow{D}_\mu H_\alpha = H_\alpha^\dagger 
(i D_\mu H_\alpha) - ( i D_\mu H_\alpha^\dagger ) H_\alpha$. The anomaly coefficients after these rotations can be determined through the general result in Eq.~\eqref{eq:anomalousrot}, and they explicitly read $c_W = - 9 q_{Q_L} - 3 q_{L_L}$, $c_Y = - q_{Q_L} + 8 q_{u_R} + 2 q_{d_R} - 3 q_{L_L} + 6 q_{e_R}$. As we discuss later, generic values of $q_{H_u}$ and $q_{H_d}$ induce after electroweak symmetry breaking a mixing between the axion and the longitudinal $Z$ weak gauge boson. Finally, we parameterize the axion anomalous coupling to photons below the weak scale in the standard form as follows $E \left( a /v_\varphi\right) \left(e^2/32\pi^2\right) F_{\mu\nu} \widetilde{F}^{\mu\nu}$. We extract if from the couplings in Eq.~\eqref{eq:DFSZnonlinear} and we find $E = (8/3) N_{\rm DW}$.

We work in the so-called decoupling limit where the extra Higgs bosons are much heavier than the weak scale. Indeed, the 2HDM is phenomenologically constrained to be in such a region to respect LHC bounds~\cite{Gunion:2002zf,Haber:2013mia,Espriu:2015mfa,Craig:2015jba}. There are three thresholds in this case. Two of them are analogous to the KSVZ scenario: the heavy Higgs bosons $m_A$, and the QCD non-perturbative scale $\Lambda_{\rm N} \sim 2 \,{\rm GeV}$. An additional threshold is the EWPT.

\subsection{Matching at the heavy Higgs bosons threshold}

Above the EWPT, the axion field lives entirely inside the phase of $\varphi$. The linear realization of the PQ symmetry, with interactions as in Eq.~\eqref{eq:DFSZlinear}, is the most convenient option to perform the rate calculation in this phase. The single axion coupling reads
\be
\mathcal{L}_{\rm DFSZ}^{\rm linear} \supset i\frac{a}{f_a} \frac{B}{3} \left(H^T_u i \sigma^2 H_d\right) + {\rm h.c.} \, ,
\label{eq:DFSZhighECoupling}
\ee
where the factor of $(1/3)$ comes from the normalization of the Wilson coefficient of the gluon anomaly operator, $v_\varphi = N_{\rm DW} f_a$, to reproduce the convention in Eq.~\eqref{eq:LPQaxion}.

Above the mass scale $m_A$, axion production is controlled by scatterings of Higgs bosons mediated by the interactions in Eq.~\eqref{eq:DFSZhighECoupling}. At temperatures below $m_A$, the number density of heavy Higgs bosons gets Maxwell-Boltzmann suppressed and axion production is due to scatterings of SM particles (including the lighter Higgs doublet corresponding to the SM-like Higgs). The interactions mediating scatterings at low temperatures can be found by integrating out the heavy scalars
\be
\begin{split}
\left.\mathcal{L}_{\rm DFSZ}^{\rm linear}\right|_{T < m_A} = i \frac{a}{f_a} & \, \left(  - \frac{\cos\alpha\cos\beta}{3}\bar{Q}_L \widetilde{H}_{\rm SM} \, Y^{(u)} u_R -  \frac{\sin\alpha\sin\beta}{3}\bar{Q}_L H_{\rm SM} \, Y^{(d)} d_R \, + \right.  \\  & \, \left. \qquad - \frac{\sin\alpha\sin\beta}{3}\bar{L}_L H_{\rm SM} \, Y^{(e)} d_R \right) + {\rm h.c.}
\label{eq:DFSZinterEffCoup}
\end{split}
\ee
with $H_{\rm SM}$ the SM-like Higgs doublet. The Yukawa matrices $y^{(\alpha)}$ for the 2HDM appearing in Eq.~\eqref{eq:Yukawa2HDM-II} and the correspondent $Y^{(\alpha)}$ defined in Eq.~\eqref{eq:LYukawaSM} are related as follows
\be
y^{(u)} = \frac{1}{\sin\beta} Y^{(u)} \ , \qquad y^{(d)} = \frac{1}{\cos\beta} Y^{(d)} \ , \qquad  y^{(e)} = \frac{1}{\cos\beta} Y^{(e)} \ .
\ee
The mixing angle $\alpha$ between the two doublets is a temperature dependent quantity and it is defined in Eq.~\eqref{eq:2HDMmixingAngle}. As the temperature drops below $m_A$, thermal corrections to the Higgs mass matrix become sub-dominant with respect to the overall mass scale $\sqrt{B}$. Hence the mixing angle $\alpha$ is approximated by $\beta$ and the mass eigenstates coincide nearly with those in the vacuum defined in Eq.~\eqref{eq:2HDMCharged}-\eqref{eq:2HDMReal}.

The interactions in Eq.~\eqref{eq:DFSZinterEffCoup} are equivalent, via appropriate field redefinitions, to the commonly used DFSZ axion interactions with SM fields parameterized as follows
\be
\mathcal{L}_{\rm DFSZ}^{\rm non-linear} = \frac{\partial_\mu a}{f_a} \left(c_{Q_L} \bar{Q}_L\gamma^\mu Q_L + c_{u_R}\bar{u}_R\gamma^\mu u_R + c_{d_R}\bar{d}_R\gamma^\mu d_R + c_{L_L} \bar{L}_L\gamma^\mu L_L + c_{e_R}\bar{e}_R\gamma^\mu e_R\right)
\label{eq:DFSZewEffCoup}
\ee
with $c_{Q_L} - \, c_{u_R} = \cos^2\beta/3$ and $c_{Q_L} - \, c_{d_R} = c_{L_L} - \, c_{e_R} = \sin^2\beta/3$. We point out how working with a linearly realized PQ symmetry and with axion interactions in Eq.~\eqref{eq:DFSZinterEffCoup} prevents any axion mixing with the $Z$ boson. The lack of such a mixing, which for the non-linear realization in Eq.~\eqref{eq:DFSZewEffCoup} must be achieved by hand, is automatic with our procedure. 

\begin{figure}
\centering
\includegraphics[width=0.9\textwidth]{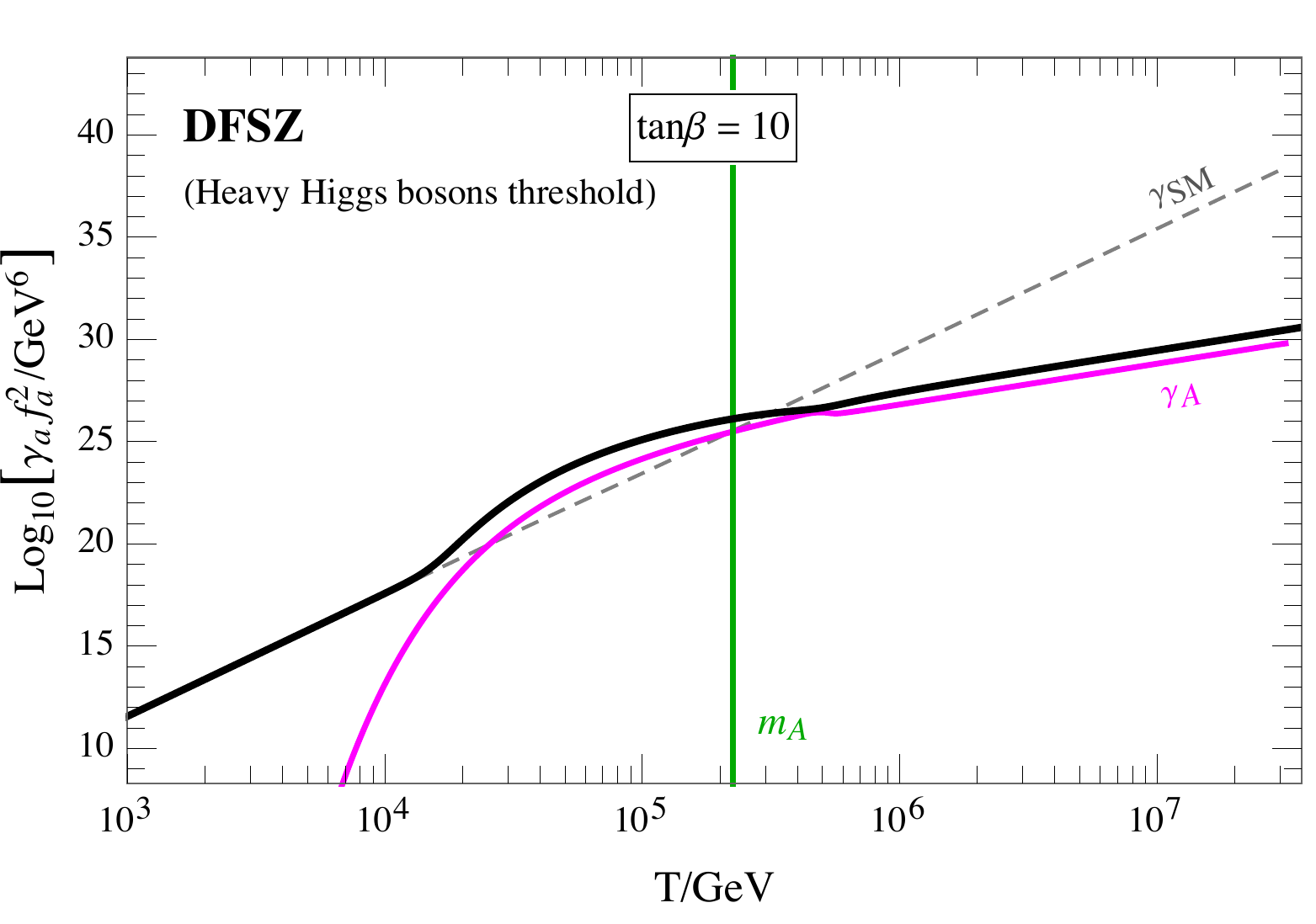}
\caption{\em DFSZ axion production rate across the heavy Higgs bosons thresholds. We set $\tan\beta=10$ and $\sqrt{2 B} = 10^5\,{\rm GeV} \simeq m_A$. The solid black line denotes the total rate.  The solid magenta line ($\gamma_A$) is the partial rate from scatterings of heavy Higgses. The dashed gray line ($\gamma_{\rm SM}$)  corresponds to the rate due to only SM particle scattering processes.}
\label{fig:DFSZhighE}
\end{figure}

The scatterings producing final state axions and their relative cross sections are summarized in App.~\ref{app:XS}, and they lead to the rate shown in Fig.~\ref{fig:DFSZhighE}. Consistently with our choice to work in the decoupling limit, we set $\sqrt{2 B} = 10^5\,{\rm GeV}$ and the resulting heavy Higgs bosons mass is around the same scale. We visualize this mass threshold with a vertical green line. The total rate is given by the solid black line. At temperatures larger than $m_A$, scatterings of heavy Higgs bosons dominate the total rate, and this is the contribution $\gamma_A$  that we denote with a solid magenta line. As expected, the magenta line drops exponentially at temperatures below $m_A$. Scatterings of SM particles control axion production below $m_A$. At temperatures much smaller than $m_A$, the rate can be evaluated either with the interactions in Eq.~\eqref{eq:DFSZinterEffCoup} or the ones in Eq.~\eqref{eq:DFSZewEffCoup}. As explained, once the temperature is much smaller than $m_A$ the temperature dependent angle $\alpha$ reaches the constant value $\beta$ and the two Lagrangians are equivalent. However, once we are not too far from $m_A$, the correct prescription is to evaluate axion production via Eq.~\eqref{eq:DFSZinterEffCoup}. For comparison, we report the rate computation obtained by using the non-linear realization in Eq.~\eqref{eq:DFSZewEffCoup} at all temperatures ($\gamma_{\rm SM}$, dashed gray line). As expected, it agrees with the full result at temperatures below $m_A$ but it is substantially different at large temperatures.

\subsection{Matching at the electroweak threshold}

We work in the decoupling limit and therefore we can match across the electroweak threshold. Axion interactions in Eq.~\eqref{eq:DFSZewEffCoup} are valid both above and below the EWPT and this is the field basis we employ to go across this threshold. We only consider the 2HDM parameter space with a smooth EWPT. If the mass of the heavy neutral pseudo-scalar $m_A$ defined in Eq.~\eqref{eq:2HDMPseudo} is much heavier than the SM Higgs and $Z$ boson, as it is the case for the decoupling limit, then the EWPT would be second order~\cite{Andersen:2017ika,Dorsch:2017nza}.

Similarly to the KSVZ scenario, the anomalous coupling to gluons mediates axion production, and the rate is given again by the expression in Eq.~\eqref{eq:F3function}. The UV origin for this interaction in the KSVZ scenario was the Yukawa operator of the heavy colored fermion $\Psi$. On the contrary, for the DFSZ scenario this operator originates from the Yukawa operators of SM quarks. The coupling $\tilde{c}^q_g(q^2)$ in the 1PI effective action receives threshold corrections from each quark as prescribed by Eq.~\eqref{eq:1PIgluonAnomaly}. The low-energy remnant once we integrate each quark can be read off Eq.~\eqref{eq:DFSZewEffCoup}: $c_g^u = \cos^2\beta/3$ and $c_g^d=\sin^2\beta/3$ for the each family of up-type quarks and down-type quarks, respectively. The production rate through the gluon anomaly vanishes above the EWPT, and it subsequently becomes more and more significant due to the accumulated effective 1PI gluon coupling.

Furthremore, quark scatterings via the couplings given in Eq.~\eqref{eq:DFSZewEffCoup} also contribute to axion production. Their cross sections are provided in App.~\ref{app:XS}. The matching across the EWPT for this class of interactions was spelled out in Ref.~\cite{Arias-Aragon:2020shv}. Axion production via SM fermion scatterings requires a chirality flip.
Above the EWPT, a chiral flipping can occur only via the Yukawa interactions in Eq.~\eqref{eq:DFSZinterEffCoup} so that only fermion scatterings with components of the Higgs doublet contribute to the axion production. On the contrary, after spontaneous electroweak symmetry breaking, the same Yukawa interactions provide quark masses that allow for chirality flips also for scatterings with gauge bosons, with gluons dominating the rate because of the hierarchy among the gauge coupling constants.

\subsection{Matching at the QCD threshold}

The procedure to investigate DFSZ axion production below the QCDPT is analogous to the one discussed in Sec.~\ref{sec:KSVZlowthreshold} for the KSVZ scenario. The leading order axion coupling to the strong sector in the KSVZ scenario originates only in the gluon anomalous term, whereas there are the additional axion interactions to the quark currents in the DFSZ scenario as given by Eq.~\eqref{eq:DFSZewEffCoup}. In other words, the effective axion interactions to the current of the light quarks ($u,d$, and $s$ below $\Lambda_{\rm N}$) can be written as Eq.~\eqref{eq:KSVZlowEqCoup} with the replaced coefficients
\be
\begin{split}
c_u & = \frac{m_u^{-1}}{2 \, {\rm Tr}\left[M_q^{-1}\right]} - \frac{\cos^2\beta}{6} \,,\\
c_d & = \frac{m_d^{-1}}{2 \, {\rm Tr}\left[M_q^{-1}\right]} - \frac{\sin^2\beta}{6}\,,\\ 
c_s & = \frac{m_s^{-1}}{2 \, {\rm Tr}\left[M_q^{-1}\right]} - \frac{\sin^2\beta}{6} \, .
\end{split}
\ee
Here, the first element for each coefficients comes from the gluon anomaly in common with the KSVZ scenario and the second one comes from the PQ charge of SM quarks.

Through the same matching procedure discussed in App.~\ref{app:PP}, we find the effective axion couplings to hadrons. We report here the ones to pions, which dominate the rate, and they are still given by the operator in Eq.~\eqref{eq:KSVZapi} but with the replaced coefficient 
\be
c_{a\pi\pi\pi}^{\rm DFSZ} = c_{a\pi\pi\pi}^{\rm KSVZ} - \frac{\cos 2\beta}{9} \ .
\ee

\begin{figure}
\centering
\includegraphics[width=0.9\textwidth]{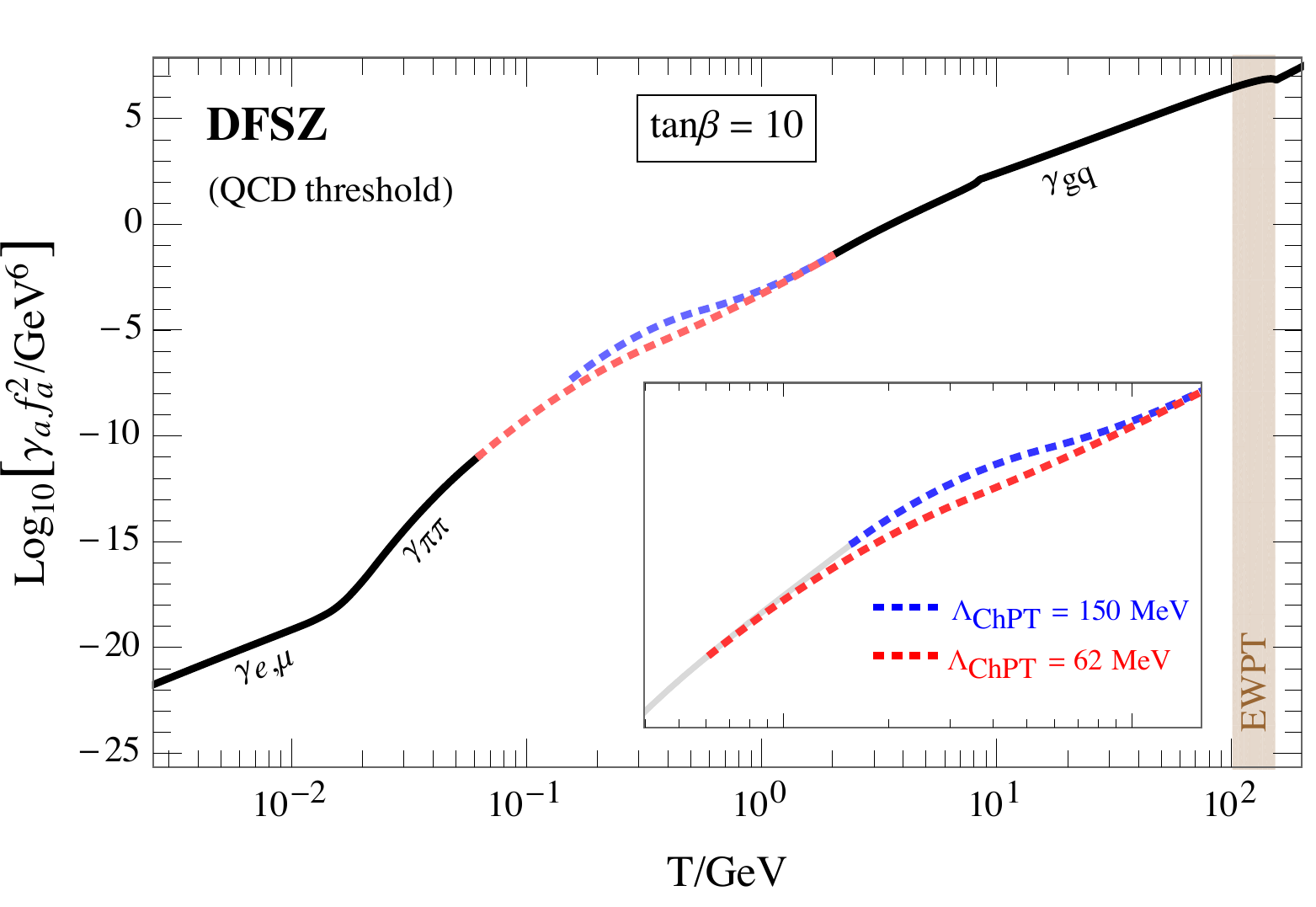}
\caption{\em DFSZ axion production rate across the QCDPT. We set $\tan\beta=10$. QCD is perturbative for $T>\Lambda_{\rm N} = 2\,{\rm GeV}$ and the rate is controlled by both gluon and quark scatterings. Right below confinement, but still above $T\sim 10\,{\rm MeV}$, pion scatterings ($\gamma_{\pi\pi}$) dominate. Below $T\lesssim 10\,{\rm MeV}$, lepton scatterings ($\gamma_{e,\mu}$) is the only production channel available. We interpolate for the two values $\Lambda_{\rm ChPT} = 62\,{\rm MeV}$ (dashed red) and $150\,{\rm MeV}$ (dashed blue). The brown region identifies the EWPT.}
\label{fig:DFSZlowE}
\end{figure}

Fig.~\ref{fig:DFSZlowE} shows the numerical result for the axion production rate across the QCDPT. The total rate is denoted by the solid black line. At temperatures right below the confinement scale, pion scatterings ($\gamma_{\pi\pi}$) dominate axion production. As discussed already for the KSVZ scenario, this evaluation for the rate is trustworthy only up to the cutoff $\Lambda_{\rm ChPT}$, and we interpolate the axion production rate between $\Lambda_{\rm ChPT}$  and $\Lambda_{\rm N} = 2 \, {\rm GeV}$. The dashed red and dashed blue lines correspond to the interpolations for $\Lambda_{\rm ChPT} = 62\,{\rm MeV}$ and $150\,{\rm MeV}$, respectively. Unlike the KSVZ scenario, these two interpolations give slightly different results for the DFSZ case. We will discuss the impact of the interpolation on cosmological observables in the next section.

The DFSZ axion also interacts with leptons via the effective couplings in Eq.~\eqref{eq:DFSZewEffCoup}. Perturbation theory can be employed at all temperatures for production via leptons since they do not carry color charge. The relevant scattering processes together with their cross sections are summarized in App.~\ref{app:XS}. As shown in Fig.~\ref{fig:DFSZlowE}, when the universe cools down much below $\Lambda_{\rm ChPT}$ (i.e., $T\ll m_\pi$), the pion contribution to the axion production rate diminishes exponentially and lepton scatterings ($\gamma_{e,\mu}$) become eventually dominant.

The bump arising near the EWPT (brown region) in the high temperature region of Fig.~\ref{fig:DFSZlowE} is the combination of several effects. Below the EWPT, the production rate is the sum of two contributions: thermal gluon scatterings via the axion anomalous coupling with a rate $\gamma_{gg} \propto T^6$, and bottom quark scatterings with gluons with a rate scaling as $\gamma_{b} \propto T^4$ (the different scaling is because the bottom mass provides the chirality flip). As we go above the weak scale, the thermal gluon scattering rate $\gamma_{gg}$ switches off, and top quark scatterings become available. However, we do not have fermions masses and therefore bottom and top quark scatterings (comparable since of $\tan\beta = 10$) lead to the scaling as $\gamma_{b,t} \propto T^6$. 

\subsection{Summary: production rate for the DFSZ axion}

\begin{figure}
\centering
\includegraphics[width=0.9\textwidth]{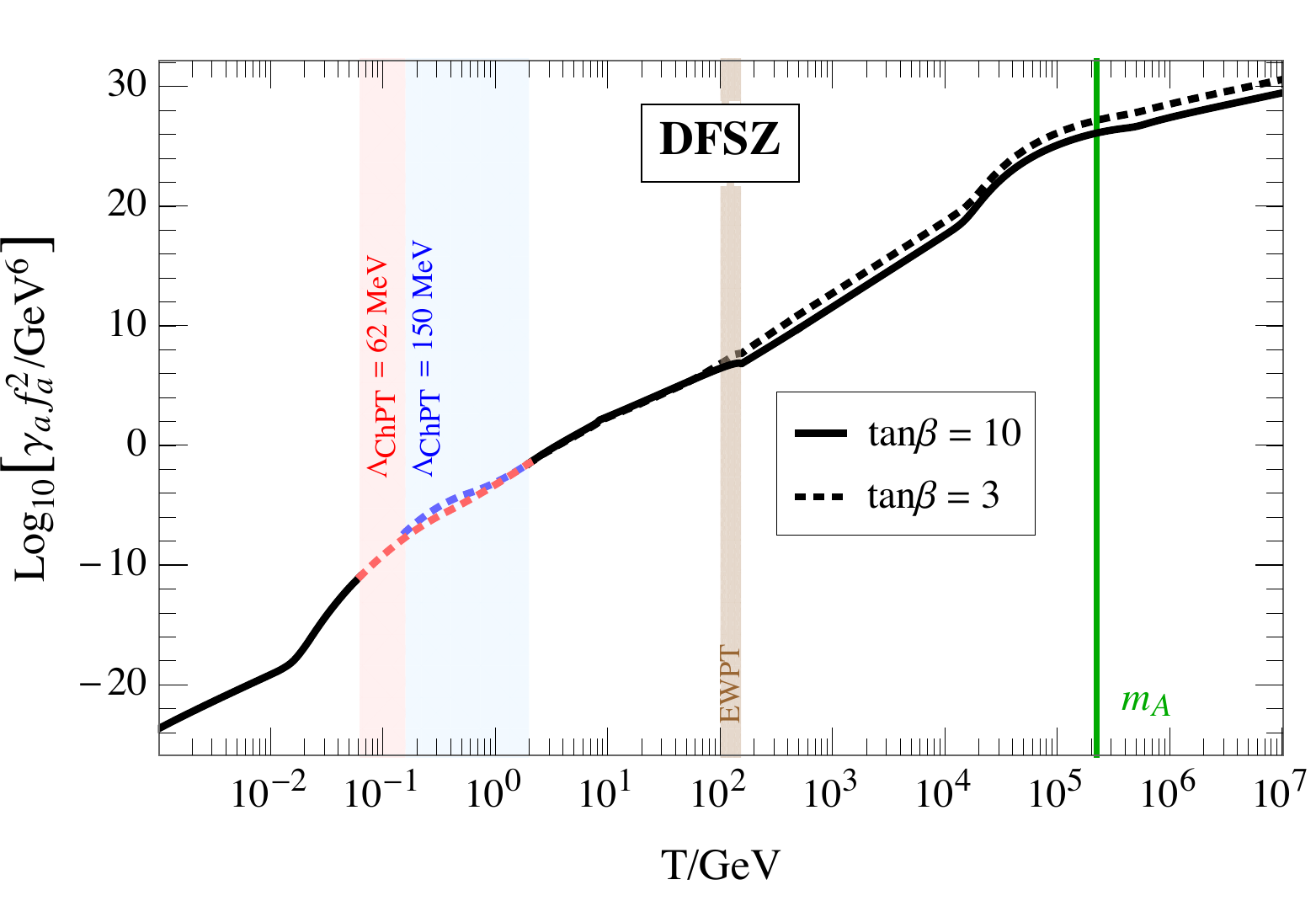}
\caption{\em DFSZ axion production rate for $m_A \simeq \sqrt{5}\times 10^5\,{\rm GeV}$, $\tan\beta= (10, 3)$ (solid, dashed).}
\label{fig:DFSZTot}
\end{figure}

We summarize the production rate for the DFSZ axion in the whole temperature range in Fig.~\ref{fig:DFSZTot}. Besides providing results for $\tan\beta = 10$ (solid black), as done already for the previous figures, we show the rate also for $\tan\beta = 3$ (dashed black). To ease the comparison, we set $B$ to reproduce the same heavy Higgs boson mass in the two cases, $m_A \simeq \sqrt{B\tan\beta} = \sqrt{5}\times 10^{5}\,{\rm GeV}$. Similarly to the KSVZ case, axion production is controlled by renormalizable interactions above the heaviest threshold, the heavy Higgs boson mass in the DFSZ scenario, and the rate consequently scales as $\gamma_A \propto T^2$. At temperatures below $m_A$ but still above the EWPT, axion production processes proceed via dimension 5 operators coupling the axion to SM fermions (with top and bottom dominating), and therefore the rate scales as $\gamma_{t,b} \propto T^6$.  Axion couplings to top quarks exhibit the $\tan^2\beta$ suppression discussed previously, and as a consequence the production rate with $\tan\beta = 3$ at temperatures between $m_A$ and the EWPT is relatively larger than the one for $\tan\beta = 10$. Below the EWPT, quark scatterings with gluons are dominant but top quarks quickly disappear from the bath, and the production rate is almost independent on $\tan\beta$. Contrarily to the previous case, below the weak scale the SM fermion scattering rate scales as $\gamma_{t,b} \propto T^4$, and in this region it dominates over the thermal gluon scattering that becomes active below the top quark mass with scaling $\gamma_{gg} \propto T^6$. Pion scatterings dominate below the QCDPT and before we hit the Maxwell-Boltzmann suppression. Another important difference with respect to the KSVZ scenario is that the production is active even below the QCDPT since the axion couples to leptons, and interactions with muons and electrons give a rate with the scaling $\gamma_{e, \mu} \propto T^4$.


\section{QCD Axion Dark Radiation}
\label{sec:results}

Scatterings of thermal bath particles produce axions in the final state, and the typical energy involved in each process is the bath temperature: the produced axions carry a kinetic energy much larger than their mass and therefore they are ultra-relativistic. What happens next? Initially, there are not enough axions to give the inverse (axion destruction) process and to ensure ultimately thermal equilibrium; axions just free-streams with their momentum decreasingly as the inverse scale factor. If axion production is efficient, we produce enough of them to thermalize with the primordial bath until the universe gets too cold and diluted, and they decouple with a relativistic thermal abundance exactly as neutrinos do.

A natural and useful application of the rates computed in Secs.~\ref{sec:KSVZ} and \ref{sec:DFSZ} is keeping track of the axion abundance. Our conceptual starting point is an early universe going through an inflationary expansion, and inflaton decays generate the thermal bath afterward. Our only assumption is that inflationary reheating ends at high scales, and in particular the primordial thermal bath of relativistic SM particles dominates the energy budget earlier than the EWPT. However, we do not commit to any explicit hypothesis about axion production during inflationary reheating, but we consider two opposite cases in our analysis: we end reheating with no axions whatsoever, or we begin the radiation dominated era with a full thermal axion abundance. These two extremes cover all the options in between. 

The quantitative tool to track the axion abundance is the Boltzmann equation
\be
\frac{d n_a}{d t} + 3 H n_a = \mathcal{C}_a \ .
\label{eq:BE1}
\ee
Here, $n_a$ is the axion number density and $t$ is the cosmic time. The number density dilution due to the Hubble expansion is accounted for by the term on the left-hand side proportional to the Hubble parameter $H$; in the absence of interactions, this is the only effect changing the axion number density. If number changing processes happen at an appreciable rate, we have to include their effects through the collision term $\mathcal{C}_a$ on the right-hand side. 

We focus here on collisions producing one axion in the final state. For this class of processes, which is by far dominant as a consequence of the tiny couplings, the general expression for the collision term takes the form
\be
\mathcal{C}_a = \gamma_a \left( 1 - \frac{n_a}{n_a^{\rm eq}} \right) \ .
\ee
Here, the function $\gamma_a$ is the total axion production rate which is the sum of several contributions, one for each process that we account for. If we consider thermal gluon scatterings the rate is given by the expression in Eq.~\eqref{eq:F3function}. For a generic binary collision 
\be
\mathcal{B}_i \, \mathcal{B}_j \; \rightarrow \; \mathcal{B}_k \, a 
\ee
with $\mathcal{B}_i$ a bath particle (SM or beyond the SM), the associated rate reads
\be
\gamma_{i j \rightarrow k a} = n_i^{\rm eq} n_j^{\rm eq} \langle \sigma_{\mathcal{B}_i \mathcal{B}_j \rightarrow \mathcal{B}_k a} v_{\rm rel} \rangle\,.
\label{eq:rateij}
\ee
Initial state bath particles are in thermal equilibrium, the scattering cross section is multiplied by the Moeller velocity, and the brackets denote a thermal average over all possible initial states. We use the Maxwell-Boltzmann statistics for the equilibrium density of bath particles since quantum degeneracy effects lead to negligible corrections
\begin{eqnarray}
n_{i}^{\rm eq}=\frac{g_i m_i^2 T}{2\pi^2} K_2\left(\frac{m_i}{T}\right)  \ .
\end{eqnarray}
The particle mass and internal degrees of freedom are denoted by $m_i$ and $g_i$, respectively, and the second kind modified Bessel functions are denoted by $K_i(x)$. The explicit expression for the thermally averaged cross section reads~\cite{Gondolo:1990dk}
\begin{eqnarray}
\label{rateinter}
\gamma_{i j \rightarrow k a}  = \frac{g_i g_j T}{32 \pi^4} \int_{s_{\rm min}}^{\infty}ds \frac{\lambda(s,m_i,m_j)}{\sqrt{s}}\sigma_{i j \rightarrow k a}(s) K_1\left(\frac{{\sqrt{s}}}{T}\right) \ .
\end{eqnarray}
The integral accounts for all the possible squared center of mass energies $s$ in the collision with cross section $\sigma_{i j \rightarrow k a}(s)$. The lower integration extreme corresponds to the kinematical threshold $s_{\rm min}= {\rm Max}\left[(m_i+m_j)^2,m_k^2\right]$, and the K{\"a}ll{\'e}n function $\lambda$ is defined as follows
\be
\lambda(s,m_i,m_j)\equiv\left[s-(m_i+m_j)^2\right]\left[s-(m_i-m_j)^2\right]\,.
\ee

It is convenient to rewrite the Boltzmann equation in terms of dimensionless quantities. We trade $n_a$ with the comoving number density $Y_a = n_a / s_R$, where $s_R$ is the entropy density of the thermal bath. Other than being dimensionless, the comoving number density is advantageous because it scales out the effect of the Hubble expansion and therefore it varies only if number changing processes are in action. Likewise, we replace the time evolution variable $t$ with the dimensionless inverse temperature $x = M / T$. Here, the choice for the scale $M$ is purely conventional. Upon using the general result provided in Eq.~\eqref{eq:xvstder} of App.~\ref{app:CO}, which is a consequence of entropy conservation, we trade $t$ with $x$ and rewrite the Boltzmann equation in terms of dimensionless quantities 
\be
\frac{d Y_a}{d \log x} = \left( 1 - \frac{1}{3} \frac{d \log g_{*{s}}}{d \log x} \right) \frac{\gamma_a(x)}{H(x) s_R(x)} \left( 1 - \frac{Y_a}{Y_a^{\rm eq}} \right) \ . 
\label{eq:BE2}
\ee

Our ultimate goal is to quantify how axions contribute to $\Delta N_{\rm eff}$. Regardless of the details of axion production, there will be a point where Hubble expansion takes over the production, and this can happen because of two reasons. Particles participating in axion production can be massive, and as the bath temperature decreases number densities get exponentially suppressed. Even if production is mediated by massless particles, the universe gets too cold and diluted eventually to give appreciable interactions within a Hubble time. Such a freeze-out of interactions happens when the bath temperature was $T_{\rm F.O.}$, long before the CMB formation, and the axion comoving density freezes to a constant value
\be
Y_a(T \leq T_{\rm F.O.}) = Y_a^\infty = {\rm constant} \ .
\label{eq:Yainfty}
\ee
Finding such an asymptotic value is the goal of our Boltzmann equation analysis. Once we have it, we evaluate $\Delta N_{\rm eff}$ via the general relation in Eq.~\eqref{eq:DeltaNeffgeneral} that for the axion reads
\be
\Delta N_{\rm eff} = \frac{4}{7} \left( \frac{11}{4} \right)^{4/3} \; 
\left[\frac{\frac{2 \pi^4}{45 \zeta(3)} \, g^{\rm SM}_{*s}(T_{\rm CMB}) \, Y_a^\infty}{1 - \frac{2 \pi^4}{45 \zeta(3)} \, Y_a^\infty} \right]^{4/3} \ .
\label{eq:YvsNeffAxion}
\ee
Here, $g^{\rm SM}_{*s}(T_{\rm CMB})$ is the SM contribution to the effective number of entropic degrees of freedom, and the second term in the denominator accounts for the axion contribution to the energy density. As explained in App.~\ref{app:CO}, this correction can be at most $1 / (1 + g^{\rm SM}_{*s}(T_{\rm F.O.}))$ and therefore becomes more relevant for late axion production. The SM contribution cannot be less than approximately 4, hence the correction can be at most $25 \%$. However, axions are produced well above the MeV scale for most of the parameter space we explore, and our complete results are well described by the approximated expression
\be
\Delta N_{\rm eff} \simeq \frac{4}{7} \left( \frac{11}{4} \right)^{4/3} \; 
\left[ \frac{2 \pi^4}{45 \zeta(3)} \, g^{\rm SM}_{*s}(T_{\rm CMB}) \, Y_a^\infty \right]^{4/3} \ .
\label{eq:YvsNeffAxionApprox}
\ee

\subsection{KSVZ axion}
\label{subsec:NeffKSVZ}

\begin{figure}
\centering
\includegraphics[width=0.9\textwidth]{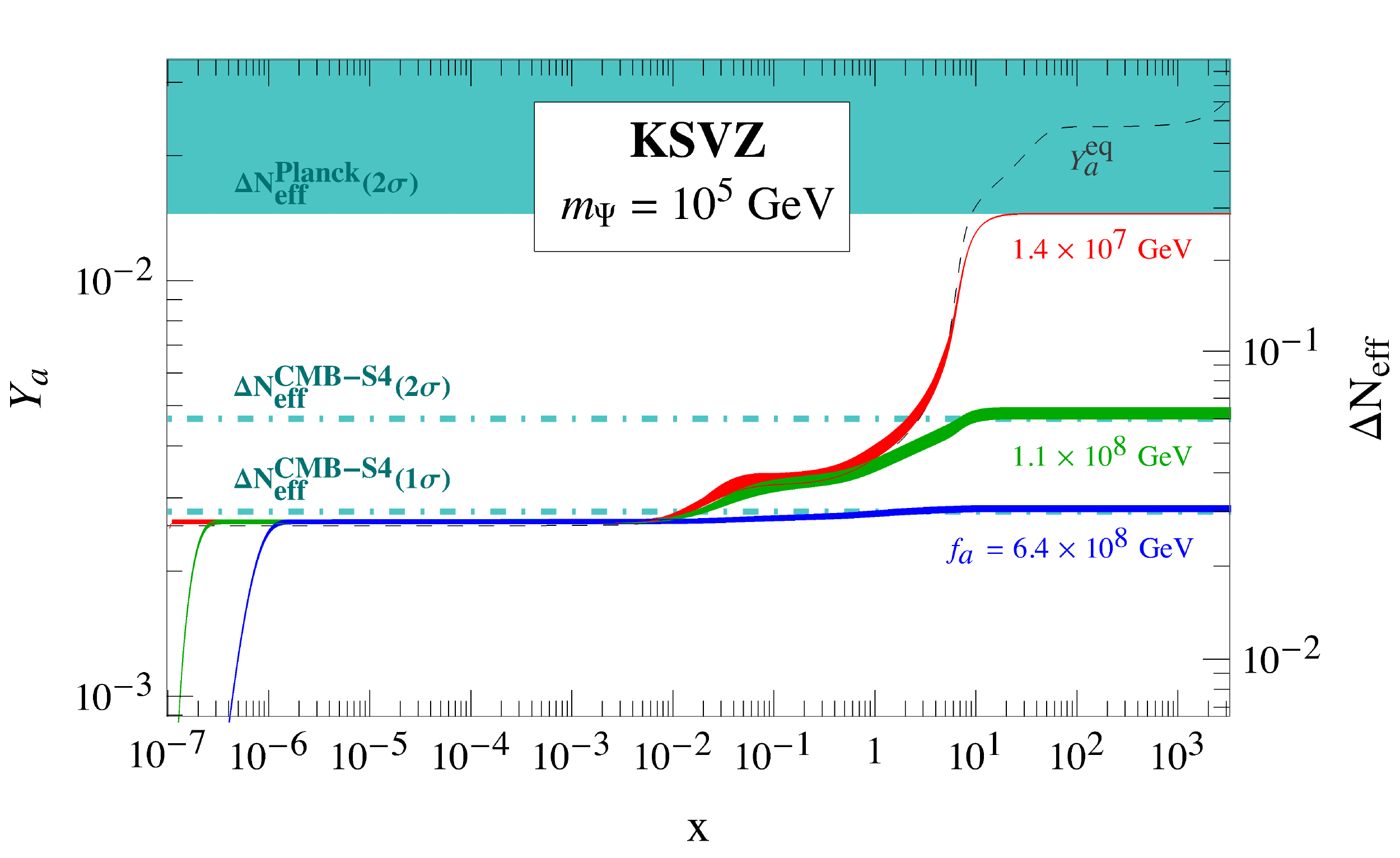}
\caption{\em Axion comoving number density $Y_a$ as a function of $x = {\rm GeV} / T$ for the KSVZ axion. The three colors for the solid lines correspond to different axion decay constants $f_a$. Each solution is presented as a band whose width corresponds to two different treatments of the thermal bath relativistic degrees of freedom. The 
dashed black line describes the equilibrium distribution. The vertical axis on the right identifies the contribution to $\Delta N_{\rm eff}$.}
\label{fig:KSVZ_Yvsx}
\end{figure}

The production rate for the KSVZ axion, with a smooth treatment of the heavy colored PQ fermion and the QCDPT thresholds, is shown in Fig.~\ref{fig:KSVZTot} as a function of the temperature. We feed the Boltzmann equation in Eq.~\eqref{eq:BE2} with this rate, and we show results from numerical integrations in Fig.~\ref{fig:KSVZ_Yvsx}. We set the mass of the heavy PQ fermion to $m_\Psi = 10^5 \, {\rm GeV}$, as done already in Sec.~\ref{sec:KSVZ}, and we run our code starting from an initial temperature $T_i = 10^7 \, {\rm GeV}$. Thus we always go across the $\Psi$ threshold. Furthermore, we set $n_a(T_i) = 0$ as the initial condition to produce this figure, and this choice has no impact on our final results as we discuss below. We employ the dimensionless combination $x = M / T$ as the evolution variable, and we set $M = 1 \, {\rm GeV}$ to have the QCDPT around the region $x \simeq 1$. 

Different colors denote different choices for the axion decay constant $f_a$ which is the only free parameter within the KSVZ framework. As a matter of fact, each solution is not a line but rather a band whose width is due to the two different datasets~\cite{Drees:2015exa,Saikawa:2018rcs} that we employ for the temperature evolution of the effective relativistic bath degrees of freedom. Finally, the dashed black line denotes the axion equilibrium comoving density whose analytical expression is rather simple, $Y^{\rm eq}_a(x) =  45 \, \zeta(3) / (2 \pi^4 \, g_{*s}(x))$, and whose temperature dependence is only due to the change in the effective entropic degrees of freedom of the thermal bath. The three solid lines reach the thermal equilibrium distribution rather quickly, as early as $T \simeq 10^6 \, {\rm GeV}$, and therefore setting the initial condition $n_a(T_i) = 0$ has no impact on the final results for the values of $f_a$ chosen in Fig.~\ref{fig:KSVZ_Yvsx}.

At small enough temperatures (i.e., large $x$), the number density reaches an asymptotic value in agreement with Eq.~\eqref{eq:Yainfty}. We include a second vertical axis on the right of Fig.~\ref{fig:KSVZ_Yvsx} to identify the corresponding value of $\Delta N_{\rm eff}$ as quantified by Eq.~\eqref{eq:YvsNeffAxion}. This contribution increases as we go to lower values of $f_a$, consistently with the picture that larger axion couplings keep physical processes efficient at lower temperatures (see Eq.~\eqref{eq:DeltaNeffIntro} and Fig.~\ref{fig:neffdec}). 

Our choices for the axion decay constant correspond to generating a $\Delta N_{\rm eff}$ equal to the Planck bound at $2 \sigma$ (red), $2\sigma$ and $1\sigma$ for CMB-S4 surveys (green and blue, respectively). Astrophysical constraints bound the axion decay constant from below~\cite{Fischer:2016cyd,Chang:2018rso,Carenza:2019pxu}. We impose the bound from SN1987A provided by the recent Ref.~\cite{ParticleDataGroup:2020ssz} that for the KSVZ axion results in $f_a \gtrsim 1.4 \times 10^8 \, {\rm GeV}$. Neutron star cooling provides bounds in the same ballpark~\cite{Hamaguchi:2018oqw,Leinson:2021ety}. Thus current Planck bounds on $\Delta N_{\rm eff}$ are sentitive to KSVZ axions with $f_a$ one order of magnitude below the stellar exclusion bound, and future CMB-S4 surveys will probe the range $f_a \sim (10^8, 10^9) \, {\rm GeV}$ that is still not in conflict with any experimental constraint.

\begin{figure}
\centering
\includegraphics[width=0.9\textwidth]{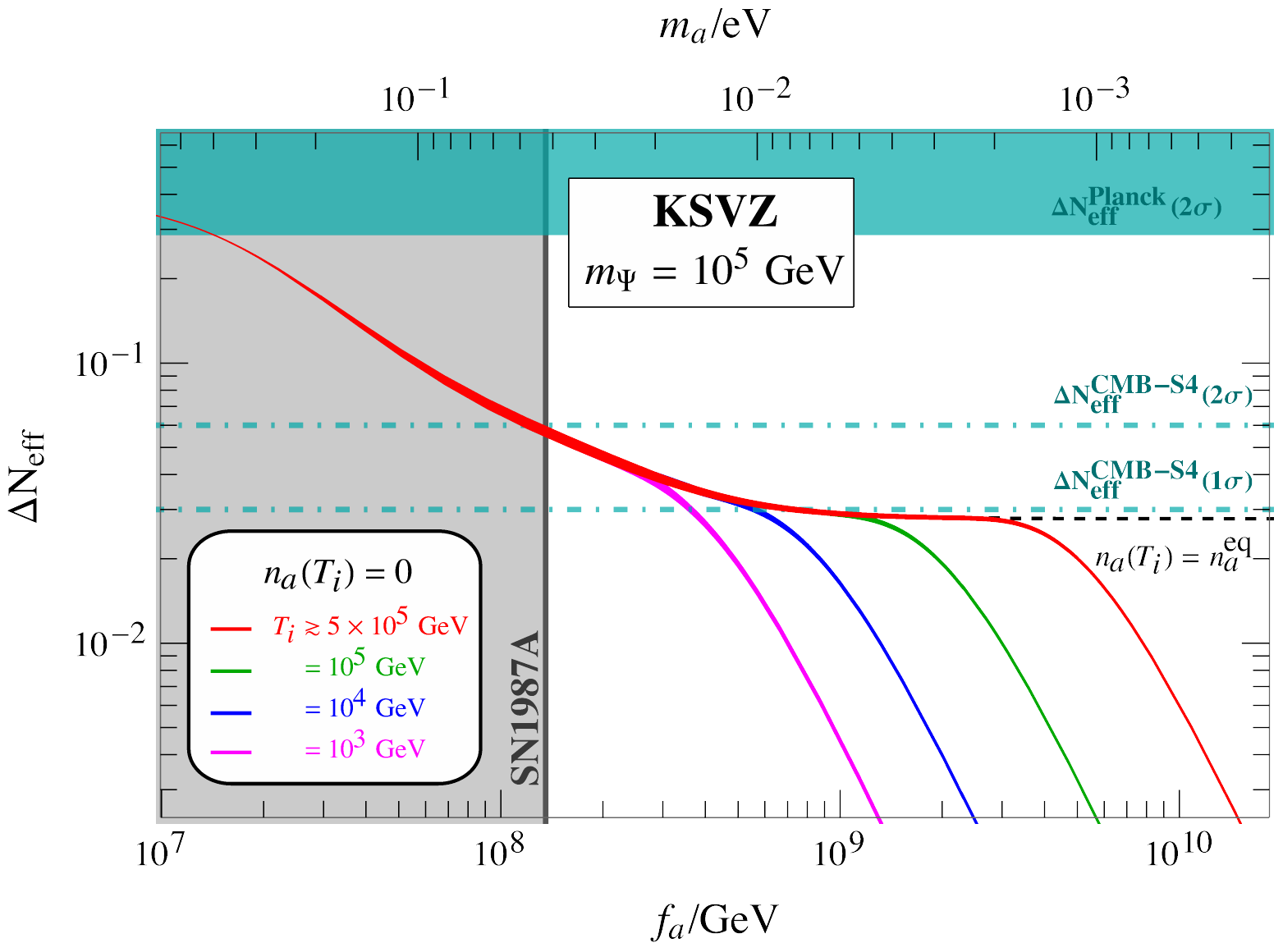}
\caption{\em Contribution to $\Delta N_{\rm eff}$ as a function of the axion decay constant $f_a$ for the KSVZ axion.}
\label{fig:KSVZ_Neffvsfa}
\end{figure}

We investigate how $\Delta N_{\rm eff}$ depends on $f_a$ in Fig.~\ref{fig:KSVZ_Neffvsfa}. We solve the Boltzmann equation again with initial condition $n_a(T_i) = 0$, but we consider a few different values for $T_i$ corresponding to the different solid colored band (whose width quantifies our uncertainties due to different treatments of the bath). The red line shows $\Delta N_{\rm eff}$ as a function of $f_a$ for any value of $T_i$ much larger than the PQ fermion mass that we keep $m_\Psi = 10^5 \, {\rm GeV}$ as in the previous plots. Axion production at temperatures above $m_\Psi$ is controlled by a renormalizable coupling, and the rate normalized by the $\Psi$ number density scales as $\gamma_a / n_\Psi \propto T$ as long as $\Psi$ are relativistic. This has to be compared with the Hubble expansion rate that scales as $H \simeq T^2 / M_{\rm Pl}$. Thus axion production is most efficient at low temperatures, and most axions coming from $\Psi$ scatterings are created at temperatures around $m_\Psi$.\footnote{This ``IR domination'' is the same as the one for dark matter \textit{freeze-in}~\cite{Hall:2009bx}.} On the contrary, below the heavy PQ fermion we have a rate normalized by the bath number density scaling as $\gamma_a / n_{\mathcal{B}} \propto T^3$, and this temperature behavior is stronger than the one for the Hubble rate: axion production is most efficient in the UV at the highest temperature available $T_i$. This explains the different results for large values of $f_a$: axions do not have enough interaction strength to thermalize in the early universe, and smaller initial temperatures lead to smaller $\Delta N_{\rm eff}$ because at low temperatures the production is less efficient. At low enough values of the axion decay constant thermalization is achieved, and all colored lines coincide: the resulting prediction for $\Delta N_{\rm eff}$ does not depend on $T_i$.

We provide in Fig.~\ref{fig:KSVZ_Neffvsfa} also the prediction for $\Delta N_{\rm eff}$ once we set the thermal equilibrium distribution as initial condition for the axion number density at temperatures above the weak scale (dashed gray line). Large axion decay constants, $f_a \gtrsim 3 \times 10^9 \, {\rm GeV}$, lead to $\Delta N_{\rm eff} \simeq 0.027$ which is the value associated with a spin-0 particle that was once in thermal equilibrium and decoupled above the weak scale (green line in Fig.~\ref{fig:neffdec} at large $T_D$). In this range of $f_a$, we set the initial abundance to the equilibrium value by hand and interactions are completely harmless. Things are different as we approach lower $f_a$ since couplings get stronger and they can keep the axion in equilibrium below the weak scale. The resulting prediction for $\Delta N_{\rm eff}$ coincides with the solid red band for $f_a \lesssim 3 \times 10^9 \, {\rm GeV}$ regardless of the initial value of $T_i$ as long as we keep $T_i \gtrsim 1 \, {\rm TeV}$ (for lower values of $T_i$ the expected $\Delta N_{\rm eff}$ would be larger, see Eq.~\eqref{eq:DeltaNeffIntro}). This result can be understood from the plot in Fig.~\ref{fig:KSVZ_Yvsx}: for the axion decay constant range we are interested in, axions always reach equilibrium long before the time when the bath temperature gets to the TeV scale. Thus in the physical region of our interest where the signal is detectable, $10^8 \, {\rm GeV} \lesssim f_a \lesssim 10^9 \, {\rm GeV}$, our predictions for $\Delta N_{\rm eff}$ do not depend on the initial condition for the axion number density. 

\subsection{DFSZ axion}
\label{subsec:NeffDFSZ}

\begin{figure}
\centering
\includegraphics[width=0.9\textwidth]{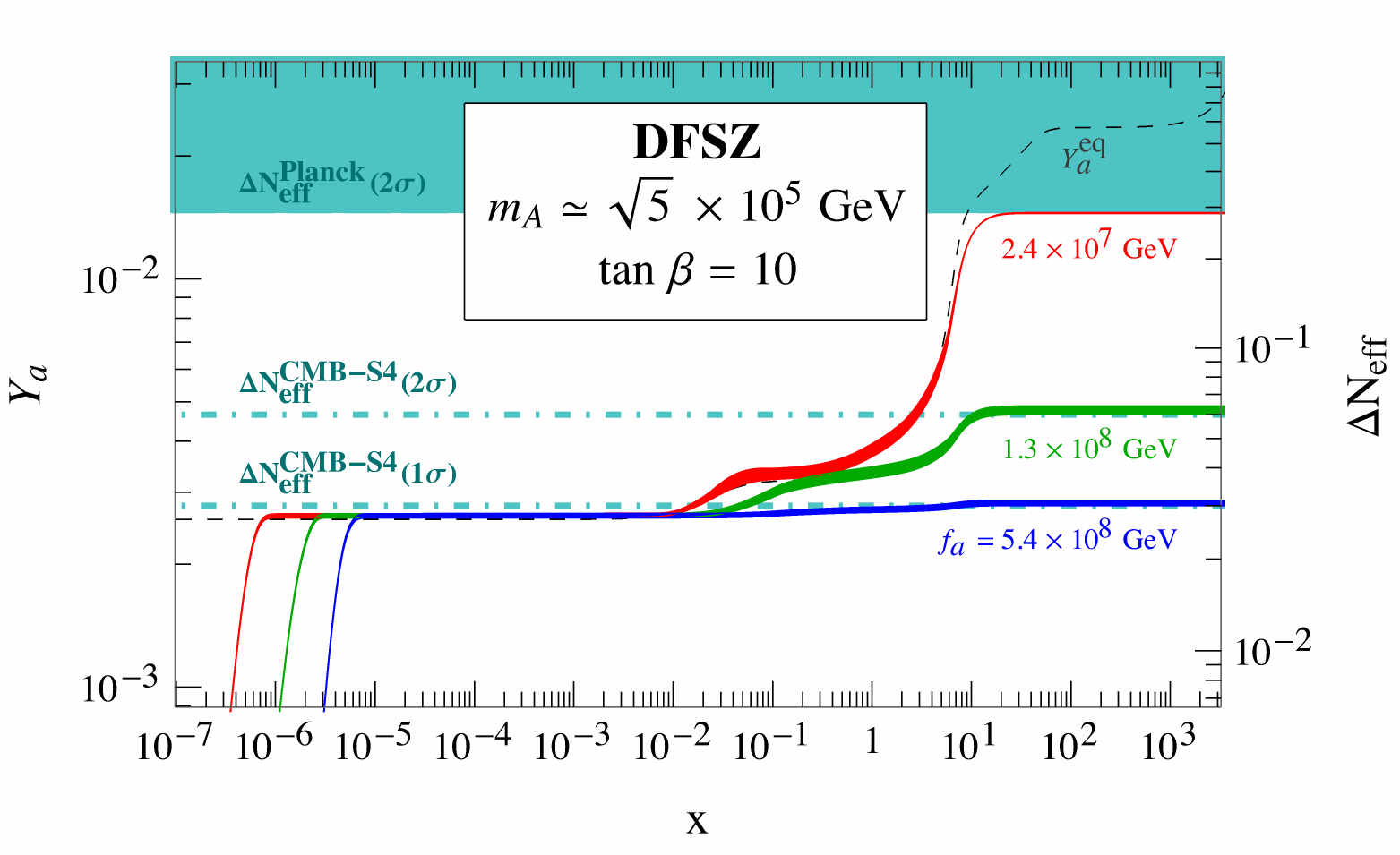}
\caption{\em Axion comoving number density $Y_a$ as a function of $x = {\rm GeV} / T$ for the DFSZ framework. Notation as in Fig.~\ref{fig:KSVZ_Yvsx}.}
\label{fig:DFSZ_Yvsx}
\end{figure}

We now turn to the DFSZ framework. The production rate, this time with smooth treatments of three different mass thresholds, is shown in Fig.~\ref{fig:DFSZTot} as a function of the temperature. Exactly as we just did for the KSVZ axion, we feed the Boltzmann equation in Eq.~\eqref{eq:BE2} with this rate, find the asymptotic value of the axion comoving number density and quantity the correspondent contribution to $\Delta N_{\rm eff}$. We fix the model parameters to the same values as in Sec.~\ref{sec:DFSZ}, $\tan\beta = 10$ and $m_A \simeq \sqrt{2 B} = 10^5 \, {\rm GeV}$, and we run our Boltzmann code again starting from an initial temperature $T_i = 10^7 \, {\rm GeV}$; this ensures that we pass again all the mass thresholds in the scenario under investigation. The numerical output of the differential equation integrations is shown in Fig.~\ref{fig:DFSZ_Yvsx}. The axion number density reaches its equilibrium value rather quickly, and therefore setting its initial value to zero does not impact our final results. At low temperatures, consistently with our previous discussion, the comoving number density settles to a constant value. We choose again three numerical values for the axion decay constant leading to $\Delta N_{\rm eff}$ equal to the Planck bound and the projected sensitivities of future CMB experiments. They are in the same ballpark as the ones for the KSVZ scenario. 

Astrophysical bounds are more severe for this case. Data from SN 1987A~\cite{Carenza:2019pxu} constrain again the axion decay constant, but this time the numerical value associated to the bound is slightly different because the DFSZ axion couples also to quarks. For the $\tan\beta$ chosen in this analysis, we find $f_a \gtrsim 1.9 \times 10^8 \, {\rm GeV}$. However, this is not the leading bound since the DFSZ axion couples to electrons as well. Studies of red giants~\cite{Viaux:2013lha} and white dwarfs~\cite{MillerBertolami:2014rka} provide competitive bounds, with the one coming from the latter slightly stronger. For the DFSZ parameter chosen in our analysis, this corresponds to the bound on the axion decay constant $f_a \gtrsim 5.2 \times 10^8 \, {\rm GeV}$. Thus future CMB-S4 surveys will probe a rather small region of the DFSZ parameter space that is still not excluded. 

\begin{figure}
\centering
\includegraphics[width=0.9\textwidth]{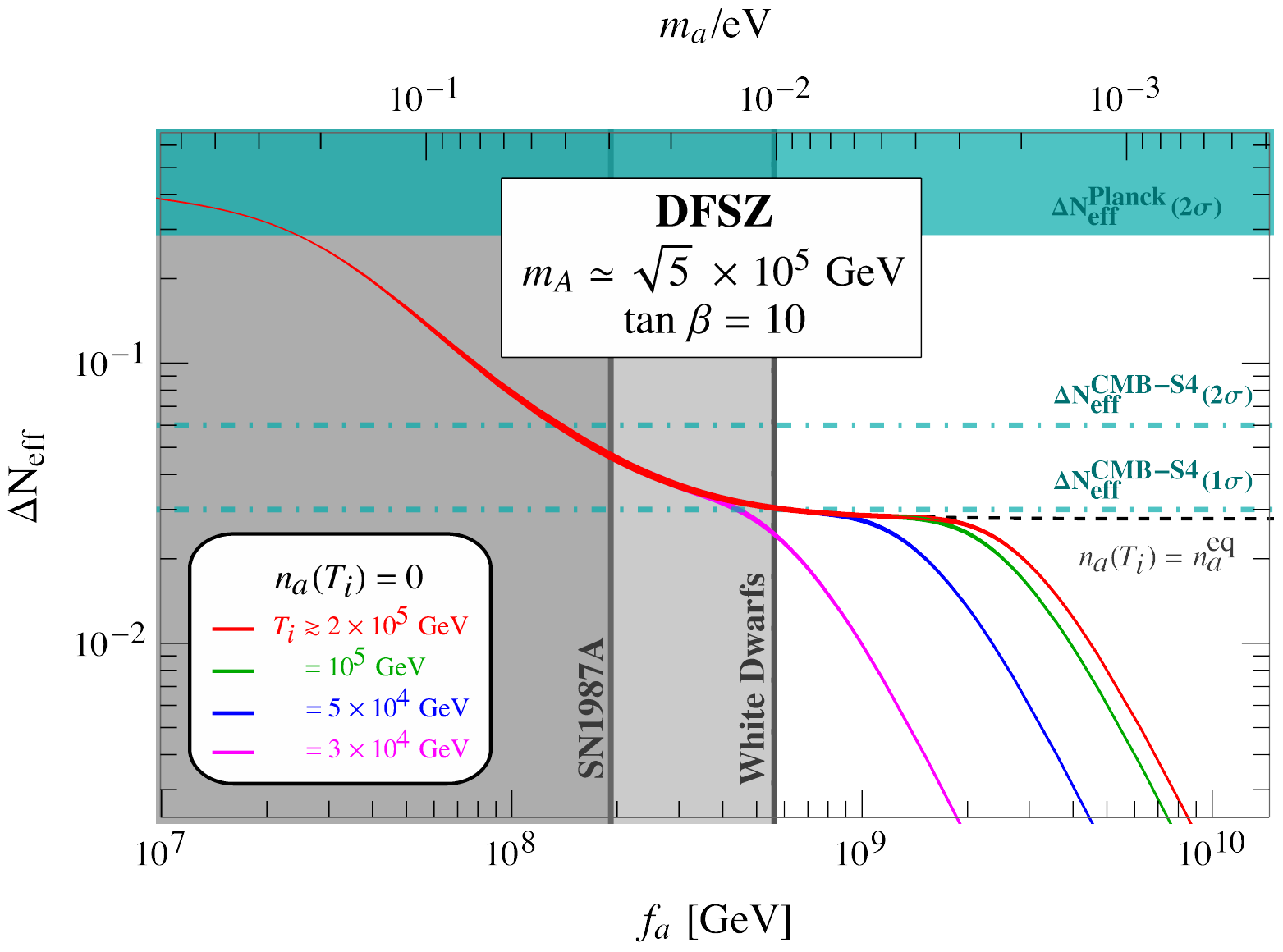}
\caption{\em Contribution to $\Delta N_{\rm eff}$ as a function of the axion decay constant $f_a$ for the DFSZ axion.}
\label{fig:DFSZNefffa}
\end{figure}

We quantify the last statement in Fig.~\ref{fig:DFSZNefffa} where we explore how $\Delta N_{\rm eff}$ depends on the axion decay constant. Solid colored bands provide the prediction obtained with vanishing initial axion abundance at various initial temperatures $T_i$. For comparison, we report also the prediction for the case when we begin the Boltzmann equation evolution with axions already in thermal equilibrium. As it was the case for the KSVZ axion, the predictions differ only at large values of the axion decay constant. This figure shows manifestly how the severe astrophysical constraints rule out most of the region where the signal in $\Delta N_{\rm eff}$ is detectable in the future. Even if we consider large values of $T_i$ and small enough axion decay constants, $f_a \lesssim 2 \times 10^9 \, {\rm GeV}$, the signal is barely within the reach of CMB-S4 surveys.

\begin{figure}
\centering
\includegraphics[width=0.9\textwidth]{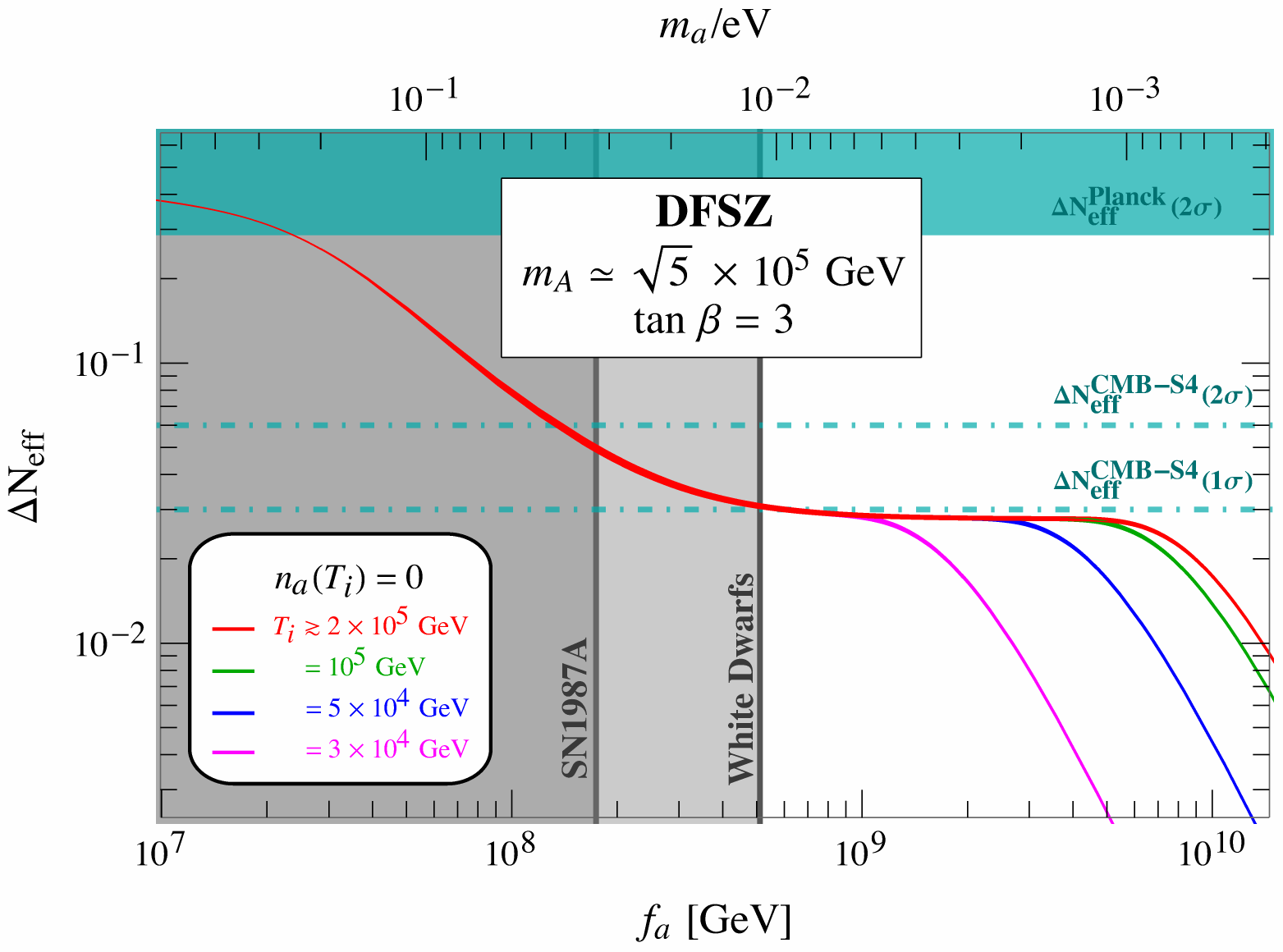}
\caption{\em Same as Fig.~\ref{fig:DFSZNefffa} but for $\tan\beta = 3$.}
\label{fig:DFSZNefffa2}
\end{figure}

Finally, we explore how our predictions depend on $\tan\beta$ in Fig.~\ref{fig:DFSZNefffa2} where we show $\Delta N_{\rm eff}$ as a function of $f_a$ for $\tan\beta=3$. We set the heavy Higgs boson mass $m_A$ to the same value as for Fig.~\ref{fig:DFSZNefffa}, and we update stellar bounds consistently to account for the different axion couplings. As explained in Sec.~\ref{sec:DFSZ}, the production rates for $\tan\beta =3$ and $10$ differ only above the EWPT with the former enhanced by approximately one order of magnitude. If axions are in thermal equilibrium until the time when the bath temperature is of the order of the weak scale there will be no difference between the two cases. This is manifest from a comparison between Figs.~\ref{fig:DFSZNefffa} and \ref{fig:DFSZNefffa2}: the resulting hot DFSZ axion abundances with $f_a \lesssim \mathcal{O}(10^9)\,{\rm GeV}$ are very similar for $\tan\beta = 10$ and $3$. However, we notice an effect at larger values of $f_a$ where it takes more effort for the axion to thermalize, and the predicted amount of dark radiation is enhanced for $\tan\beta = 3$. Thus $\Delta N_{\rm eff}$ does depend on $\tan\beta$ at large values of $f_a$, and smaller $\tan\beta$ makes the signal detectable for a wider range of axion decay constant values.

\subsection{Interplay with inflationary reheating}
\label{subsec:reheating}

Our analysis so far relied upon the assumption that the energy density of the universe was dominated by a gas of relativistic particles at the time of axion production. This is the extrapolation of how we ``look at'' our universe at the time of BBN, and it is worth keeping in mind that it is an extrapolation not supported by any observation. Within the inflationary paradigm, this extrapolation must come to an end because back enough in time the energy budget was controlled by the vacuum energy of the inflaton field. Inflaton decays populate the thermal bath eventually, and the highest temperature $T_R$ ever achieved during the radiation dominated epoch is known as the reheating temperature. Thus throughout our numerical Boltzmann analysis we have always implicitly assumed the hierarchy $T_R > T_i$. 

Our predictions are insensitive to the dynamics of inflation as long as the reheating temperature is high enough. For both of the frameworks under investigation, axion production is mediated by renormalizable couplings above the highest mass threshold; we have a Yukawa interactions with the $\Psi$ fermion and a scalar potential cubic term with the Higgs fields $H_u$ and $H_d$ in the KSVZ and DFSZ framework, respectively. The renormalizability of the axion couplings ensures that production is most efficient in the IR, and therefore around the mass of the heavy particles. Thus all we need is a reheat temperature larger than the heavy thresholds: $T_R > m_\Psi$ and $T_R > m_A$ for the KSVZ and the DFSZ framework, respectively. As a matter of fact, inflaton decays can provide an additional source for axion dark radiation. However, we consider also this option since in our analysis we accounted for the two extreme situations, $n_a(T_i) = 0$ and $n_a(T_i) = n^{\rm eq}_a(T_i)$, and this ensures that we cover all possible options. As we have already explained, our final predictions in the parameter space region where the signal is detectable do not depend on the initial conditions. 

What happens for lower values of $T_R$? Inflationary dynamics can play a relevant role only for $T_R < m_\Psi$ or $T_R < m_A$. In this regime, we solve the coupled Boltzmann equations describing inflationary reheating~\footnote{The second equation is valid only if radiation has the equation of state $p_R = \rho_R / 3$. Strictly speaking, this is only valid above the weak scale. For corrections to this simplified description see Ref.~\cite{Drees:2017iod}.}
\begin{align}
\label{eq:BEinf1} & \, \frac{d \rho_\phi}{dt} + 3 H \rho_\phi  = - \Gamma_\phi \rho_\phi \ , \\ 
\label{eq:BEinf2} & \, \frac{d \rho_{\rm R}}{dt} + 4 H \rho_{\rm R} = \Gamma_\phi \rho_\phi \ .
\end{align}
This system describes the evolution of the inflaton ($\phi$) and the radiation bath ($R)$ energy densities.  Inflaton decays, with a rate $\Gamma_\phi$, deplete the former and enhance the latter. The Hubble expansion rate, which allows us to understand when axions are produced most efficiently once we compare it with the production rate, is given by the Friedmann equation and it has the following scaling 
\be
H = \frac{\sqrt{\rho_\phi + \rho_R}}{\sqrt{3} M_{\rm Pl}} \simeq  \frac{T^2}{M_{\rm Pl}} \left\{ 
\begin{array}{ccl}
1 & $\quad\qquad$ & ~~~~~~~~T < T_R \\
\left(T  / T_R\right)^2 & $\quad\qquad$  & T_R \leq T \leq T_{\rm MAX} 
\end{array}
\right. \ .
\ee
Below the reheat temperature $T_R$ we have the typical scaling for a radiation dominated epoch. The reheat temperature is connected to the inflaton decay width through the relation $T_R \simeq \sqrt{\Gamma_\phi M_{\rm Pl}}$. Although this is defined as the highest temperature ever achieved by the thermal bath during the radiation dominated epoch, this is not the highest temperature achieved by the thermal bath in general. The bath itself exists even for temperatures larger than $T_R$ as a sub-dominant component since the decaying inflaton is still dominating the energy budget. The highest temperature ever achieved is usually denoted by $T_{\rm MAX}$ and it scales as $T_{\rm MAX} \simeq \left(M_{\rm Pl} \Gamma_\phi E_I^2 \right)^{1/4}$, with $E_I^4$ the constant energy density driving the inflationary expansion. Thus the bath temperature spans a potentially large range between $T_{\rm MAX}$ and $T_R$ before becoming the dominant energy component. During this phase, which corresponds to an early matter domination with the energy budget controlled by inflaton oscillations, the Hubble rate is proportional to $T^4$.

If we consider renormalizable interactions, production of particles via scatterings is efficient at low temperatures for a radiation dominated universe. This is the case also during inflationary reheating given the higher power of the temperature appearing in the scaling for the Hubble parameter. One can be more quantitative and state that in the range $5 \leq d \leq 8$, with $d$ the mass dimension of the operator mediating scatterings, the production is dominated at small temperatures during reheating and therefore maximized at $T_R$~\cite{Co:2015pka,Garcia:2017tuj,Chen:2017kvz,Calibbi:2021fld}. In particular, for $d < 8$ the final abundance does not depend on $T_{\rm MAX}$. Only for $d > 8$ particle production is UV dominated also during inflationary reheating, and therefore the resulting abundance is sensitive to $T_{\rm MAX}$ and to the details of reheating.

For the axion frameworks studied in this work, we never go above mass dimension 5 and therefore we are never sensitive to the peculiarities of inflationary reheating. Whether we are below the mass of $\Psi$ for the KSVZ or the mass of $A$ for the DFSZ, axion interactions are mediated by dimension 5 effective operators and the production is maximed at $T_R$. This is the scenario investigated by Ref.~\cite{Salvio:2013iaa} with the gluon and top quark couplings dominating the production rate for the KSVZ and DFSZ axion, respectively.


\section{Conclusions}
\label{sec:conclusions}

The PQ mechanism, where the $\theta$ parameter of QCD is promoted to a dynamical field, is undeniably one of the most elegant solutions to the strong CP problem. A plenitude of UV complete candidate models provides viable realizations of PQ symmetry breaking, but they all share the same low-energy residual: an approximate Nambu-Goldstone boson. Such a field, known as the axion, features the model-independent coupling to gluons given in Eq.~\eqref{eq:LPQaxion} as well as model-dependent interactions with other SM particles. Given the Nambu-Goldstone nature of the axion, its couplings to visible matter are suppressed by the large PQ breaking scale and this makes axion detection very challenging. Notwithstanding these difficulties, the field of axion experimental searches has been literally blossoming in the recent decade, and the present time is rather unique for the quest for axions.  

In spite of the rich set of options for axion couplings, all the terrestrial searches are sensitive to a handful of them: the ones to light quarks and gluons that in turn describe coupling to nuclei, the one to electrons, and the one to photons. An effective low-energy theory with only these interactions, with any UV completion that can be matched onto it, is enough to capture the phenomenology of axion searches.

We focused on an experimental signature to which all axion couplings can potentially contribute. The physics is the one of thermal axion production in the early universe, and the experimental manifestation is the presence of additional radiation that we infer from the CMB anisotropy spectrum. The net signal is accumulated through the expansion history with axions produced from bath particle collisions possibly at all temperature scales. Trustworthy predictions are possible only upon knowing axion couplings to all SM particles and to the model-dependent beyond the SM degrees of freedom specific to each theory.

The effect we consider is quantified by an additional contribution to the effective number of neutrinos species $\Delta N_{\rm eff}$. Bounds on the amount of axion dark radiation from the Planck data are already quite remarkable, and prospects provided by CMB-S4 surveys make this signal rather intriguing for the future. This population of relativistic axions can also leave an imprint on cosmological structure through baryon acoustic oscillations (BAO), and this effect provides an additional constraint on $\Delta N_{\rm eff}$  ~\cite{Green:2019glg,Xu:2021rwg,Baumann:2019keh}. Future large scale structure surveys will provide an improved BAO measurement, and this determination will be complementary to the CMB anisotropy spectrum. Furthermore, Ref.~\cite{Dror:2021nyr} suggested recently how such a cosmic axion background could be detected even with experiments in our terrestrial laboratories, although the signal is more sensitive to non-thermal energy spectra. These impressive projections combined with the top-down motivation make reliable theoretical predictions for the amount of axion dark radiation of the utmost importance.

This work addressed the presence of mass thresholds through the expansion history. As we showed with two explicit examples, the KSVZ and the DFSZ frameworks, passing through them alters the production rate significantly. For the KSVZ scenario, the rate changes significantly across the heavy fermion mass because renormalizable axion interactions become non-renormalizable, and this changes the rate temperature dependence. An analogous threshold is due to the heavy Higgs bosons in the DFSZ scenario, and in such a case the EWPT is also an important threshold where axion production mediated by fermion scattering changes drastically its temperature behavior as a consequence of chirality flips induced by fermion masses~\cite{Arias-Aragon:2020shv}. Finally, the QCDPT is common to both frameworks, and the matching procedure is far from being straightforward~\cite{DEramo:2021psx}.

The central results of our analysis are predictions for $\Delta N_{\rm eff}$ as a function of the axion decay constant $f_a$. They are explicitly presented in Fig.~\ref{fig:KSVZ_Neffvsfa} for the KSVZ framework, and in Figs.~\ref{fig:DFSZNefffa} and \ref{fig:DFSZNefffa2} for the DFSZ framework. Scatterings of thermal bath particles lead to a detectable signal in the future for both cases. For the KSVZ axion, the less severe stellar bounds allow for a stronger signal, as large as the $2 \sigma$ sensitivity of future CMB-S4 surveys and for values of the axion decay constant $f_a \lesssim 3 \times 10^9 \, {\rm GeV}$. On the contrary, white dwarf bounds for the DFSZ axion allow for a signal only detectable at $1 \sigma$ in the future. The range of testable axion decay constants for the KSVZ axion depends on the specific value of $\tan\beta$. For the two representative cases we analyzed, $\tan\beta = 10$ and $3$, the signal is testable for axion decay constants satisfying the upper bounds $f_a \lesssim 2 \times 10^9 \, {\rm GeV}$ and $f_a \lesssim 6 \times 10^9 \, {\rm GeV}$, respectively.

We assumed a radiation dominated universe through our analysis, but the production rates in Figs.~\ref{fig:KSVZTot} and \ref{fig:DFSZTot} are independent of the cosmological history. They can be employed to investigate axion production for alternative scenarios such as late inflationary reheating. Our methodology can also be extended to other microscopic realizations besides the two frameworks studied here. Flavor-violating axion couplings, with the production rate controlled by decays of bath particles instead of scatterings, are of particular interest. Plausible origins for the flavor violation can be loop corrections to axion couplings~\cite{Choi:2017gpf,Chala:2020wvs,Bauer:2020jbp,Choi:2021kuy,Bonilla:2021ufe}, or they can even be present at tree-level as a consequence of the PQ charge assignments~\cite{Ema:2016ops,Calibbi:2016hwq}. Predicting $\Delta N_{\rm eff}$ for specific axion UV complete models, along the lines of the analysis presented here, would be a piece of useful information to discriminate among them.

\paragraph{Acknowledgments.} The authors thank L. Di Luzio, S. H. Lim, T. Opferkuch, J. Schaffner-Bielich, C. S. Shin, L. Tolos for useful discussions. This work is supported by the research grants: ``The Dark Universe: A Synergic Multi-messenger Approach'' number 2017X7X85K under the program PRIN 2017 funded by the Ministero dell'Istruzione, Universit\`a e della Ricerca (MIUR); ``New Theoretical Tools for Axion Cosmology'' under the Supporting TAlent in ReSearch@University of Padova (STARS@UNIPD). The authors also supported by Istituto Nazionale di Fisica Nucleare (INFN) through the Theoretical Astroparticle Physics (TAsP) project. F.D. acknowledges support from the European Union's Horizon 2020 research and innovation programme under the Marie Sk\l odowska-Curie grant agreement No 860881-HIDDeN. 


\appendix


\section{Conventions and Useful Results I: Particle Physics}
\label{app:PP}

We collect in this Appendix our notations and conventions for SM fields and Lagrangian. Starting from the electroweak symmetric phase, we describe the SM gauge group and the matter content. We present spectrum and interactions, above and below the weak scale, for the case of a minimal Higgs sector with just one scalar weak doublet. The DFSZ framework features two Higgs doublets, and we discuss spectrum and interactions for this case as well. We quantify the effect of anomalous chiral rotations that are necessary to perform changes of field basis. Finally, we provide basic notions of ChPT and we introduce the formalism that, among several applications, allows us to determine axion couplings to hadrons.

\subsection*{Standard Model with minimal Higgs sector}

At energies above the Fermi scale, the theory enjoys a full $SU(3)_c \times SU(2)_L \times U(1)_Y$ gauge symmetry, and the Lagrangian takes the form
\be
\mathcal{L}_{\rm SM} = \mathcal{L}_{\rm gauge} + \mathcal{L}_{\rm fermion} + \mathcal{L}_{\rm Higgs} + \mathcal{L}_{\rm Yukawa} \ .
\label{eq:SMLag321}
\ee
The first term contains gauge boson kinetic terms 
\be
\mathcal{L}_{\rm gauge} = - \frac{1}{4} G^{A \, \mu\nu} G^{A}_{\mu\nu} - \frac{1}{4} W^{I \, \mu\nu} W^{I}_{\mu\nu} - 
\frac{1}{4} B^{\mu\nu} B_{\mu\nu} \ ,
\label{eq:Lgauge}
\ee
and they are constructed by employing the field strengths defined as follows
\be
\begin{split}
G^{A}_{\mu\nu} = & \, \partial_\mu G^A_\nu - \partial_\nu G^A_\mu + g_s \, f^{ABC} G^B_\mu G^C_\nu \ , \\
W^{I}_{\mu\nu} = & \, \partial_\mu W^I_\nu - \partial_\nu W^I_\mu + g \, \epsilon^{IJK} W^J_\mu W^K_\nu \ , \\
B_{\mu\nu} = & \, \partial_\mu B_\nu - \partial_\nu B_\mu 
\end{split}
\ee
with $f^{ABC}$ and $\epsilon^{IJK}$ the structure constants of the non-Abelian groups $SU(3)_c$ and $SU(2)_L$, respectively. The indices\ $A = 1, \ldots, 8$ and $I = 1, 2, 3$ run over the adjoint representations. Fermion and scalar kinetic terms are expressed via the gauge covariant derivative
\be
D_\mu = \partial_\mu - i g_s \frac{\lambda^A}{2} G^A_\mu - i g \frac{\sigma^I}{2} W^I_\mu - i g^\prime Y B_\mu \ .
\label{eq:Dmu}
\ee
The non-Abelian generators are Gell-Mann ($\lambda^A$) and Pauli ($\sigma^I$) matrices, and they act on color and weak-isospin indices (if any), respectively. The Abelian part has instead a term proportional to the hypercharge $Y$ of the field the covariant derivative acts on. The SM matter fields with their quantum numbers are listed in Tab.~\ref{tab:SMmatter}. Upper case fermions denote weak doublets whereas lower case fermions are singlet under the weak-isospin group, and the index $i$ runs over the three fermion generations. Fermion kinetic terms read
\be
\mathcal{L}_{\rm fermions} = \bar{Q}_{L} \, i \slashed{D} Q_{L} + \bar{u}_{R} \, i \slashed{D} u_{R} + 
\bar{d}_{R} \, i \slashed{D} d_{R} + \bar{L}_{L} \, i \slashed{D} L_{L}  + \bar{e}_{R} \, i \slashed{D} e_{R} \ .
\label{eq:Lfermions}
\ee
We use a compact notation where the sum over the three different generations is understood. The Lagrangian for the Higgs field has a canonically normalized kinetic term and the most general renormalizable scalar potential leading to electroweak symmetry breaking
\be
\mathcal{L}_{\rm scalar} = \squared{D_\mu H} - V_{\rm SM}(H) = \squared{D_\mu H} + \mu^2 H^\dag H - \frac{\lambda}{4} (H^\dag H)^2 \ .
\label{eq:LHiggs}
\ee
Finally, the Yukawa part of the Lagrangian (also in a compact matrix form) reads
\be
\mathcal{L}_{\rm Yukawa} = 
 - \bar{Q}_L \, \widetilde{H} \, Y^{(u)} u_R - \bar{Q}_L \, H \, Y^{(d)} d_R - \bar{L}_L \, H \, Y^{(e)} e_R + {\rm h.c.} \ ,
\label{eq:LYukawaSM}
\ee
where $\widetilde{H} = i \sigma^2 H^*$. In the most general fermion basis, $Y^{(u,d,e)}$ are generic $3\times 3$ matrices in flavor space, and they can be diagonalized upon performing bi-unitary rotations
\be
Y^{(\psi)} = U^\dag_{\psi_L} \hat{Y}^{(\psi)} U_{\psi_R}  \ .
\ee
Here, $\psi = u, d, e$ and the matrices $Y^{(\psi)}$ are diagonal. Hence we redefine fermion fields by performing the following unitary rotations in flavor space
\be
Q_L \rightarrow U_{u_L} Q_L \ , \quad  u_R \rightarrow U_{u_R} u_R \ , \quad  d_R \rightarrow U_{d_R} d_R \ , \quad
L_L \rightarrow U_{e_L} L_L \ , \quad  e_R \rightarrow U_{e_R} e_R \ .
\ee
After these operations, the Yukawa Lagrangian in Eq.~\eqref{eq:LYukawaSM} takes the form
\be
\mathcal{L}_{\rm Yukawa} = 
- \bar{Q}_L \, \widetilde{H} \, \hat{Y}^{(u)} u_R - \bar{Q}_L \, H \, V_{\rm CKM}  \, \hat{Y}^{(d)} d_R - \bar{L}_L \, H \, \hat{Y}^{(e)} e_R + {\rm h.c.} \ , 
\label{eq:LYukawaSMdiag}
\ee
where we introduce the Cabibbo–Kobayashi–Maskawa (CKM) matrix $V_{\rm CKM} \equiv U^\dag_{u_L} U_{d_L}$.

\setlength{\tabcolsep}{10pt}
\renewcommand{\arraystretch}{1.1}
\begin{table} 
\begin{center}
\begin{tabular}{ c || c c c c c  c}
 &  $Q_{L i}$ &  $u_{R i}$ & $d_{R i}$ & $L_{L i}$ &$e_{R i}$ & $H$ \\ 
\hline  \hline 
$SU(3)_c$ & $\bs 3$ & $ \bs 3$ & $ \bs 3$ & $\bs 1$ & $\bs 1$ & $\bs 1$  \\
$SU(2)_L$ & $\bs 2$ & $\bs 1$ & $\bs 1$ & $\bs 2$ & $\bs 1$ & $\bs 2$  \\
$U(1)_Y$  & $+1/6$ & $+2/3$ & $-1/3$ & $-1/2$ & $-1$ & $+1/2$ 
\end{tabular}
\end{center}
\caption{\em SM matter fields and gauge quantum numbers ($i$ runs over the three generations).}
\label{tab:SMmatter}
\end{table}

\subsection*{Electroweak symmetry breaking}

The scalar potential in Eq.~\eqref{eq:LHiggs} has an electroweak symmetry breaking minimum where the Higgs field gets a vacuum expectation value (vev) 
\be
\vev{H} = \frac{1}{\sqrt{2}} \left(\begin{array}{c} 0 \\ v \end{array} \right) \ , \qquad \qquad \qquad \qquad v^2 = \frac{4 \mu^2}{\lambda} = (246 \, {\rm GeV})^2 \ .
\label{eq:HvevSM}
\ee
As a consequence, electroweak gauge bosons acquire mass terms and mix among each other. The complete gauge boson mass spectrum reads
\be
\begin{split}
W^\pm_\mu = & \, \frac{W^1_\mu \mp i \, W^2_\mu}{\sqrt{2}}  \ , \qquad \quad \qquad\qquad 
m_{W^\pm} = \frac{g}{2} v \ , \\
\label{eq:Zboson} Z_\mu = & \,  c_w W^3_\mu - s_w B_\mu  \ ,  \qquad \qquad \qquad 
m_Z = \frac{\sqrt{g^2 + g^{\prime 2}}}{2} v \ , \\
A_\mu = & \,  s_w W^3_\mu + c_w B_\mu \ , \qquad \qquad \qquad m_\gamma = 0  \ ,
\end{split}
\ee
where the photon remains massless and we define the weak mixing angle
\be
\label{eq:swdef} 
(s_w, c_w) = \frac{1}{\sqrt{g^2 + g^{\prime 2}}} (g^\prime, g)  \ .
\ee
Likewise, the Yukawa operators in Eq.~\eqref{eq:LYukawaSMdiag} provide fermion masses once the Higgs field gets a vev. The presence of the CKM matrix makes fermion mass terms still not diagonal, and the unitary rotation $d_L \rightarrow V_{\rm CKM} \, d_L$ on the left-handed down quarks brings us to the physical states. The fermion mass eigenvalues are equal to the diagonal entries of the Yukawa matrices multiplied by $v / \sqrt{2}$.

Once electroweak symmetry is broken, the Lagrangian schematically reads again as in Eq.~\eqref{eq:SMLag321} but this time the four contributions have different ingredients. The piece for gauge bosons contains again kinetic terms for gauge bosons, but this time with field strengths of the electroweak mass eigenstates, and also mass terms for the $W$ and $Z$ bosons. Fermions fill vector-like representations of the gauge group, and gauge interactions are conveniently expressed in terms of the covariant derivative
\be
D_\mu = \partial_\mu - i g_s \frac{\lambda^A}{2} G^A_\mu - i \frac{g}{\sqrt{2}} \left( \frac{\sigma^+}{2} W^+_\mu  + \frac{\sigma^-}{2} W^-_\mu \right) 
- i \frac{g}{c_w} \left(\frac{\sigma^3}{2} - s_w^2 Q \right) Z_\mu - i e Q A_\mu \ .
\label{eq:Dcov31}
\ee
Here, we introduce the Pauli matrices $\sigma^\pm \equiv \sigma^1 \pm i \sigma^2$ that are used to define the ladder operators for the weak isospin group. Furthermore, we define (minus) the electron charge $e \equiv g g^\prime / g_w > 0$ and the electric charge generator $Q \equiv \sigma^3 /2 +Y$. The same covariant derivative describes gauge interactions for the radial model of the Higgs field, the Higgs boson $h$, and this is the only scalar appearing in the Yukawa interactions with fermions. 

\subsection*{Two Higgs Doublet Model}

The Lagrangian for a theory with two weak doublets $H_u$ and $H_d$ is richer than the one given in Eq.~\eqref{eq:LHiggs}, and it takes the schematic form
\be
\mathcal{L}_{\rm scalars}^{\text{(2HDM)}} = \squared{D_\mu H_u} + \squared{D_\mu H_d} - V_{\rm 2HDM}(H_u, H_d)   \ .
\ee
We assign opposite hypercharges to the scalars, $Y_{u,d} = \pm 1/2$, and only operators with an even number of Higgs fields can appear in the scalar potential since they are weak doublets. Considering only renormalizable operators, we can only have terms with two of four Higgs fields built from the only three quadratic gauge invariant combinations: $H_u^\dag H_u$, $H_d^\dag H_d$, and $H^T_u i \sigma^2 H_d$. The first two options cannot carry a net PQ charge whereas the last one can. Keeping in mind that we are interested in the DFSZ framework, we consider theories where the combination $H_u H_d$ has a non-vanishing PQ charge. 

We consider the renormalizable and gauge invariant scalar potential 
\be
\begin{split}
\label{pot2hd}
V_{\rm 2HDM}(H_u, H_d) = & \, - \mu_u^2 H_u^\dag H_u - \mu_d^2 H_d^\dag H_d + \left(B  \, H^T_u i \sigma^2 H_d + {\rm h.c.}\right) + \\ &  \frac{\lambda_1}{4} (H_u^\dag H_u)^2 +  \frac{\lambda_2}{4} (H_d^\dag H_d)^2 +  \lambda_3 (H_u^\dag H_u) (H_d^\dag H_d) + \\ &  \lambda_4 (H^T_u i \sigma^2 H_d) (H^T_u i \sigma^2 H_d)^\dag \ .
\end{split}
\ee
The only operator that is not invariant under PQ is the quadratic one proportional to $B$. Within the DFSZ framework, it arises once the PQ breaking scalar gets a vev, and the phase of $B$ contains the axion field above electroweak symmetry breaking. Upon redefining the Higgs doublets, it is possible to take the PQ-breaking coefficient $B$ to be real and positive.

We search for an EWSB minimum where only the neutral components acquire vevs
\be
\begin{split}
\vev{H_u} = &\, \frac{1}{\sqrt{2}} \left(\begin{array}{c} 0 \\ v_u \end{array} \right) = 
\frac{1}{\sqrt{2}} \left(\begin{array}{c} 0 \\ v \sin\beta \end{array} \right)  \ ,  \\ 
\vev{H_d} = & \, \frac{1}{\sqrt{2}} \left(\begin{array}{c} v_d \\ 0 \end{array} \right) = \frac{1}{\sqrt{2}} \left(\begin{array}{c} v \cos\beta \\ 0  \end{array} \right)  \ .
\end{split}
\label{eq:Hvev2HDM}
\ee
The vevs satisfy the constraint $v^2 = v_u^2 + v_d^2 = (246 \, {\rm GeV})^2$, and we define the angle $\beta$ as $\tan\beta \equiv v_u / v_d$. The minimum conditions are
\be
\begin{split}
\mu _u^2 = & \, \frac{v^2}{2} \left[\frac{\lambda _1}{2} \sin^2\beta + (\lambda _3 + \lambda_4) \cos^2\beta  \right]  - \frac{B}{\tan\beta} \ ,  \\
\mu _d^2 = & \,  \frac{v^2}{2} \left[ \frac{\lambda _2}{2} \cos^2\beta + (\lambda _3 + \lambda_4) \sin^2\beta  \right] - B \tan\beta \ .
\end{split}
\ee

We expand around this EWSB vacuum and we determine the mass spectrum. The Higgs doublets can be decomposed in terms of neutral and charged scalar components
\be
H_u = \left(\begin{array}{c} H_u^+ \\ \frac{v_u + S_u + i P_u}{\sqrt{2}} \end{array} \right) \ , \qquad \qquad \qquad 
H_d = \left(\begin{array}{c} \frac{v_d + S_d + i P_d}{\sqrt{2}} \\ H_u^- \end{array} \right) \ .
\label{eq:Higgs2HDM}
\ee
For neutral fields, we distinguish between CP-even and CP-odd scalars. The squared mass matrices for the charged, pseudo-scalar and scalar fields read respectively
\be
\begin{split}
m_C^2 = & \, \left(
\begin{array}{cc}
 B / \tan\beta - \frac{1}{2} \lambda _4 v^2 \cos^2\beta & B - \frac{1}{4} \lambda _4 v^2 \sin2\beta \\
 B - \frac{1}{4} \lambda _4 v^2 \sin2\beta & B \tan\beta - \frac{1}{2} \lambda _4 v^2 \sin^2\beta \\
\end{array}
\right) \  , \\
m^2_P = & \, B \, \left(\begin{array}{cc}
1 / \tan\beta & 1 \\
1 & \tan\beta \end{array} \right) \ , \\
m^2_S = & \, \left(
\begin{array}{cc}
B / \tan\beta + \frac{1}{2} \lambda _1 v^2 \sin ^2\beta & - B  + \frac{1}{2}  \left(\lambda _3+\lambda _4\right) v^2 \sin (2 \beta ) \\
- B + \frac{1}{2}  \left(\lambda _3+\lambda _4\right) v^2 \sin (2 \beta ) & B \tan\beta + \frac{1}{2} \lambda _2 v^2 \cos^2\beta \\
\end{array}
\right) \ .
\end{split}
\ee
The first two matrices have vanishing determinant, and this ensures massless Goldstones to provide longitudinal components for the $W$ and $Z$ gauge bosons. The masses of the heavy charged ($H^\pm$) and pseudo-scalar ($A$) Higgs bosons can be found from the trace
\be
\begin{split}
m_{H^\pm}^2 = & \, {\rm Tr} [m_C^2] =  \frac{2 B}{\sin(2 \beta)}  - \frac{\lambda_4 v^2}{2} \ , \\
m_A^2 = & \,  {\rm Tr} [m_P^2] = \frac{2 B}{\sin(2 \beta)} \ .
\end{split}
\ee
The CP-even Higgs bosons, the SM-like $h$ and the heavier $H$, are both massive. We provide here the mass eigenvalues in the decoupling limit which is valid when $B \gg v^2$
\be
\begin{split}
m_h^2 \simeq & \, \left[\lambda _1 \sin^4\beta  + \lambda _2 \cos^4\beta + \left(\lambda _3+\lambda _4\right) \sin^2(2\beta) \right] \frac{v^2}{2} \ , \\
m_H^2 \simeq & \,  \frac{2 B}{\sin(2 \beta)} + \left[ \frac{\lambda _1 + \lambda _2}{4} - (\lambda_3 + \lambda_4) \right] \sin^2(2 \beta ) \frac{v^2}{2}    \ .
\end{split}
\ee

The charged mass eigenstates result in
\be
\left( \begin{array}{c} G^+ \\ H^+ \end{array}\right) = 
\left( \begin{array}{cc}
\sin\beta & - \cos\beta  \\
\cos\beta & \sin\beta 
\end{array} \right) 
\left( \begin{array}{c} H_u^{+}  \\ H_d^{-\,\dag} \end{array}\right) \ ,
\label{eq:2HDMCharged}
\ee
where $G^+$ is the charged Goldstone eaten up by the $W$ boson. Likewise, if we dub $G^0$ the Goldstone eaten by the $Z$ boson, we have the pseudo-scalar mass eigenstates
\be
\left( \begin{array}{c} G^0 \\ A \end{array}\right) = 
\left( \begin{array}{cc}
\sin\beta & - \cos\beta  \\
\cos\beta & \sin\beta 
\end{array} \right) 
\left( \begin{array}{c} P_u  \\ P_d \end{array}\right) \ .
\label{eq:2HDMPseudo}
\ee
Finally, in the decoupling limit ($B \gg v^2$), we have the CP-even mass eigenstates
\be
\left( \begin{array}{c} h \\ H \end{array}\right) = 
\left( \begin{array}{cc}
\sin\beta & \cos\beta  \\
- \cos\beta & \sin\beta 
\end{array} \right) 
\left( \begin{array}{c} S_u  \\ S_d \end{array}\right) \ .
\label{eq:2HDMReal}
\ee

\subsection*{Anomalous chiral rotations}

Chiral, axion dependent, rotations on fermion fields can be useful to find a new field basis better suited for the specific framework under investigation. Here, we state our conventions for these field transformations and we quantify gauge anomaly effects. For a generic Dirac fermion $\chi$, before accounting for any interaction, the theory has a $U(1)_V$ symmetry where both left- and right-handed Weyl components are rotated with the same phase, and we have a conserved Noether's vector current $J^{(V)}_\mu = \bar{\chi} \gamma_\mu \chi$. If the fermion is massless ($m_\chi = 0$) then the theory has also the $U(1)_A$ symmetry, where left- and right-handed Weyl components are rotated with opposite phases, and the resulting Noether's current results in $J^{(A)}_\mu = \bar{\chi} \gamma_\mu \gamma^5 \chi$. On the contrary, if the fermion field is massive, we have that the axial current has a non-vanishing divergence
\be
\left. \partial^\mu J^{(A)}_\mu\right|_{\rm classical}  = 2 m_\chi \, \bar{\chi} i \gamma^5 \chi \ .
\ee 

The result above is valid at the classical level. Once we include quantum corrections, the axial current can be non-conserved if the fermion carries gauge charges, even if the fermion itself is massless. It is possible to derive its divergence by different methods, like evaluating the Green function of the axial current with gauge bosons in perturbation theory (triangle diagrams)~\cite{Adler:1969gk,Bell:1969ts}, or via the Jacobian of the path integral measure~\cite{Fujikawa:1979ay}.

We start from the well-known QED result 
\be 
\left. \partial^\mu J^{(A)}_\mu\right|_{\rm anomaly} = - \frac{e^2}{8 \pi^2} \, F_{\mu\nu} \widetilde{F}^{\mu\nu} \ . 
\label{eq:ABJ}
\ee
The generalization to a non-Abelian gauge theory, such as QCD, does not require any new calculation,  all we need to do is adding a group theory factor to the QED expression in \Eq{eq:ABJ} (the triangle diagrams have the same Lorentz structure)
\be
\left. \partial^\mu J^{(A)}_\mu\right|_{\rm anomaly}  = - \frac{g_s^2}{8\pi^2} \, {\rm Tr} \left[t^C t^D \right] \, G^C_{\mu\nu} \widetilde{G}^{D \mu\nu}   = 
- \frac{g_s^2}{16 \pi^2} \, G^C_{\mu\nu} \widetilde{G}^{C \mu\nu}   \ ,
\label{eq:axialQCD}
\ee
where $t^C$ are generators of the color group normalized as ${\rm Tr} \left[t^C t^D \right] = \delta^{CD} / 2 $.

We have all the tools to quantify the effects of performing chiral rotations. For a generic massless Dirac fermion $\chi$, charged under a representation $R$ of the $SU(3)_c$ gauge group, we perform the local axial rotation 
\be
\chi \; \rightarrow \; \exp[i \alpha_A(x) \gamma^5] \chi \ .
\label{eq:axialROT}
\ee
As a consequence, the Lagrangian changes due to both classical and quantum effects. The former is straightforward whereas the latter can be found by computing the change in the path integral measure~\cite{Fujikawa:1979ay}. We do not need to reproduce the derivation since we know that this contribution must reproduce the equation of motion in \Eq{eq:axialQCD}. Thus we have
\be
- \partial_\mu \alpha_A \; \bar{\chi} \gamma^\mu \gamma^5 \chi  + \left. \Delta \mathcal{L}_\chi  \right|_{\rm anomaly}  =
\alpha_A \partial_\mu J^{(A)\, \mu} + \left. \Delta \mathcal{L}_\chi  \right|_{\rm anomaly}  = 0 \ .
\ee
Upon comparing we find how chiral rotations alter the Lagrangian at the quantum level
\be
\left. \Delta \mathcal{L}_\chi  \right|_{\rm anomaly}  = 2 \alpha_A\, \times \, \frac{\alpha_s}{8 \pi} \, G^C_{\mu\nu} \widetilde{G}^{C \mu\nu} \ .
\label{eq:anomalousrot}
\ee
This is valid for an axial rotation of a Dirac fermion as in Eq.~\eqref{eq:axialROT}. If we only rotate one Weyl component the result is half the one above and with the appropriate sign.

\subsection*{Rudimental ChPT}

We review basic notions of Chiral Perturbation Theory (ChPT)~\cite{Weinberg:1978kz,Gasser:1983yg,Gasser:1984gg} needed to study axion couplings. Our starting point is the QCD Lagrangian with $N_f = 3$ quark flavors
\be
\mathcal{L}_{\rm QCD} = - \frac{1}{4} G^{\mu\nu} G_{\mu\nu} + \overline{q} i \slashed{D} q - \left[ \bar{q}_R M_q q_L  + {\rm h.c.} \right] \ .
\label{eq:QCDLag}
\ee
For the ease of notation, we introduce the quark vector in flavor space $q = (u \; d\;  s)^T$ where each entry is a Dirac field with left and right-handed components. Chiral projectors, defined in the usual way $P_{L,R} = (1 \mp \gamma^5) / 2$, extract the different quark chiralities: $q_{L,R} = P_{L,R} \; q$. The quark mass matrix in the mass eigenbasis reads $M_q = \text{diag}\left(m_u, m_d, m_s \right)$.

If we neglect the quark mass matrix, left- and right-handed quarks are decoupled and the QCD Lagrangian in \Eq{eq:QCDLag} is invariant under independent rotations of the two fermion chiralities: the theory has a global $U(3)_L \times U(3)_R$ symmetry. The vectorial part of the symmetry group where both chiralities are rotated by the same angle, namely the baryon number $U(1)_V$ and the isospin $SU(3)_V$, are good approximate symmetries of Nature; the former is broken by gauge anomalies whereas the latter is only broken by quark mass differences and electroweak interactions. The axial part $U(1)_A \times SU(3)_A$ is spontaneously broken by the quark condensate, and we do not expect mixed parity multiplets in the hadronic spectrum but rather Goldstone bosons associated to the broken axial generators. However, there is no Goldstone boson associated to the broken $U(1)_A$ symmetry since $m_{\eta^\prime} \gg m_\pi$. This was dubbed as the $U(1)_A$ problem of QCD~\cite{Weinberg:1975ui} and it was solved only thanks to a complete understanding of the rich structure of the QCD vacuum~\cite{tHooft:1976rip,tHooft:1976snw}. We do not have a ninth Goldstone boson $\eta^\prime$ in the spectrum because the $U(1)_A$ is not even an approximate symmetry of the QCD Lagrangian in the massless limit since it is anomalous.

Concerning the spontaneous breaking of the $SU(3)_A$ part, we employ a non-linear sigma model to describe the associated Goldstone bosons. Keeping only terms with two derivatives, which correspond to quadratic terms in the exchanged momentum $\mathcal{O}(p^2)$, we have the low-energy chiral Lagrangian 
\be
\mathcal{L}_{\rm ChPT} =  \frac{f_\pi^2}{4} \, {\rm Tr} \left[ \partial^\mu U^\dag \, \partial_\mu U \right]  \ , \qquad \qquad \qquad U = \exp\left[ i \, \frac{\pi^a \, \lambda^a}{f_\pi} \right]  \ ,
 \label{eq:chiralL}
\ee
where $\lambda^a$ are the $SU(3)$ Gell-Mann matrices and $f_\pi \simeq 93 \, {\rm MeV}$. The Goldstone bosons $\pi^a$, with $a = 1, \ldots, 8$, enter through the unitary matrix $U$ that under a generic chiral rotation transforms as $U \rightarrow L \, U \, R^\dag$. We pick the basis
\be
\pi^a \, \lambda^a = \left( \begin{array}{ccc} 
\pi^0 + \frac{\eta}{\sqrt{3}} & \sqrt{2} \pi^+ & \sqrt{2} K^+ \\
\sqrt{2} \pi^- & - \pi^0 + \frac{\eta}{\sqrt{3}} & \sqrt{2} K^0 \\
\sqrt{2} K^- & \sqrt{2} \; \bar{K}^0 & - 2 \frac{\eta}{\sqrt{3}} 
\end{array}\right) \ .
 \label{eq:pionoctet}
\ee
This choice is convenient since these fields are physical eigenstates once we introduce chiral symmetry breaking quark masses.

The Lagrangian in \Eq{eq:chiralL} holds for exact chiral symmetry in the high-energy theory. Quark masses, which break chiral symmetry, are easily incorporated by applying the formalism of \Refs{Gasser:1983yg,Gasser:1984gg} for matrix elements of currents in the chiral effective theory. Furthermore, this method also allows us to derive axion couplings to the Goldstone octet in Eq.\eqref{eq:pionoctet}. We review this method starting from the QCD Lagrangian written as follows
\be
\mathcal{L}_{{\rm QCD} + \mathcal{S}} =- \frac{1}{4} G^{\mu\nu} G_{\mu\nu} + \bar{q} \, i \slashed{D} q - 
\left[ \bar{q}_L (s + ip) q_R + {\rm h.c.} \right] 
- \bar{q}_L \, l^\mu  \gamma_\mu q_L - \bar{q}_R \, r^\mu \gamma_\mu q_R \ .
\label{eq:QCDwithCurrents}
\ee
Here, we include four different external sources $\mathcal{S}$: scalar $s$, pseudo-scalar $p$, vector left $l^\mu$, and vector right $r^\mu$. The QCD Lagrangian with quark mass terms in \Eq{eq:QCDLag} is recovered for $l^\mu = r^\mu = p = 0$ and $s = M_q$. Setting the spin-one sources to a non-vanishing value allows us to deal with coupling to vector bosons, such as the photon, as well as the spin-one axion currents. Finally, the pseudo-scalar current $p$ also plays an important role to determine axion couplings. 

The Lagrangian in Eq.~\eqref{eq:QCDwithCurrents} has a \textit{local} $SU(2)_L \times SU(2)_R$ symmetry where left- and right-handed quarks transform separately
\be
q_L(x) \; \rightarrow \; L(x) q_L(x) \ , \qquad \qquad \qquad q_R(x) \; \rightarrow \; R(x) q_R(x) \ ,
\ee 
and the sources also transform as follows
\be
\begin{split}
s(x) + i p(x) \; \rightarrow & \; L(x) \left[ s(x) + i p(x) \right] R(x)^\dag \ , \\
l_\mu(x) \; \rightarrow & \; L(x) \, l_\mu(x) \, L(x)^\dag + i \partial_\mu L(x) \, L(x)^\dag  \ , \\ 
r_\mu(x) \; \rightarrow & \;  R(x) \, r_\mu(x) \, R(x)^\dag + i \partial_\mu R(x) \, R(x)^\dag \ .
\end{split}
\ee
In the above equations, we restore the explicit dependence on the space-time location $x$ to emphasize that the transformation is local. The spin-one sources transform as gauge fields so we define the covariant derivative
\be
D_\mu  U = \partial_\mu U + i l_\mu U - i U r_\mu \ ,
\ee
and we can match the Lagrangian in \Eq{eq:QCDwithCurrents} onto a low-energy chiral Lagrangian
\be
\mathcal{L}_{{\rm ChPT} + \mathcal{S}} = \frac{f_\pi^2}{4} \, {\rm Tr} \left[ D^\mu U^\dag \, D_\mu U \right] +
 \mu \, \frac{f_\pi^2}{2}  {\rm Tr} \left[(s+ip) U^\dag + U (s - i p)  \right] \ .
 \label{eq:chiralLsources}
\ee
Here, $\mu$ is a dimensionful parameter that we determine by the requirement of reproducing the meson masses
\be
\begin{split}
\mathcal{L}_{{\rm G.B. mass}} = & \,  \left.\mathcal{L}_{{\rm ChPT} + \mathcal{S}}\right|_{s = M}^{p=l^\mu=r^\mu=0} =  \mu \, \frac{f_\pi^2}{2}  {\rm Tr} \left[ M (U^\dag + U) \right] =  \\ &
- \frac{\mu}{2} \left[ (m_u + m_d) \pi^0 \pi^0 + 2 \frac{m_u  -m_d}{\sqrt{3}}  \pi^0 \eta + \frac{m_u + m_d + 4 m_s}{3}  \eta^2 \right] + \\ & - \mu \left[ (m_u + m_d ) \pi^+ \pi^- + (m_d + m_s) K^0 \overline{K^0} + (m_u + m_s) K^+ K^- \right]  \ .
\end{split}
\ee
If we look at the pion mass, and we neglect the mixing with the $\eta$, we find $m_\pi^2 = \mu (m_u + m_d) $.


\section{Conventions and Useful Results II: Thermal Corrections}
\label{app:TH}

We discuss thermal corrections in this Appendix. We analyze axion production via thermal gluon scattering, and we compute thermal masses for the DFSZ Higgs sector. 

\subsection*{Thermal corrections to axion production via gluon loops}

The axion production rate can be expressed in terms of the axion self-energy~\cite{Weldon:1990iw,Gale:1990pn}
\be
\gamma_a = - 2\int \frac{d^3 p_a}{2 E_a \left(2\pi\right)^3} f_{\rm BE} (E_a)\,{\rm Im} \,\Pi_a = \int \frac{d^3 p_a}{2 E_a \left(2\pi\right)^3} \Pi_a^{<} \, .
\label{eq:AxionProductionRate}
\ee
Here, $f_{\rm BE}(E_a)$ is the Bose-Einstein distribution and $\Pi_a^{<}$ denotes the non time-ordered axion two-point function. At finite temperature, the expansion parameter is not anymore $\alpha_s / (4\pi)$ but rather the gauge coupling constant $g_s$ because of collinear enhancements~\cite{Braaten:1991dd}, and this require in principle the resummation of infinite processes involving many particles. However, as explained by Ref.~\cite{Salvio:2013iaa}, such an enhancement is absent for the axion anomalous interaction given in Eq.~\eqref{eq:LPQaxion}, and we can safely consider only binary scatterings. Nevertheless, even if one restricts to binary collisions there are still IR divergences to take care of. The production rate with IR divergence properly accounted for can be found in Ref.~\cite{Salvio:2013iaa} but only at high temperatures, $T \gtrsim 10^4 \, {\rm GeV}$. We extend this study to lower temperatures.

Only one diagram contributes to the production rate, the one-loop axion two-point functions with virtual gluons. Within the context of the optical theorem, this contribution could be interpreted as the thermal gluon decay~\cite{Salvio:2013iaa}; using the cutting rules, we could identify diagrammatically as  $g_{\rm th} \rightarrow g_{\rm th} + a$ where $g_{\rm th}$ indicates the thermal excitation of the gluon field in a medium. This is the reason why Ref.~\cite{Salvio:2013iaa} dubbed it the `decay'' diagram. We evaluate the it with the resummed thermal gluon propagators in the loop, and we work at the leading order in $g_s$ because we can restrict to binary collisions. At this point, all we need to derive a proper thermal gluon propagator.

For a generic gauge theory, thermal effects induce the following correction to the gluon two-point function in momentum space
\be
\Delta \mathcal{L}_{\rm thermal} = - \frac{1}{2}G_\mu^A \Pi^{\mu\nu} G_\nu^A \, .
\ee
The thermal self energy $\Pi^{\mu\nu}$ is a function of the external momentum $K^\mu = (\omega, k  \hat{k})$. Here and thereafter, capital characters denote four-vectors whereas lower-case characters their components. The four-vector $K$ has a spatial component of size $k$ and direction $\hat{k}$. 

There are two\footnote{If we consider axion production from thermal scatterings of $SU(2)_L$ weak or $U(1)_Y$ hypercharge gauge bosons there is an additional one from Higgs doublets~\cite{Rychkov:2007uq}.} sources for the gluon self-energy at one-loop: gluon self-interactions and gauge interactions of colored fermions (i.e., quarks)
\be
\Pi^{\mu\nu} = \Pi_{G}^{\mu\nu} + \sum_q \Pi_{q}^{\mu\nu} \ .
\ee
The pure gauge contribution reads
\be
\begin{split}
\Pi_{G}^{\mu\nu} = & \, \frac{g_s^2}{2} C_2(G) \int \frac{d^3 p}{2 p \left(2\pi\right)^3} f_{\rm BE}(p) \\
& \times \left[ \frac{g^{\mu\nu} \left(4 p\cdot k - 2k^2\right) - 8 p^\mu p^\nu + 2k^\mu k^\nu - 5 p^\mu k^\nu - 3 p^\nu k^\mu }{2p\cdot k +k^2}\right.  \\
& \quad \left.  +  \frac{g^{\mu\nu} \left(4 p\cdot k + 2k^2\right) + 8 p^\mu p^\nu - 2k^\mu k^\nu - 3 p^\mu k^\nu - 5 p^\nu k^\mu }{2p\cdot k - k^2} \right] \, ,
\end{split}
\ee
where $C_2(G)$ denotes the quadratic Casimir operator for the adjoint representation and $f_{\rm B.E.}(p)$ is the Bose-Einstein distribution. The contribution from a quark $q$ with mass $m_q$ results in 
\be
\begin{split}
\Pi_{q}^{\mu\nu} = & \, 2 g_s^2 \int \frac{d^3 p}{2E_p\left(2\pi\right)^3}\left[f_{\rm FD}(E_p) + \bar{f}_{\rm FD}(E_p)  \right]  \\
& \times \left[ \frac{g^{\mu\nu} p\cdot k - 2p^\mu p^\nu - p^\mu k^\nu - p^\nu k^\mu }{2p\cdot k +k^2}  +  \frac{g^{\mu\nu} p\cdot k + 2p^\mu p^\nu - p^\mu k^\nu - p^\nu k^\mu }{2p\cdot k - k^2} \right] \ ,
\end{split}
\ee
where $E_p = \sqrt{p^2 + m_q^2}$, and $f_{\rm FD}(E_p)$ and $\bar{f}_{\rm FD}(E_p)$ are the Fermi-Dirac distributions for the quark $q$ and the anti-quark $\bar{q}$, respectively.

We decompose $\Pi^{\mu\nu}$ into its longitudinal ($\rm L$) and transverse ($\rm T$) components~\cite{Laine:2016hma,Bellac:2011kqa} 
\begin{eqnarray}
\pi_{\rm L} & = & -\frac{\omega^2 - k^2}{k^2}\Pi^{00} \, , \\
\pi_{\rm T} & = & -\frac{1}{2}\pi_{\rm L} + \frac{1}{2}g_{\mu\nu}\Pi^{\mu\nu} \ .
\label{eq:exex}
\end{eqnarray}
The longitudinal and transverse gauge contributions read
\begin{eqnarray}
\left. \pi_{\rm L} \right|_{\rm G} & = & -  C_2(G) \frac{g_s^2}{2\pi^2}  \left(\frac{\omega^2 - k^2}{k^2}\right) \int dp  f_{\rm BE} \left(p\right) \left[2pL + \frac{M}{k}-\frac{k}{4}L_- \right] \, ,\\
\left. \pi_{\rm T} \right|_{\rm G} & = & -\frac{1}{2}\left.\pi_{\rm L}\right|_{\rm V} + C_2(G)\frac{g_s^2}{2\pi^2}  \int dp  f_{\rm BE} \left(p\right)  \left[2pL+\frac{5(\omega^2-k^2)}{8k}L_- \right] \, ,
\end{eqnarray}
where we set $E_p = p$ and
\begin{eqnarray}
L & \equiv & 1- \frac{\omega}{k} \log\left[\frac{\omega_+}{\omega_-}\right] \, ,\\
L_\pm &\equiv & \log\left[\frac{p+\omega_+}{p+\omega_-}\right] \pm  \log\left[\frac{p-\omega_+}{p-\omega_-}\right] \, , \\
M &\equiv & \left(p+\omega_+ \right)\left(p+\omega_- \right) \log\left[\frac{p+\omega_+}{p+\omega_-}\right] - \left(p-\omega_+ \right)\left(p-\omega_- \right)  \log\left[\frac{p-\omega_+}{p-\omega_-}\right]
\end{eqnarray}
with $\omega_\pm  \equiv  (\omega \pm k)/2$. We assume a negligible particle/antiparticle asymmetry for quarks, and consistently we set $f_{\rm FD}(E_p) = \bar{f}_{\rm FD}(E_p)$. We find
\begin{align}
\left. \pi_{\rm L} \right|_{\rm q} = & \, - \frac{g_s^2}{8\pi^2} \left(\frac{\omega^2 - k^2}{k^2}\right) \int dp \frac{p^2}{E_p} f_{\rm FD}(E_p)  \left[8 - \frac{(\omega^2 - k^2) + 4 E_p^2 +4E_p\omega}{pk}N_+ \right. \nonumber \\ & \qquad \qquad \qquad \qquad\qquad\qquad\qquad \qquad \left. + \frac{(\omega^2 - k^2) + 4 E_p^2 - 4E_p\omega}{pk}N_- \right] \ , \\
\left. \pi_{\rm T} \right|_{\rm q} = & \,  -\frac{1}{2}\left.\pi_{\rm L}\right|_{\rm q} + \frac{g_s^2}{8\pi^2} \int dp \frac{p^2}{E_p} f_{\rm FD}(E_p)  \left[ 8 - \frac{2 m_q^2 + (\omega^2 - k^2)}{pk} \left(N_+ - N_-\right)\right] \ ,
\end{align}
where
\be
N_\pm = \log \left[\frac{2(E_p \omega + p k)\pm (\omega^2-k^2)}{2(E_p \omega - p k)\pm (\omega^2-k^2)}\right] \, .
\ee
For a massless quark ($m_q^2=0$), we recover the analytic expressions in the literature~\cite{Salvio:2013iaa}.

With the gluon self-energy at our disposal, we can evaluate the spectral densities
\begin{eqnarray}
\rho_{\rm T} & = & -2\,{\rm Im}\,\frac{1}{\omega^2 - k^2 - \pi_{\rm T}} \, ,\\
\rho_{\rm L} & = & -2\,{\rm Im}\,\frac{\omega^2-k^2}{k^2}\frac{1}{\omega^2 - k^2 - \pi_{\rm L}} \, .
\end{eqnarray}
The imaginary part must be extracted according to the rescription $\omega \rightarrow \omega + i0^+$. In order to deal with technical difficulties in the numerical analysis, we take into account the spectral densities rewritten as
\be
\rho_i(k) = 2 \pi Z_i \left(k\right) \delta\left(\omega^2 - \omega_i^2\left(k\right)\right) + \rho_i^{\rm cont} \, ,
\ee
where $i ={\rm \left(T,L\right)}$ and $Z_i$ indicates the residues at the poles of $\omega = \pm \omega_i\left(k\right)$ which are located in the time-like region ($|\omega| > k$), and we consider the contribution from the continuum parts $\rho_i^{\rm cont}$ only in the space-like region ($|\omega| < k$)~\cite{Rychkov:2007uq}.

\begin{figure}
\centering
 \includegraphics[width=.49\linewidth]{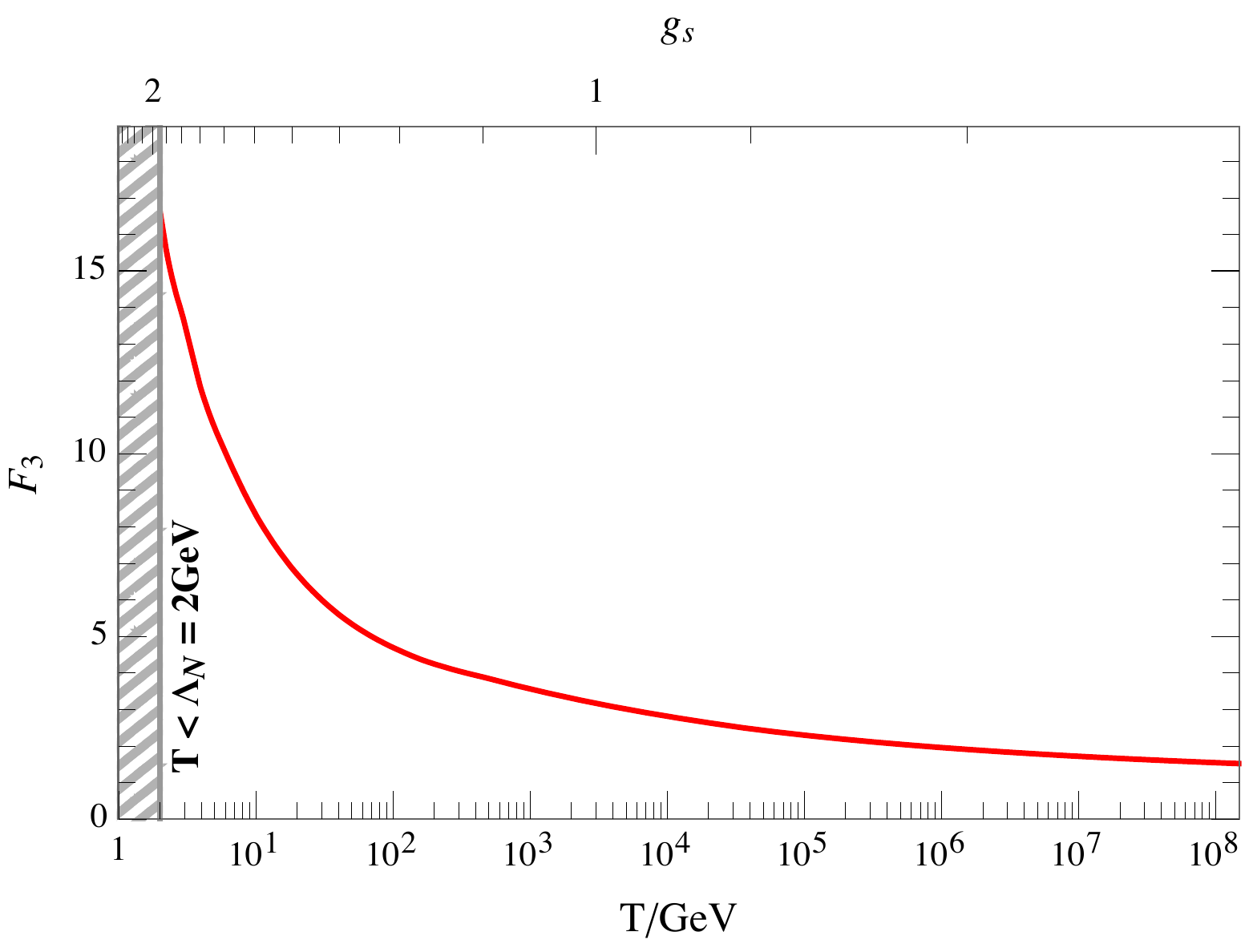}
  \includegraphics[width=.49\linewidth]{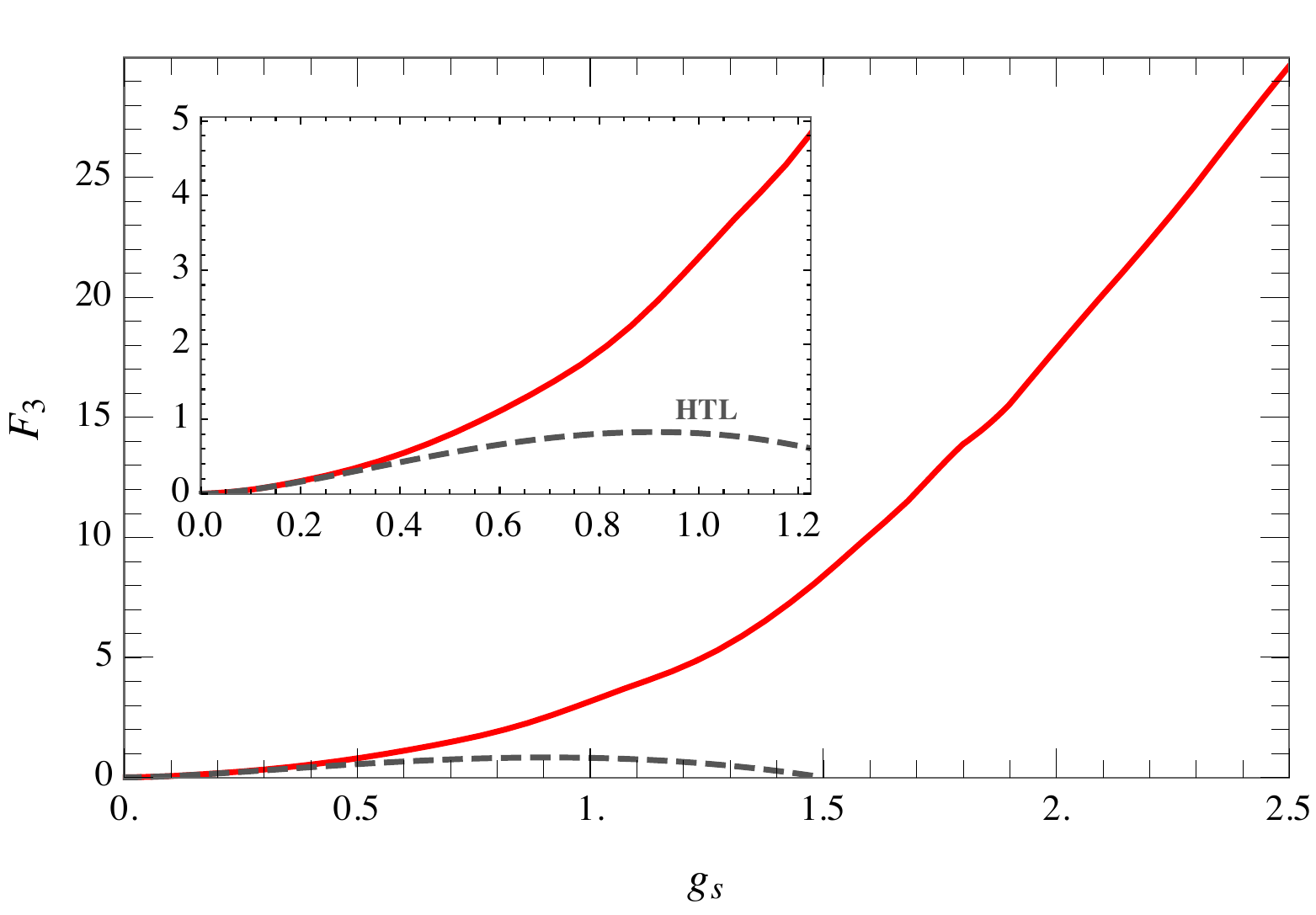}
\caption{\em The control function $F_3$ defined in Eq.~\eqref{eq:F3function} as a function of the temperature (left) and the strong coupling $g_s$ (right). The dashed gray line on the right panel corresponds to the Hard Thermal Loop (HTL) approximation~\cite{Braaten:1991dd,Graf:2010tv}, and it is valid only for small couplings $g_s < 0.5$.}
\label{fig:F3}
\end{figure}

We express the axion two-point function in terms of spectral densities
\begin{eqnarray}
\Pi_a^{<} & = & \frac{d_g}{4\pi^3} \left(\frac{g_s^2}{32\pi^2 f_a}\right)^2 \frac{1}{p_a}\int_{-\infty}^{\infty} d k_0 \int_0^{\infty} dk \int_{\left| k-p_a\right|}^{k+p_a} dq  \, f_B \left(k_0\right) f_B \left(E_a - k_0\right) \nonumber \\
&& \times \Bigg\{ \left(\rho_T\left(k\right)\rho_L\left(q\right)+\rho_L\left(k\right)\rho_T\left(q\right)\right) \left[\left(k+q\right)^2- p_a^2\right] \left[p_a^2-\left(k - q\right)^2\right]  \nonumber \\
&& + \rho_T\left(k\right)\rho_T\left(q\right)\left[\left(\frac{k_0^2}{k^2}+\frac{q_0^2}{q^2}\right)\left(\left(k^2- p_a^2+q^2\right)^2 + 4 k^2 q^2\right)+8k_0 q_0\left(k^2 + q^2-p_a^2\right)\right] \Bigg\} \, , \nonumber \\
&& 
\label{eq:nonTimeOrderedAxionPropagator}
\end{eqnarray}
where $d_g = 8$ is the dimension of the $SU(3)$ strong gauge group and the four momentum of the gluons are $K^\mu = (k_0, k\hat{k})$ and $Q^\mu = P_a^\mu - K^\mu = (q_0, q \hat{q})$. 
We integrate numerically Eq.~\eqref{eq:AxionProductionRate} where we use Eq.~\eqref{eq:nonTimeOrderedAxionPropagator} for the axion self-energy. We employ the `\texttt{RunDec}'~\cite{Chetyrkin:2000yt} code to account for the running of the strong coupling constant up to four loops. The numerical result of the control function $F_3$ defined in Eq.~\eqref{eq:F3function} is shown in Fig.~\ref{fig:F3}; the left and right plots illustrate the value of $F_3$ in the function of temperature $T$ and the strong coupling $g_s$, respectively, including the decoupling of quarks at different temperatures.

\subsection*{Thermal masses for the electroweak sector}

The tree-level scalar potential for the 2HDM is given in Eq.~\eqref{pot2hd}. The thermal evolution of electroweak sector in the 2HDM is the subject of Refs.~\cite{Cline:1996mga,Cline:2011mm,Basler:2016obg,Carrington:1991hz}, and one-loop thermal corrections to the potential read
\be
V_{\rm th}\left(H_i, T \right) = \frac{T^4}{2\pi^2} \sum_{b,f}\Bigg[n_{b} J_B\left[m_b^2(H_i)/T^2\right] - n_{f} J_F\left[m_f^2(H_i)/T^2\right]\Bigg] \,.
\label{eq:vtherm}
\ee
Here, we include contributions from bosons and fermions, and we denote their (Higgs fields dependent) masses $m_b(H_i)$ and $m_f(H_i)$, respectively. The relative minus sign is due to fermion fields running in the loop. The dimensionless quantities $n_b$ and $n_f$ count the number of internal degrees of freedom. The function $J_{F}$ and $J_{B}$ are defined as follows~\cite{Laine:2016hma,Bellac:2011kqa}
\begin{eqnarray}
J_B\left[m_b^2/T^2\right] & = & \int_0^\infty dx x^2 \log\left[1-e^{-\sqrt{x^2 + m_b^2/T^2}}\right] \,, \\ 
J_F\left[m_f^2/T^2\right] & = & \int_0^\infty dx x^2 \log\left[1 + e^{-\sqrt{x^2 + m_f^2/T^2}}\right]  \, .
\end{eqnarray}
At high temperatures, more specifically for $m_{b,f} / T < 1.8$~\cite{Dorsch:2013wja}, we can approximate 
\begin{eqnarray}
J_B\left[m_b^2/T^2\right] & \approx &  -\frac{\pi^4}{45} + \frac{\pi^2}{12}\frac{m_b^2}{T^2} - \frac{\pi}{6} \left(\frac{m_b^2}{T^2}\right)^{3/2} - \frac{1}{32}\frac{m_b^4}{T^4} \log\frac{m_b^2}{a_bT^2} + \mathcal{O}\left( \frac{m_b^6}{T^6}\right) \, , \\  
J_F\left[m_f^2/T^2\right] & \approx  &  \frac{7\pi^4}{360} - \frac{\pi^2}{24}\frac{m_f^2}{T^2}  - \frac{1}{32}\frac{m_f^4}{T^4} \log\frac{m_f^2}{a_f T^2} + \mathcal{O}\left( \frac{m_f^6}{T^6}\right) \, ,
\end{eqnarray}
where $a_f= \pi^2 \exp(3/2-2\gamma_E)$ and $a_b= 16a_f$ with the Euler-Mascheroni number $\gamma_E$.

In the electroweak symmetric phase, the global minimum is located with the Higgs fields at the origin, $v_{u,d}(T>T_{\rm EWPT}) = 0$. The thermal corrections to the quadratic fluctuations of the Higgs fields around the origin read
\begin{eqnarray}
\label{eq:therm-higgs}
V^{T}_{\rm 2HDM} = \delta \mu_u^2\left(T\right) H_u^\dagger H_u + \delta \mu_d^2\left(T\right) H_d^\dagger H_d\,,
\end{eqnarray}
where
\begin{align}
\delta \mu_u^2\left(T\right) = & \, \frac{T^2}{24}\left( \frac92 g^2 + \frac32 {g^\prime}^2 +6\frac{\sum_i \left(\hat{Y}_i^{(u)}\right)^2}{\sin^2\beta}+ 3 \lambda_{1}+4 \lambda_{3}+2 \lambda_{4} \right) \ , \\ 
\delta \mu_d^2\left(T\right) = &  \,  \frac{T^2}{24}\left( \frac92 g^2 + \frac32 {g^\prime}^2 +6\frac{\sum_i \left(\hat{Y}_i^{(d)}\right)^2}{\cos^2\beta}+2\frac{\sum_i \left(\hat{Y}_i^{(e)}\right)^2}{\cos^2\beta} + 3 \lambda_{2}+4 \lambda_{3}+2 \lambda_{4} \right) \ .
\end{align}
The factors of $\cos\beta$ or $\sin\beta$ come from the Yukawa interactions to reproduce the SM fermion spectrum. One can obtain the thermally corrected mass matrices for the Higgs sector by $\partial^2 \left(V_{\rm 2HDM} + V_{\rm 2HDM}^T \right)/\partial H_{i}\partial{H_{j}}$ where $V_{\rm 2HDM}$ denotes the zero-temperature and tree-level scalar potential in Eq.~\eqref{pot2hd}. These corrected masses will be used in the calculation of cross sections for axion production above the weak scale (see App.~\ref{app:XS} for more details).


\section{Conventions and Useful Results III: Cross Sections}
\label{app:XS}

We present results for the cross section for each binary collisions producing hot axions in the final state. With the only exception of thermal gluon scatterings, which we evaluated in the previous Appendix, these are the processes that we have to account for. We express each cross section as a function of the (squared of the) center of mass energy, and we evaluate the thermal average as prescribed by Eq.~\eqref{rateinter}. The interaction rate in Eq.~\eqref{eq:rateij} is what we need to incorporate into our Boltzmann equation analysis. We employ the {\tt FeynCalc} package to check all analytical expressions for cross sections \cite{Shtabovenko:2016sxi,Shtabovenko:2020gxv}.

\subsection*{KSVZ axion above the heavy fermion threshold}

At temperatures larges than $m_\Psi$ the dominant processes for axion production are scatterings of the heavy colored PQ fermion through the interaction in Eq.~\eqref{eq:KSVZhighELagrangian}. We set the axion decay constant $f_a$ to normalize the gluon anomalous coupling, and we find the cross sections
\begin{align}
& \, \sigma_{\Psi + \bar{\Psi} \rightarrow g \, a}  =  \frac{g_s^2}{9\pi f_a^2} \frac{\frac{m_{\Psi}^2}{s}}{1-4 \frac{m_{\Psi}^2}{s}}  \arctanh\left[\sqrt{1-4 \frac{m_{\Psi}^2}{s}}\right] ] \ , \\
& \, \sigma_{\Psi + g \rightarrow \Psi +   a}  = \sigma_{\bar{\Psi} + g \rightarrow \bar{\Psi} +   a} =  \frac{g_s^2}{192\pi f_a^2}  \frac{\frac{m_{\Psi}^2}{s}}{1-\frac{m_{\Psi}^2}{s}} \left( 4 \frac{m_{\Psi}^2}{s} - \frac{m_{\Psi}^4}{s^2} -3 -2 \log \left[ \frac{m_{\Psi}^2}{s} \right] \right) \ .
\end{align}

\subsection*{DFSZ axion above the heavy Higgs bosons threshold}

In this phase, we find it convenient to work with the linear realization of PQ symmetry with the only axion interaction in Eq.~\eqref{eq:DFSZhighECoupling}. Quark-antiquark annihilations have cross sections
\begin{align}
\label{eq:XSDFSZhighEuuha} \sum_{ij} \sigma_{\bar{Q}_L^i + u_R^j \rightarrow H_d^\dagger + a} = & \,  \sum_{ij}\sigma_{Q_L^i + \bar{u}_R^j \rightarrow H_d + a}  =  \frac{B^2}{576\pi f_a^2 s^2}\frac{\sum_i \left| \hat{Y}^{(u)}_i \right|^2}{\sin^2\beta} \frac{ 1-\frac{m_{H_d}^2}{s}}{\left(1-\frac{m_{H_u}^2}{s}\right)^2} \ , \\  
\label{eq:XSDFSZhighEddha} \sum_{ij} \sigma_{\bar{Q}_L^i + d_R^j \rightarrow H_u^\dagger + a} = & \, \sum_{ij}\sigma_{Q_L^i + \bar{d}_R^j \rightarrow H_u + a}  =  \frac{B^2}{576\pi f_a^2 s^2} \frac{\sum_i \left| \hat{Y}^{(d)}_i \right|^2}{\cos^2\beta} \frac{1-\frac{m_{H_u}^2}{s}}{\left(1-\frac{m_{H_d}^2}{s}\right)^2} \ .
\end{align}
The indices $i,j$ run over quark flavors, $m_{H_{u,d}}$ are the thermally corrected masses of the each Higgs doublet, and $\hat{Y}^{(u,d)}$ are the diagonalized Yukawa matrices appearing in Eq.~\eqref{eq:LYukawaSMdiag} for up-quarks and down-quarks, respectively. The expressions in Eqs.~\eqref{eq:XSDFSZhighEuuha} and \eqref{eq:XSDFSZhighEddha} have poles at the mass of the Higgs boson different from the one on the external state, and this can lead to divergences when we integrate over the phase space. We regularize such an unpleasant behavior with Breit-Wigner corrections to internal propagators, and we use the decay widths
\begin{align}
\Gamma_{H_u} \simeq \left(3 \sum \frac{|\hat{Y}^{(u)}|^2}{\sin^2\beta}\right) \frac{m_{H_u}}{8\pi}  \ , \qquad \Gamma_{H_d} \simeq \left(3 \sum 
\frac{|\hat{Y}^{(d)}|^2}{\cos^2\beta} + \sum \frac{|\hat{Y}^{(l)}|^2}{\cos^2\beta}\right) \frac{m_{H_d}}{8\pi} \ .
\end{align}

When Higgs fields appear in the initial state we have the cross sections
\begin{align}
& \sum_{ij}\sigma_{Q_L^i + H_d^\dagger \rightarrow u_R^j +  a}    =   \sum_{ij}\sigma_{u_R^j + H_d \rightarrow Q_L^i +  a} = \sum_{ij}\sigma_{\bar{Q}_L^i + H_d \rightarrow \bar{u}_R^j +  a}    =   \sum_{ij}\sigma_{\bar{u}_R^j + H_d^\dagger \rightarrow \bar{Q}_L^i +  a}  \nonumber \\ \label{eq:XSDFSZhighEuhua}
 =  & \frac{B^2}{288\pi f_a^2 s^2} \frac{\sum_i \left| \hat{Y}^{(u)}_i \right|^2}{\sin^2\beta} \frac{1}{\left(1-\frac{m_{H_d}^2}{s}\right)^2}  \left(2 \arccoth\left[\frac{1-\frac{m_{H_d}^2-2m_{H_u}^2}{s}}{1-\frac{m_{H_d}^2}{s}}\right] - \frac{1-\frac{m_{H_d}^2}{s}}{1-\frac{m_{H_d}^2-m_{H_u}^2}{s} }\right)  \ , \\
& \sum_{ij}\sigma_{Q_L^i + H_u^\dagger \rightarrow d_R^j +  a}    =   \sum_{ij}\sigma_{d_R^j + H_u \rightarrow Q_L^i +  a}=\sum_{ij}\sigma_{\bar{Q}_L^i + H_u \rightarrow \bar{d}_R^j +  a}    =   \sum_{ij}\sigma_{\bar{d}_R^j + H_u^\dagger \rightarrow \bar{Q}_L^i +  a} \nonumber  \\
 =  & \frac{B^2}{288\pi f_a^2 s^2} \frac{\sum_i \left| \hat{Y}^{(d)}_i \right|^2}{\cos^2\beta}   \frac{1}{\left(1-\frac{m_{H_u}^2}{s}\right)^2}  \left(2 \arccoth\left[\frac{1-\frac{m_{H_u}^2-2m_{H_d}^2}{s}}{1-\frac{m_{H_u}^2}{s}}\right] - \frac{1-\frac{m_{H_u}^2}{s}}{1-\frac{m_{H_u}^2-m_{H_d}^2}{s} }\right)   \ ,
\label{eq:XSDFSZhighEdhda}
\end{align}
for scatterings mediated by up-type and down-type quark Yukawa interactions, respectively. The expressions in Eqs.~\eqref{eq:XSDFSZhighEddha} and Eq.~\eqref{eq:XSDFSZhighEdhda} describe lepton scatterings with $\hat{Y}^{(d)} \rightarrow \hat{Y}^{(l)}$.

The doublets $H_u$ and $H_d$ are not mass eigenstate but we provide a simple two-step procedure to convert the cross sections above into the ones for mass eigenstates.~\footnote{The axion coupling in Eq.~\eqref{eq:DFSZhighECoupling} possesses the $SO(2)$ symmetric property for the Higgs doublets, and the cubic vertex can be written as $i B (a/f_a) H_i^\dagger \epsilon^{ij} H_j$ with $H_j = \left(H_u , i \sigma^2 H_d^* \right)$ and $\epsilon^{12} = -\epsilon^{21}=1$.} 

\begin{enumerate}
\item We introduce the temperature dependent mixing angle $\alpha$ 
\be
\left(\begin{tabular}{c}
$H_u$ \\ $i\sigma^2 H_d^*$
\end{tabular} \right)
 = \left(\begin{tabular}{cc}
$\sin\alpha$ & $\cos\alpha$ \\ $-\cos\alpha$ & $\sin\alpha$
\end{tabular} \right) \left(\begin{tabular}{c}
$H_1$ \\ $H_2$
\end{tabular} \right) \, ,
\label{eq:2HDMmixingAngle}
\ee
where $H_1$ and $H_2$ denote the lighter and the heavier physical states, respectively. As expected, we recover $\alpha\approx \beta$ at low temperatures, $T \ll (2B/\sin 2\beta)^{1/2}$.

\item We can replace the interaction states $H_{u,d}$ in Eqs.~\eqref{eq:XSDFSZhighEuuha}-\eqref{eq:XSDFSZhighEdhda} with the physical eigenstates $H_{1,2}$ through the rotation above. As an example
\begin{eqnarray}
\sum_{ij} \sigma_{\bar{Q}_L^i + u_R^j \rightarrow H_1^\dagger + a} & = & \cos^2\alpha \frac{ B^2}{576\pi f_a^2s^2}\frac{\sum_i \left| \hat{Y}^{(u)}_i \right|^2}{\sin^2\beta}\frac{\left( 1-m_{H_1}^2/s\right)}{\left(1-m_{H_2}^2/s\right)^2} \, , \\
\sum_{ij} \sigma_{\bar{Q}_L^i + u_R^j \rightarrow H_2^\dagger + a} & = & \sin^2\alpha\frac{B^2}{576\pi f_a^2s^2}\frac{\sum_i \left| \hat{Y}^{(u)}_i \right|^2}{\sin^2\beta} \frac{\left( 1-m_{H_2}^2/s\right)}{\left(1-m_{H_1}^2/s\right)^2} \, .
\end{eqnarray}
\end{enumerate}

Furthermore, there are the additional contributions to the axion production from gauge boson scatterings. In this case, since the Higgs doublets are identical in terms of the gauge charge assignment  there are no mixing angles appearing in the cross section for physical states. After straightforward calculations, we find the cross section for the case of the initial gauge boson state 
\be
 \sigma_{H_i + V_\mu \rightarrow H_{j\neq i}^{\dagger} +  a}   =   \sigma_{H_i^{\dagger} + V_\mu \rightarrow H_{j\neq i} +  a}
=  \frac{g_V^2 B^2}{144\pi f_a^2 s^2}  \frac{ \left(1+s_j\right)\arctanh\left[1-2\frac{s_j}{1+s_j}\right]+s_j-1}{\left(1-s_i\right)^3}  
\ee
with $s_i \equiv m_{H_i}^2/s$ and $g_{V}$ the corresponding gauge coupling. If the gauge boson in the final state we find 
\begin{eqnarray}
&&\sigma_{H_i + H_{j\neq i} \rightarrow V_\mu +  a}    =   \sigma_{H_i^\dagger + H_{j\neq i}^\dagger \rightarrow V_\mu +  a}   \nonumber \\
& =  & \frac{d_V g_V^2 B^2}{288\pi f_a^2s^2}  \frac{1}{1-2\left
(s_i+s_j\right)+ \left(s_i-s_j\right)^2}  \Bigg(-2 \sqrt{1-2\left
(s_i+s_j\right)+ \left(s_i-s_j\right)^2} \nonumber  \\
&& + \arctanh\left[\frac{\sqrt{1-2\left
(s_i+s_j\right)+ \left(s_i-s_j\right)^2}}{1 +\left(s_i-s_j\right)}\right]   + \arctanh\left[\frac{\sqrt{1-2\left
(s_i+s_j\right)+ \left(s_i-s_j\right)^2}}{1 - \left(s_i-s_j\right)}\right]\nonumber \\
&& + \left(1- 2s_i-2s_j\right) \arccoth\left[\frac{1 - \left(s_i+s_j\right)}{\sqrt{1-2\left
(s_i+s_j\right)+ \left(s_i-s_j\right)^2}}\right] \Bigg) \ .
\end{eqnarray}

\subsection*{DFSZ axion below the heavy Higgs bosons and above the EWPT}

As the universe cools down further below the heavy Higgs boson masses, such heavy degrees of freedom are integrated out and the bath contains effectively only SM particles. We employ here the non-linear realization of the PQ symmetry with axion couplings given in Eq.~\eqref{eq:DFSZewEffCoup}. Matrix elements of SM fermion scatterings depend only on the combinations~\cite{Arias-Aragon:2020shv}
\be
\begin{split}
\hat{c}_u & = - c_{Q_L} + c_{u_R} \ , \\
\hat{c}_d & = - c_{Q_L} + c_{d_R} \ , \\
\hat{c}_e & = - c_{L_L} + c_{e_R}
\end{split} 
\label{eq:combinationEW}
\ee
for the up-type quarks $u=(u,c,t)$, the down-type quarks $d=(d,s,b)$, and the charged-lepton $e=(e,\mu,\tau)$, respectively. The chirality flip mentioned in the main text is such that only processes with the components of the complex Higgs doublet contribute to the rate. Thus the rate will be dominated by third generation SM fermions since their interaction strength with the Higgs field is proportional to the Yukawa couplings. We parameterize the Higgs field $H^T = (\chi_+ \; , \chi_0)$, where each doublet component is a complex scalar field, and we also introduce $\chi_{-} \equiv \chi_+^\dag$ and $\chi_0^c \equiv \chi_0^\dag$. The scattering cross sections take a particular simple form once we ignore CKM factors, which lead only to few percent corrections since the rate is controlled by third generation fermions. If both initial state particles are SM fermions 
\begin{align}
\sigma_{f \bar{f} \rightarrow \chi_0 a} = & \, \sigma_{f \bar{f} \rightarrow \chi^c_0 a} = \frac{\hat{c}^2_f \, y_f^2}{64 \pi f_a^2} \ , \\
\sigma_{f \bar{f^\prime} \rightarrow \chi_+ a} = & \, \sigma_{f^\prime \bar{f} \rightarrow \chi_- a} = \frac{\hat{c}^2_f \, y_f^2 + \hat{c}_{f^\prime} \, y_{f^\prime}^2}{64 \pi f_a^2} \ .
\end{align}
Here, $f$ is a generic SM fermions and $y_f$ the associated Yukawa coupling in the basis where such a coupling is diagonal. The fermion $f^\prime$ is the weak-isospin partner of $f$. If a scalar appears in the initial state we have
\begin{align}
\sigma_{f \chi_0 \rightarrow f a} = & \, \sigma_{\bar{f} \chi^c_0 \rightarrow \bar{f} a} = \sigma_{f \chi_0^c \rightarrow f a} =
\sigma_{\bar{f} \chi_0 \rightarrow \bar{f} a} = \frac{\hat{c}^2_f \, y_f^2}{64 \pi f_a^2} \ ,\\
\sigma_{f \chi_- \rightarrow f^\prime a} = & \, \sigma_{\bar{f} \chi_+ \rightarrow \bar{f^\prime} a} = \frac{\hat{c}^2_f \, y_f^2 + \hat{c}_{f^\prime} \, y_{f^\prime}^2}{64 \pi f_a^2} \ .
\end{align}

\subsection*{DFSZ axion below the EWPT and above the QCDPT}

Below the EWPT, SM fermions and gauge bosons acquire a finite mass. We perform calculations in this phase with the PQ symmetry non-linearly realized, and  cross sections still depend only on the same combinations in Eq.~\eqref{eq:combinationEW}. We report here explicit expressions for quark scattering cross sections, the lepton case is a straightforward generalization. Here, we generalize the results provided by Ref.~\cite{Arias-Aragon:2020shv} by accounting also for flavor-violating processes whose contributions lead to corrections proportional to CKM factors.

We begin with quark/antiquark annihilations. For final state gluons we have
\be
\sigma_{q+\bar{q} \rightarrow g + a} = \frac{\hat{c}_{q}^2 g_s^2}{9\pi f_a^2} \frac{m_{q}^2/s}{1-4m_{q}^2/s}\arctanh\left[\sqrt{1-\frac{4m_{q}^2}{s}}\right] \ .
\ee
If we replace the gluon with the SM Higgs boson, we find
\be
\sigma_{q_i+\bar{q}_i \rightarrow h + a} = \frac{\hat{c}_{q}^2 \left| \hat{Y}^{(q)}_i\right|^2}{64\pi f_a^2} \frac{1-\frac{m_h^2}{s}}{1-\frac{4m_{q_i}^2}{s}} \left(\sqrt{1-\frac{4m_{q_i}^2}{s}}-\frac{4m_{q_i}^2}{s}\arctanh\left[\sqrt{1-\frac{4m_{q_i}^2}{s}}\right]\right)  \ .
\ee
Quarks can also annihilate to weak gauge bosons. For final state $Z$ bosons we have
\begin{align}
\sigma_{u_i+\bar{u}_i \rightarrow Z + a} = & \, \frac{\hat{c}_{u}^2 \left| \hat{Y}^{(u)}_i\right|^2}{576\pi f_a^2} \frac{1-\frac{m_Z^2}{s}}{\frac{m_Z^2}{s}\sqrt{1-\frac{4m_{u_i}^2}{s}}} \Bigg(9\frac{m_Z^2}{s}\left(1-2\frac{m_Z^2}{s}\right) \nonumber \\
& \left. +4 \frac{17 \frac{m_Z^4}{s^2}+32\frac{m_W^4}{s^2}-\frac{m_Z^2}{s}\left(9\frac{m_{u_i}^2}{s}+40\frac{m_W^2}{s}\right)}{\sqrt{1-\frac{4m_{u_i}^2}{s}}}\arctanh\left[\sqrt{1-\frac{4m_{u_i}^2}{s}}\right]\right)  \ , \\
\sigma_{d_i+\bar{d}_i \rightarrow Z + a} = & \, \frac{\hat{c}_{d}^2 \left| \hat{Y}^{(d)}_i\right|^2}{576\pi f_a^2} \frac{1-\frac{m_Z^2}{s}}{\frac{m_Z^2}{s}\sqrt{1-\frac{4m_{d_i}^2}{s}}} \Bigg(9\frac{m_Z^2}{s}\left(1-2\frac{m_Z^2}{s}\right) \nonumber  \\
& \left. +4 \frac{5 \frac{m_Z^4}{s^2}+8\frac{m_W^4}{s^2}-\frac{m_Z^2}{s}\left(9\frac{m_{d_i}^2}{s}+4\frac{m_W^2}{s}\right)}{\sqrt{1-\frac{4m_{d_i}^2}{s}}}\arctanh\left[\sqrt{1-\frac{4m_{d_i}^2}{s}}\right]\right)  
\end{align}
for up and down quarks, respectively. Quark/antiquark annihilations to the charged weak gauge boson $W^\pm$ can be flavor-changing processes, and their cross sections read
\be
\begin{split}
& \sigma_{u_i+\bar{d}_j \rightarrow W^+ + a}  = \sigma_{d_j+\bar{u}_i \rightarrow W^- + a}  =  \left| V_{\rm CKM}^{ij}\right|^2 \frac{g^2 }{128\pi f_a^2} \frac{1-\frac{m_W^2}{s}}{\frac{m_W^2}{s} \left[1-\frac{(m_{u_i}+m_{d_j})^2}{s} \right] \left[1-\frac{(m_{u_i}-m_{d_j})^2}{s} \right]}  \\
&\times \left[ \rule{0cm}{1cm} \left(\hat{c}_u^2 \frac{m_{u_i}^2}{s}+\hat{c}_d^2 \frac{m_{d_j}^2}{s}\right)\left(1-\frac{2m_W^2}{s}\right)\sqrt{1-\frac{\left(m_{u_i}+m_{d_j}\right)^2}{s}}\sqrt{1-\frac{\left(m_{u_i}-m_{d_j}\right)^2}{s}} + \right. \\
&  +2\hat{c}_u^2\frac{m_{u_i}^2}{s} \frac{2m_W^2-m_{u_i}^2+m_{d_j}^2}{s} \arccoth\left[\frac{1+\frac{m_{u_i}^2-m_{d_j}^2}{s}}{\sqrt{1-\frac{\left(m_{u_i}+m_{d_j}\right)^2}{s}}\sqrt{1-\frac{\left(m_{u_i}-m_{d_j}\right)^2}{s}}}\right] +\\
& + 2\hat{c}_d^2\frac{m_{d_j}^2}{s} \frac{2m_W^2+m_{u_i}^2-m_{d_j}^2}{s} \arccoth\left[\frac{1-\frac{m_{u_i}^2-m_{d_j}^2}{s}}{\sqrt{1-\frac{\left(m_{u_i}+m_{d_j}\right)^2}{s}}\sqrt{1-\frac{\left(m_{u_i}-m_{d_j}\right)^2}{s}}}\right] +\\
& \left. \rule{0cm}{1cm} + 4\hat{c}_u\hat{c}_d\frac{m_{u_i}^2m_{d_j}^2}{s^2}  \arccoth\left[\frac{1-\frac{m_{u_i}^2+m_{d_j}^2}{s}}{\sqrt{1-\frac{\left(m_{u_i}+m_{d_j}\right)^2}{s}}\sqrt{1-\frac{\left(m_{u_i}-m_{d_j}\right)^2}{s}}}\right] \right] \ .
\end{split}
\ee

We switch to quark or antiquark scattering with SM bosons. For a gluon we have
\be
\sigma_{q_i + g \rightarrow q_i + a} = \sigma_{\bar{q}_i + g \rightarrow \bar{q}_i + a} = \frac{\hat{c}_{q}^2 g_s^2}{192\pi f_a^2} \frac{\frac{m_{q_i}^2}{s}}{1-\frac{m_{q_i}^2}{s}} \left[ 4 \frac{m_{q_i}^2}{s} - \frac{m_{q_i}^4}{s^2} -3 -2 \log \left[ \frac{m_{q_i}^2}{s} \right] \right] \ .
\ee
For a SM Higgs boson in the initial state we find
\be
\begin{split}
& \sigma_{q_i+ h \rightarrow q_i + a} = \sigma_{\bar{q}_i+ h \rightarrow \bar{q}_i + a}
=  \frac{\hat{c}_{q}^2 \left| \hat{Y}^{(q)}_i\right|^2}{64\pi f_a^2} \frac{1-\frac{m_{q_i}^2}{s}}{\left(1-\frac{\left(m_{h}+m_{q_i}\right)^2}{s}\right)\left(1-\frac{\left(m_{h}-m_{q_i}\right)^2}{s}\right)} \\
&\times  \left( \rule{0cm}{1cm} \left(1-\frac{m_h^2-m_{q_i}^2}{s}\right) \sqrt{1-\frac{\left(m_{h}+m_{q_i}\right)^2}{s}}\sqrt{1-\frac{\left(m_{h}-m_{q_i}\right)^2}{s}} +\right. \\
& \left. \rule{0cm}{1cm} - 4 \frac{m_{q_i}^2}{s}  \arccoth\left[\frac{1- \frac{m_h^2-m_{q_i}^2}{s}}{\sqrt{1-\frac{\left(m_{h}+m_{q_i}\right)^2}{s}}\sqrt{1-\frac{\left(m_{h}-m_{q_i}\right)^2}{s}}}\right]  \right) \ .
\end{split}
\ee
In the case of incident $Z$ bosons, cross sections read
\be
\begin{split}
& \sigma_{u_i+ Z \rightarrow u_i + a}  = \sigma_{\bar{u}_i + Z \rightarrow \bar{u}_i + a}
= \frac{\hat{c}_{u}^2 \left| \hat{Y}^{(u)}_i\right|^2}{1728\pi f_a^2} \frac{1-\frac{m_{u_i}^2}{s}}{\frac{m_Z^2}{s}\sqrt{1-\frac{\left(m_Z+m_{u_i}\right)^2}{s}}\sqrt{1-\frac{\left(m_Z-m_{u_i}\right)^2}{s}}}  \\
&\times \left[ \rule{0cm}{1cm}  9 \frac{m_Z^2}{s} - \frac{m_Z^2-m_{u_i}^2}{s}\left(17\frac{m_Z^4}{s^2}+32\frac{m_W^4}{s^2}-40 \frac{m_Z^2m_W^2}{s^2}\right)  + \right. \\
&  - 3 \left(8\frac{m_Z^4}{s^2}+32\frac{m_W^4}{s^2}-\frac{m_Z^2}{s}\left(3\frac{m_{u_i}^2}{s}+40\frac{m_W^2}{s}\right)\right) +\\
&  + 4 \frac{17\frac{m_Z^4}{s^2}+32\frac{m_W^4}{s^2}- \frac{m_Z^2}{s}\left(9\frac{m_{u_i}^2}{s}+40\frac{m_W^2}{s}\right)}{\sqrt{1-\frac{\left(m_Z+m_{u_i}\right)^2}{s}}\sqrt{1-\frac{\left(m_Z-m_{u_i}\right)^2}{s}}}   \left.  \rule{0cm}{1cm} \arccoth\left[\frac{1- \frac{m_Z^2-m_{u_i}^2}{s}}{\sqrt{1-\frac{\left(m_{Z}+m_{u_i}\right)^2}{s}}\sqrt{1-\frac{\left(m_{Z}-m_{u_i}\right)^2}{s}}}\right] \right] \ ,
\end{split}
\ee
\be
\begin{split}
& \sigma_{d_i+ Z \rightarrow d_i + a}  = \sigma_{\bar{d}_i + Z \rightarrow \bar{d}_i + a} 
= \frac{\hat{c}_{d}^2 \left| \hat{Y}^{(d)}_i\right|^2}{1728\pi f_a^2} \frac{1-\frac{m_{d_i}^2}{s}}{\frac{m_Z^2}{s}\sqrt{1-\frac{\left(m_Z+m_{d_i}\right)^2}{s}}\sqrt{1-\frac{\left(m_Z-m_{d_i}\right)^2}{s}}}  \\
& \times \left[ \rule{0cm}{1cm} 3 \left(4\frac{m_Z^4}{s^2}-8\frac{m_W^4}{s^2}+\frac{m_Z^2}{s}\left(3+3\frac{m_{d_i}^2}{s}+4\frac{m_W^2}{s}\right)\right) + \right.\\
&- \frac{m_Z^2-m_{d_i}^2}{s}\left(5\frac{m_Z^4}{s^2}+8\frac{m_W^4}{s^2}-4 \frac{m_Z^2m_W^2}{s^2}\right) +\\
&  + 4 \frac{5\frac{m_Z^4}{s^2}+8\frac{m_W^4}{s^2}- \frac{m_Z^2}{s}\left(9\frac{m_{d_i}^2}{s}+4\frac{m_W^2}{s}\right)}{\sqrt{1-\frac{\left(m_Z+m_{d_i}\right)^2}{s}}\sqrt{1-\frac{\left(m_Z-m_{d_i}\right)^2}{s}}}   \left.  \rule{0cm}{1cm} \arccoth\left[\frac{1- \frac{m_Z^2-m_{d_i}^2}{s}}{\sqrt{1-\frac{\left(m_{Z}+m_{d_i}\right)^2}{s}}\sqrt{1-\frac{\left(m_{Z}-m_{d_i}\right)^2}{s}}}\right] \right] 
\end{split}
\ee
for up and down quarks, respectively. Likewise, flavor-chaging processes with charged weak gauge bosons $W^\pm$ give the cross section
\be
\begin{split}
& \sigma_{u_i+ W^- \rightarrow d_j + a}  = \sigma_{\bar{u}_i + W^+ \rightarrow \bar{d}_j + a} = \left| V_{\rm CKM}^{ij}\right|^2 \frac{g^2}{384\pi f_a^2} \frac{1-\frac{m_{d_j}^2}{s}}{\frac{m_W^2}{s}\sqrt{1-\frac{\left(m_W+m_{u_i}\right)^2}{s}}\sqrt{1-\frac{\left(m_W-m_{u_i}\right)^2}{s}}}  \\
&\times \left[ \rule{0cm}{1cm}  \hat{c}_u^2\frac{m_{u_i}^2}{s}\left(1- \frac{m_W^2-m_{u_i}^2}{s}- \frac{m_{d_j}^2}{s}\left(3+\frac{m_W^2-m_{u_i}^2}{s}\right)\right)  + \right. \\
&  + \hat{c}_d^2 \frac{m_{d_j}^2}{s}\left(1 + \frac{m_W^2 -2m_{u_i}^2}{s} - \frac{2m_W^2-m_W^2m_{u_i}^2-m_{u_i}^4}{s^2}\right) + 2 \hat{c}_u \hat{c}_d \frac{m_{u_i}^2m_{d_j}^2}{s^2}\left(3+ \frac{m_W^2-m_{u_i}^2}{s}\right) + \\
& -4 \hat{c}_u \frac{m_{u_i}^2}{s}\frac{2 \hat{c}_d \frac{m_{d_j}^2}{s}-\hat{c}_u\frac{2m_W^2 - m_{u_i}^2+m_{d_j}^2}{s}}{\sqrt{1-\frac{\left(m_W+m_{u_i}\right)^2}{s}}\sqrt{1-\frac{\left(m_W-m_{u_i}\right)^2}{s}}}\left.  \rule{0cm}{1cm} \arccoth\left[\frac{1- \frac{m_W^2-m_{u_i}^2}{s}}{\sqrt{1-\frac{\left(m_{W}+m_{u_i}\right)^2}{s}}\sqrt{1-\frac{\left(m_{W}-m_{u_i}\right)^2}{s}}}\right] \right] \ .
\end{split}
\label{eq:XSuwda}
\ee

One can easily derive the cross section of the scatterings of $d_j + W^+ \rightarrow u_i + a$ (equivalently, $\bar{d}_j + W^- \rightarrow \bar{u}_i + a$) from Eq.~\eqref{eq:XSuwda} with the exchange of $m_{u_i}\leftrightarrow m_{d_j}$ and $\hat{c}_u \leftrightarrow \hat{c}_d$.

\subsection*{KSVZ and DFSZ axions below the QCDPT}

The ChPT formalism describes low-energy axion interactions with the strong sector. As discussed in the main text, we trust calculations in this regime only up to $\Lambda_{\rm ChPT} \sim 100\,{\rm MeV}$. Thus axion production is dominated by pion scatterings mediated by the Lagrangian
\be
\mathcal{L}_{a\pi\pi\pi} = \frac{\partial_\mu a}{f_a} \, \frac{c_{a\pi\pi\pi}}{f_\pi} \left(\pi^0\pi^+\partial^\mu \pi^- + \pi^0\pi^-\partial^\mu \pi^+ - 2 \pi^+\pi^-\partial^\mu \pi^0\right) \ .
\ee
The dimensionless coupling $c_{a\pi\pi\pi}$ was given in the main text both for the KSVZ and the DFSZ axion, and the consequent cross sections for pion scatterings result in
\begin{align}
\sigma_{\pi^\pm  \pi^0 \rightarrow \pi^\pm  a} = & \, \frac{3s}{64\pi} \left(\frac{c_{a\pi\pi\pi}}{f_\pi f_a}\right)^2 \frac{\left(1-\frac{m_{\pi^\pm}^2}{s}\right)^3 
\left(1 - \frac{2m_{\pi^\pm}^2-m_{\pi^0}^2}{s} + \frac{\left(m_{\pi^\pm}^2-m_{\pi^0}^2\right)^2}{s^2}\right)}{\sqrt{1-\frac{(m_{\pi^0}-m_{\pi^\pm})^2}{s}}\sqrt{1-\frac{(m_{\pi^0}+m_{\pi^\pm})^2}{s}}} \ , \\ 
\sigma_{\pi^+  \pi^- \rightarrow \pi^0  a} = & \, \frac{9s}{64\pi} \left(\frac{c_{a\pi\pi\pi}}{f_\pi f_a}\right)^2 \frac{\left(1-\frac{m_{\pi^0}^2}{s}\right)^3}{\sqrt{1-\frac{4m_{\pi^\pm}^2}{s}}} \ .
\end{align}

\subsection*{DFSZ axion with leptons}

The DFSZ axion has also interactions with leptons giving cross sections~\cite{DEramo:2018vss}
\begin{align}
\sigma_{l_i \, \bar{l}_i \rightarrow \gamma \, a}  = & \, \frac{e^2}{4\pi}\frac{\hat{c}_e^2}{f_a^2}  \frac{\left(\frac{m_{l_i}^2}{s}\right) \arctanh\left(\sqrt{1-\frac{4m_{l_i}^2}{s}}\right)}{\left(1-\frac{4m_{l_i}^2}{s} \right)}  \ , \\
\sigma_{l_i \, \gamma \rightarrow l_i \,   a}  = & \, \sigma_{\bar{l_i} + \gamma \rightarrow \bar{l_i} + a} = \frac{e^2}{32\pi}\frac{\hat{c}_e^2}{f_a^2}  \frac{\left(\frac{m_{l_i}^2}{s}\right)\left[ 4\left(\frac{m_{l_i}^2}{s}\right) - \left(\frac{m_{l_i}^2}{s}\right)^2 -3 -2 \log \left( \frac{m_{l_i}^2}{s} \right) \right]}{\left(1- \frac{m_{l_i}^2}{s} \right)}  \ ,
\end{align}
where the coupling $\hat{c}_e$ is defined in Eq.~\eqref{eq:combinationEW}.


\section{Conventions and Useful Results IV: Cosmology}
\label{app:CO}

In this work, we study production of thermal axions during a radiation dominated era. We collect in this Appendix useful properties of the primordial thermal bath. The cosmological background where axion production takes place is a Friedmann-Lemaître-Robertson-Walker (FLRW) expanding universe with metric
\be
ds^2 = dt^2 - a(t)^2 \delta_{ij} dx^i dx^j \ .
\label{eq:FLRW}
\ee
The growth of the scale factor $a(t)$ is quantified by the Hubble parameter $H(t) \equiv (da/dt) / a(t)$ which in turn depends on the energy density $\rho$ of the universe via the Friedmann equation
\be
H = \frac{\sqrt{\rho}}{\sqrt{3} M_{\rm Pl}}  \ .
\ee
We use the reduced Planck mass $M_{\rm Pl} = (8 \pi G)^{-1/2} = 2.44 \times 10^{18} \, {\rm GeV}$.  Within our framework, the energy budget is dominated by a thermal bath of relativistic particles in thermal equilibrium with temperature $T$. The associated energy density scales as follows
\be
\rho_R(T) = \frac{\pi^2}{30} \, g_{*}(T) \, T^4 \ ,
\label{eq:rhor}
\ee
where $g_{*}(T)$ denotes the effective number of relativistic degrees of freedom contributing to the energy density. Another crucial property of the thermal bath is its entropy density
\be
s_R(T) = \frac{2\pi^2}{45} \, g_{*{s}}(T) \, T^3 \ .
\label{eq:sr}
\ee
Likewise, $g_{*s}(T)$ are the effective number of entropic relativistic degrees of freedom.

\subsection*{Temperature as the evolution variable}

The cosmic time $t$ appearing in the FLRW metric in Eq.~\eqref{eq:FLRW} is not the most convenient variable to describe the evolution of a radiation dominated universe. The presence of a thermal bath makes the temperature $T$ of the bath itself the most natural variable to keep track of the expansion. For a radiation dominated universe the expansion is adiabatic and the entropy in a comoving volume $s_R a^3$ does not change with time
\be
\frac{d s_R}{d t} + 3 H s_R = 0 \ .
\label{eq:entcon}
\ee
We plug the definition given in Eq.~\eqref{eq:sr} into Eq.~\eqref{eq:entcon} and we find
\be
\frac{d T}{d t} = - \frac{H T}{1 + \frac{1}{3} \frac{d \log g_{*s}(T)}{d \log T}} \ .
\ee
Given a generic function of time $\xi$, such as the axion number density $n_a$ appearing in the Boltzmann equation, we can trade easily time with temperature derivatives
\be
\frac{d \xi}{d \log T} =  - \left( 1 + \frac{1}{3} \frac{d \log g_{*s}(T)}{d \log T} \right) \frac{1}{H}  \frac{d \xi}{d t} \ .
\ee
It is often convenient to employ the dimensionless evolution variable $x \equiv M/T$, with the overall mass scale $M$ purely conventional. Thus we find another useful relation
\be
\frac{d \xi}{d \log x} = - \frac{d \xi}{d \log T}  =  \left( 1 - \frac{1}{3} \frac{d \log g_{*s}(x)}{d \log x} \right) \frac{1}{H}  \frac{d \xi}{d t} \ .
\label{eq:xvstder}
\ee

\subsection*{Temperature dependence of $g^{\rm SM}_{*}(T)$ and $g^{\rm SM}_{*{s}}(T)$}

At large temperatures all the degrees of freedom are relativistic so $g_{*}(T)$ and $g_{*{s}}(T)$ are constant. However, we consider axion production at temperatures below the weak scale where these quantities change as SM particles become non-relativistic. It is worth noting thanks to Eq.~\eqref{eq:xvstder} that not only the absolute values matter but also their temperature derivatives. This effect is particularly significant around the QCDPT.  We employ in our analysis the two different choices for the SM effective relativistic degrees of freedom.

\begin{itemize}

\item \textbf{Ref.~\cite{Drees:2015exa}.} At temperatures above the EW scale  all particles are considered free and massless and respecting the Stephan-Boltzmann law for bosons and fermions due to the crossover nature of EW transition in the SM. For massive particles around and below the EW scale when the temperature reaches each particle mass one should follow Fermi and Bose statistics. For the strongly  interacting fluid especially above $100$~MeV including the crossover QCD transition at $150$~MeV the result of lattice simulation is used  for up, down and strange quarks ($2+1$ flavors) added to the result for the charm quark at $1$~GeV. Then they matched to the free gas limit at very high temperatures. For temperatures below $100$~MeV the hadron resonance gas result for equation of state is used that matches to the lattice simulation of equation of state below the QCD transition epoch. At temperatures around $1$~MeV the result of evolution of neutrino temperature with respect to photon temperature that includes the effect of decoupling of different types of neutrinos is implemented. In this model the number effective neutrinos based on previous calculation is assumed as $N_{\rm eff}\simeq 3.046$. The rest of SM particles considered free.  Considering all these effects  improves the calculation for the extra number of relativistic particles for any given models. There are uncertainties on hadron resonance gas model, lattice simulation, and thermal effect of QCD at high temperatures, and electroweak transition.  

\item \textbf{Ref.~\cite{Saikawa:2018rcs}.} This study uses a different treatment for the electroweak and QCD transitions, hadron resonance gas model, and neutrino decoupling. Around the EW transition the thermal corrections on the Higgs field evolution including  the perturbative and nonperturbative effects for the interaction in the EW sector are used. Since around  the EW transition the change of d.o.f. is not abrupt like the QCD case, due to lesser interacting particles in the thermal bath, these corrections will have tiny effects on the final result. 
Below $120$~MeV the hadron resonance gas model and above that the QCD equation of state from a different lattice simulation for 2+1+1 flavors are used. Then it is freely matched to the perturbative QCD equation of state above $1$~GeV. Around $1$~MeV the neutrino decoupling is considered assuming $N_{\rm eff}\simeq 3.045$. Also, the negligible effects of plasma on electrons and photons are illustrated. 

\end{itemize}

\begin{figure}
\centering
 \includegraphics[width=.47\linewidth]{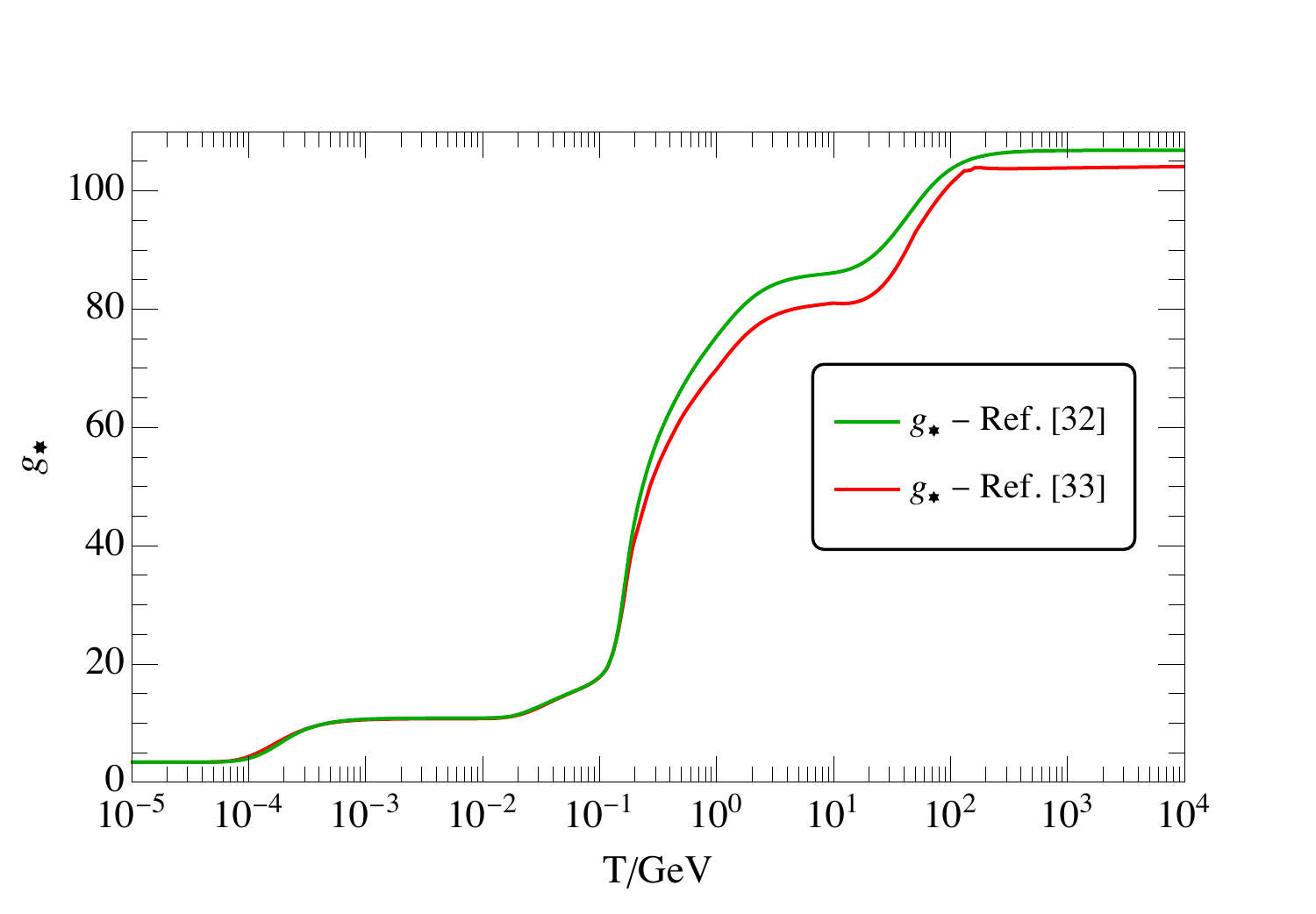} $\quad$
  \includegraphics[width=.47\linewidth]{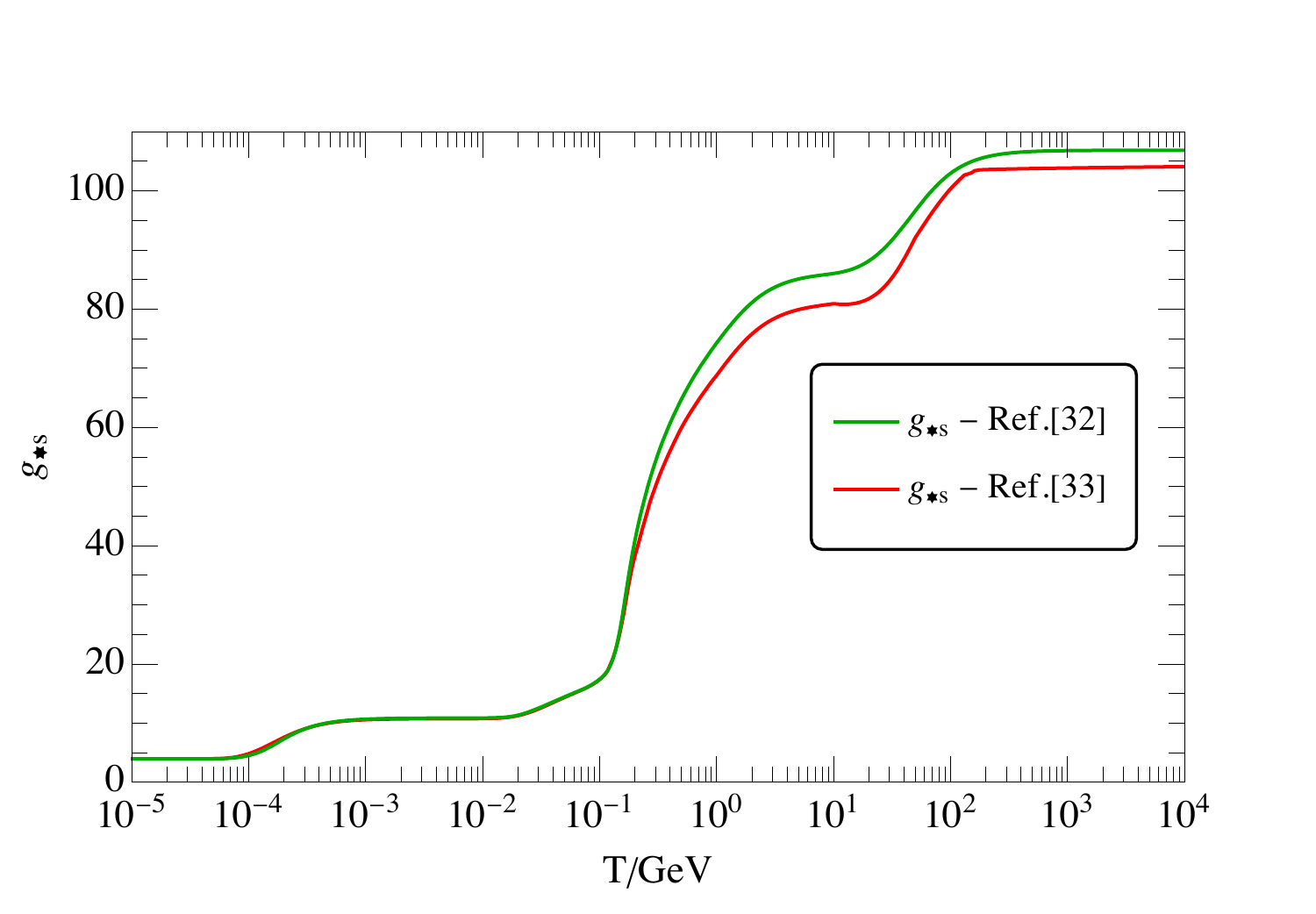}
\caption{\em Effective number of SM relativistic degrees of freedom contributing to the energy density $g_*(T)$ and to the entropy density $g_{*s}(T)$ as a function of the temperature. We report the results of Ref.~\cite{Drees:2015exa} (green solid line) and Ref.~\cite{Saikawa:2018rcs} (red solid line).}
\label{fig:dof}
\end{figure}

We compare the two different treatments in Fig.~\ref{fig:dof} where we show the temperature evolution for $g^{\rm SM}_{*}(T)$ (left panel) and $g^{\rm SM}_{*{s}}(T)$ (right panel). As it is manifest from these results, theoretical uncertainties will cause at most $10 \%$ difference in our prediction for the energy density stored in axion dark radiation. Furthermore, there are additional contributions to the effective relativistic degrees of freedom at high temperatures for the frameworks studied in this paper: the heavy PQ fermion and the extra Higgs bosons for the KSVZ and the DFSZ models, respectively. We include their effects by treating them as free particles with contributions given by the integrals in Eqs.~(2.9) and (2.10) of Ref.~\cite{Drees:2015exa}.

\subsection*{How to compute $\Delta N_{\rm eff}$}

We provide the definition for the effective numbers of neutrino species $\Delta N_{\rm eff}$ valid for a generic dark radiation candidate $\DR$. When the universe was approximately 380,000 years old, the plasma opacity to electromagnetic radiation suddenly dropped and photons free-streamed until they reached our detectors today. At this stage the bath temperature was approximately $T_{\rm CMB} \simeq 0.3 \, {\rm eV}$, and the only relativistic SM degrees of freedom were photons and neutrinos. The total energy density stored in radiation reads
\be
\rho_R(T_{\rm CMB})  = \rho_\gamma + \rho_\nu + \rho_\DR = 
\left[ 1 + \frac{7}{8} \left( N^{\rm SM}_{\rm eff}  + \Delta N_{\rm eff} \right) \left( \frac{4}{11} \right)^{4/3}\right]  \rho_\gamma  \ .
\ee
In the last equality, the effect of $\DR$ is parameterized by an effective number of additional neutrinos $\Delta N_{\rm eff}$ which we can find by direct comparison
\be
\Delta N_{\rm eff} = \frac{8}{7} \left( \frac{11}{4} \right)^{4/3} \left.\frac{\rho_{\rm DR}}{\rho_\gamma}\right|_{\rm CMB} \ . 
\label{eq:delneff}
\ee

We evaluate $\Delta N_{\rm eff}$ for a dark radiation candidate $\DR$ that reaches thermal equilibrium with the bath at early times and it decouples subsequently. Thermal equilibrium erases the memory of whatever happened at earlier times and we can neglect physics before decoupling. As long as $\Phi$ is coupled, the number and energy densities result in
\begin{align}
n_\DR(T) = & \, g_{n \DR} \frac{\zeta(3)}{\pi^2} T^3 \ , \qquad \qquad \qquad  g_{n \DR}  = g_\DR \left\{ \begin{array}{ccc} 1 &  & \text{boson} \\ 3/4 &  & \text{fermion}  \end{array}\right. \ , \\
\rho_\DR(T) = & \, g_{* \DR} \frac{\pi^2}{30} T^4 \ , \qquad \quad \qquad \qquad  g_{* \DR}  = g_\DR \left\{ \begin{array}{ccc} 1 &  & \text{boson} \\ 7/8 &  & \text{fermion}  \end{array}\right. \ .
\end{align}
Here, $g_\DR$ is a constant number accounting for the internal degrees of freedom (e.g., spin) of the particle $\DR$, and the distinction between bosons and fermions is due to the different phase-space equilibrium distributions. The Riemann $\zeta$ function appearing in the number density is approximately $\zeta(3) \simeq 1.2$. We get rid of the temperature in the equations above to find the relation between energy and number densities
\be
\rho_\DR = \frac{g_{* \DR}}{30} \left(\frac{\pi^{7/2}}{\zeta(3)} \right)^{4/3} \, \left(\frac{n_\DR}{g_{n \DR} }\right)^{4/3} \ .
\label{eq:rhovsn}
\ee
Another useful expression is the one between energy and entropy densities for the photons
\be
\rho_\gamma = 2 \times \frac{\pi^2}{30} T^4 = \frac{\pi^2}{15} \left( \frac{45}{2 \pi^2}  \frac{s_R}{g_{*s}(T)}  \right)^{4/3} \ .
\label{eq:rhogammavss}
\ee
We use Eqs.~\eqref{eq:rhovsn} and \eqref{eq:rhogammavss} to evaluate the dark radiation amount via Eq.~\eqref{eq:delneff}
\be
\Delta N_{\rm eff} = g_{* \DR} \; \frac{4}{7} \left( \frac{11}{4} \right)^{4/3} \; 
\left( \frac{2 \, \pi^4}{45 \, \zeta(3)}  \,  g_{*s}(T_{\rm CMB}) \frac{Y_\DR(T_{\rm CMB})}{g_{n \DR}}\right)^{4/3} 
\label{eq:DeltaNeffftemp}
\ee
with $Y_\DR = n_\DR / s_R$ the $\DR$ comoving number density. After decoupling, which happens at a temperature $T_D$, $\DR$'s just free-stream: the phase-space distribution keeps a thermal shape with temperature red-shifting with the scale factor as $T_\DR \propto a^{-1}$, and the number density gets diluted as $n_\DR \propto a^{-3}$. Thus the comoving number density stays constant because of entropy conservation throughout the expansion 
\be
Y_\DR(T_{\rm CMB}) = Y_\DR(T \leq T_D) = Y_\DR(T_D) = \frac{n_\DR(T_D)}{s_R(T_D)} = \frac{g_{n \DR}}{g_{*s}(T_D)} \frac{45 \, \zeta(3)}{2 \pi^4} \ .
\label{eq:YDR}
\ee
We plug this expression for the comoving yield into Eq.~\eqref{eq:DeltaNeffftemp} and we find
\be
\Delta N_{\rm eff} = g_{* \DR} \; \frac{4}{7} \left( \frac{11}{4} \right)^{4/3} \, \left( \frac{g_{*s}(T_{\rm CMB}) }{g_{*s}(T_D) } \right)^{4/3} \ .
\label{eq:DeltaNeffftemp2}
\ee
The effective number of relativistic entropic degrees of freedom includes contributions from both the SM bath and $\DR$. We define it as follows 
\be
g_{*s}(T) = g^{\rm SM}_{*s}(T) +  g^\DR_{*s}(T) \ .
\label{eq:gstarsTOT}
\ee
The SM part is illustrated in the right panel of Fig.~\ref{fig:dof}, and we quantify the additional contribution by knowing that $\Phi$'s decouple at $T_D$ and free-stream subsequently
\be
g^\DR_{*s}(T) = g_{*\DR} \left\{ \begin{array}{ccccccl}
1 & & & & & $\quad$  & T > T_D \\
\frac{g^{\rm SM}_{*s}(T)}{g^{\rm SM}_{*s}(T_D)} & & & & & $\quad$  & T \leq T_D 
\end{array} \right. \ .
\label{eq:gstarSDR}
\ee
By using this result we can find an equivalent way to express $\Delta N_{\rm eff}$ that reads
\be
\Delta N_{\rm eff} = g_{* \DR} \; \frac{4}{7} \left( \frac{11}{4} \right)^{4/3} \, \left( \frac{g^{\rm SM}_{*s}(T_{\rm CMB}) }{g^{\rm SM}_{*s}(T_D) } \right)^{4/3} \ .
\label{eq:DeltaNeffforfig1}
\ee
Unlike Eq.~\eqref{eq:DeltaNeffftemp2}, this result contains only the SM contribution to the entropic degrees of freedom that we can read off the plots in Fig.~\ref{fig:dof}. Furthermore, this relation is consistent with the temperature ratio at the CMB formation as dictated by entropy conservation
\be
\left. \frac{T_\DR}{T_\gamma}\right|_{\rm CMB} =  \left( \frac{g^{\rm SM}_{*s}(T_{\rm CMB}) }{g^{\rm SM}_{*s}(T_D)} \right)^{1/3}  \ .
\ee
The numerical result in Eq.~\eqref{eq:DeltaNeffIntro} of the introduction is a consequence of Eq.~\eqref{eq:DeltaNeffforfig1}, and the value $g^{\rm SM}_{*s}(T_{\rm CMB}) = 2 + N_{\rm eff}^{\rm SM} \times (7/11) \simeq 3.94$ that we use accounts for non-instantaneous neutrino decoupling. The output of this analysis describes the curves in Fig.~\ref{fig:neffdec}.

The case discussed above is not the most general one. Thermalization may not be achieved, and even if $\DR$'s reach thermal equilibrium the temperature $T_D$ is not the most practical variable to employ. As we do in our analysis for the QCD axion, the standard procedure is to solve the Boltzmann equation and find the asymptotic density. We conclude this Appendix with the explanation of how to use such an asymptotic value to find $\Delta N_{\rm eff}$.  The starting point is still Eq.~\eqref{eq:DeltaNeffftemp} since it does not rely upon any assumption about thermalization. The only unknown quantity in that expression is the number of effective entropic degrees of freedom at recombination $g_{*s}(T_{\rm CMB})$: the SM part is known, we need to quantify the contribution from $\DR$ in terms of $Y_\DR(T_{\rm CMB})$
\be
g^\DR_{*s}(T_{\rm CMB}) = g_{* \DR} \left. \left(\frac{T_\DR}{T_\gamma}\right)^3\right|_{\rm CMB} = 
g_{* \DR} \; \frac{2 \pi^4}{45 \zeta(3)} g_{*s}(T_{\rm CMB}) \frac{Y_\DR(T_{\rm CMB})}{g_{n \DR}} \ .
\ee
The full number of entropic relativistic degrees of freedom appearing after the last equality is given by the two contributions in Eq.~\eqref{eq:gstarsTOT}. Thus the above equation allows us to solve for $g^\DR_{*s}(T_{\rm CMB})$ and eventually for $g_{*s}(T_{\rm CMB})$, and we find our final result
\be
\Delta N_{\rm eff} = g_{* \DR} \; \frac{4}{7} \left( \frac{11}{4} \right)^{4/3} \; 
\left[\frac{\frac{2 \pi^4}{45 \zeta(3)} \, g^{\rm SM}_{*s}(T_{\rm CMB}) \, \frac{Y_\DR(T_{\rm CMB})}{g_{n \DR}}}{1 - \frac{2 \pi^4}{45 \zeta(3)} g_{* \DR} \frac{Y_\DR(T_{\rm CMB})}{g_{n \DR}}}\right]^{4/3} \ .
\label{eq:DeltaNeffgeneral}
\ee
The second term in the denominator accounts for the entropy associated to the dark radiation particle $\DR$. We estimate its relevance by looking back at the case when $\DR$'s decouple at the temperature $T_D$, and we plug the explicit equilibrium comoving density as given in Eq.~\eqref{eq:YDR}. We find that the correction results in $g_{* \DR} / (g^{\rm SM}_{*s}(T_D) + g_{* \DR} )$, and therefore it is relevant only if the dark radiation stays in thermal equilibrium until a time when its effective number of entropic degrees of freedom is comparable with the one of the SM bath.


\subsection*{Theoretical uncertainty on $\Delta N_{\rm eff}$ due to the interpolation}

\begin{figure}
\centering
 \includegraphics[width=.47\linewidth]{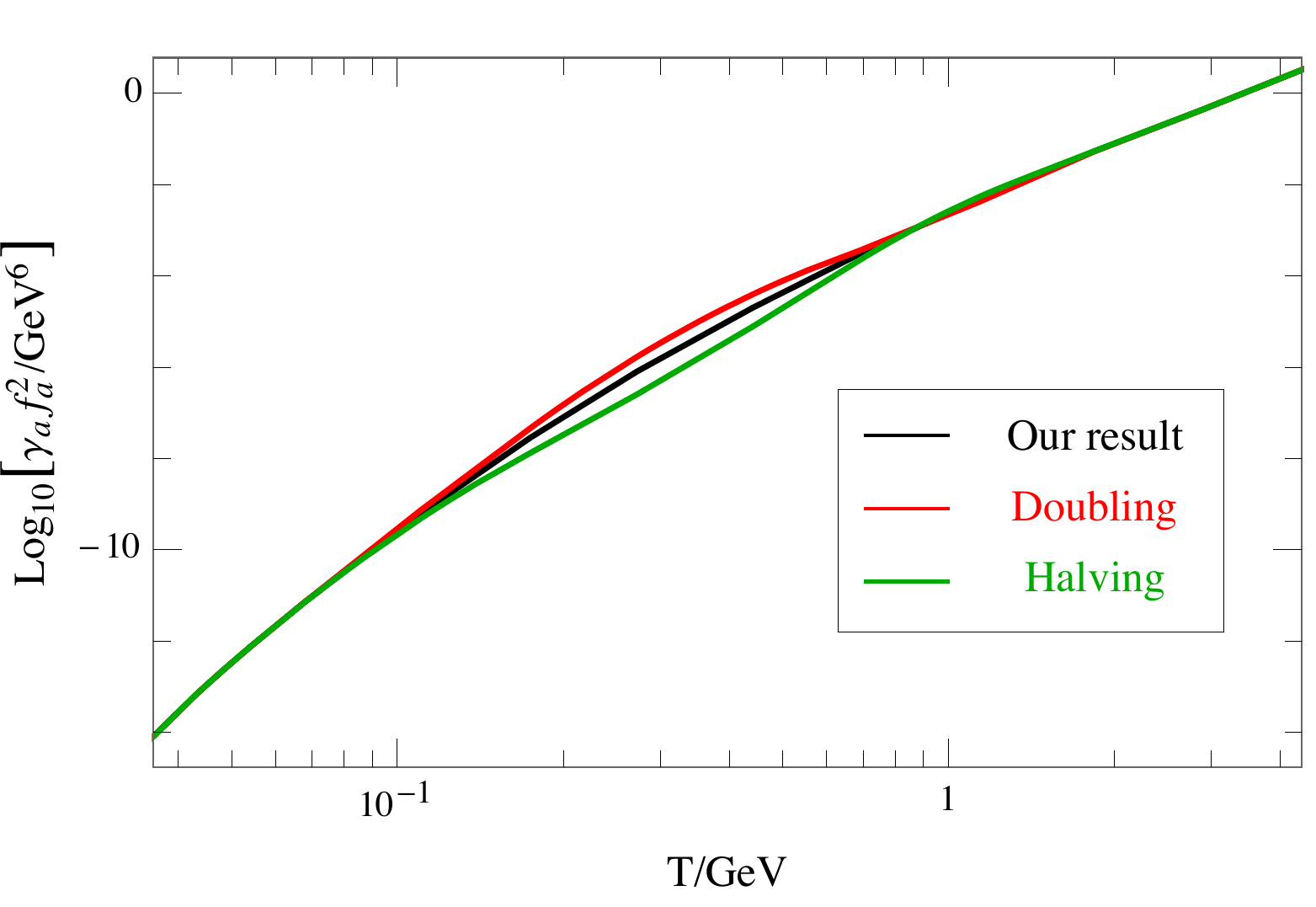} $\quad$
  \includegraphics[width=.47\linewidth]{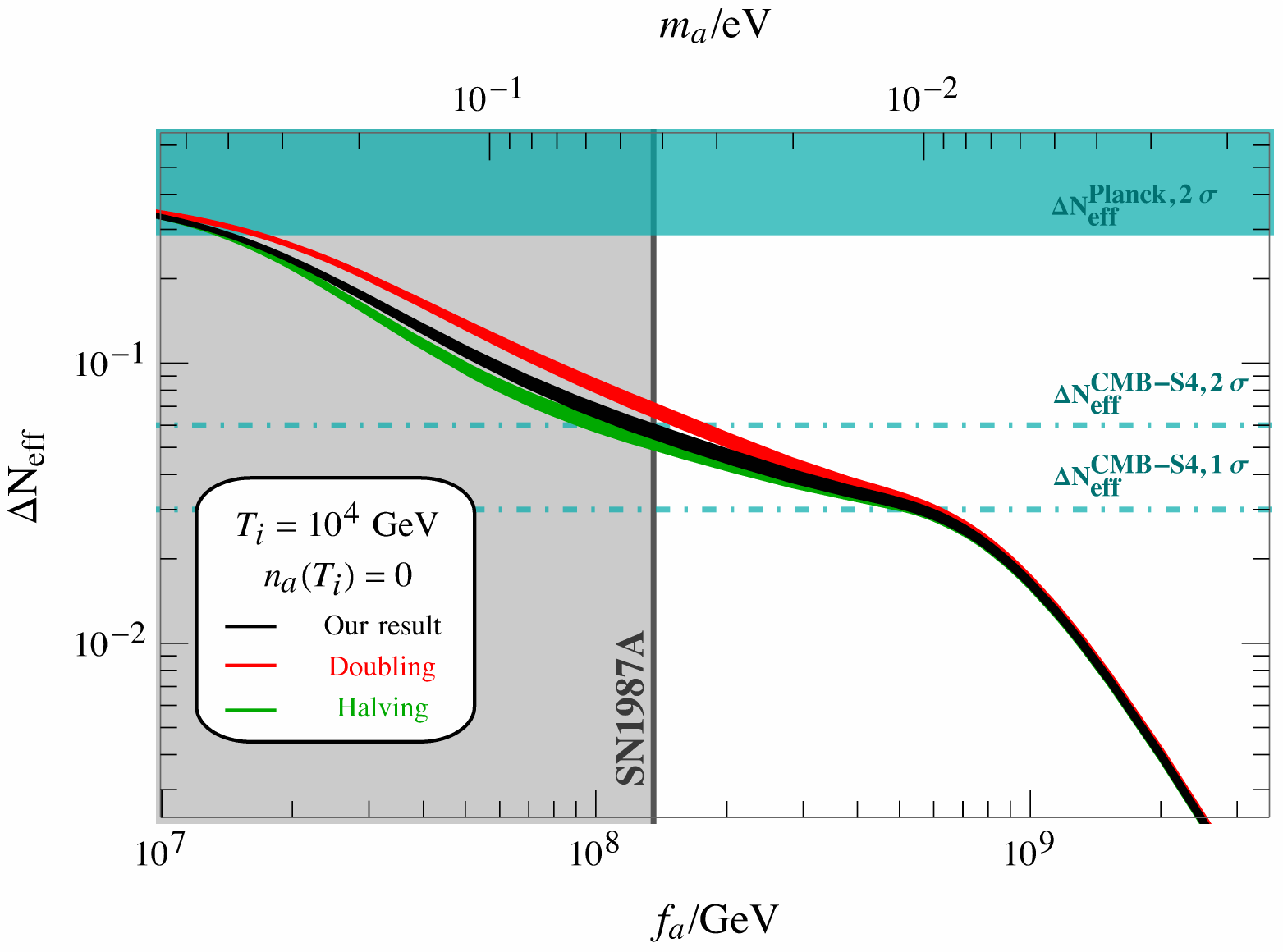}
\caption{\em Left panel: production rate for the KSVZ axion employed in our paper (black), and factor of 2 enhancement (red) or suppression (green) around the QCDPT. Right panel: predictions for $\Delta N_{\rm eff}$ for the three different cases.}
\label{fig:uncertainty}
\end{figure}

We conclude this Appendix with a discussion of theoretical uncertainties associated with our interpolation across the QCDPT. In our work, we performed a smooth interpolation for the production rate between $\Lambda_{\rm ChPT}$ and $\Lambda_{\rm N}$ with the cubic `spline' method, motivated by the fact that the QCDPT is a crossover leading to mild shifts of thermal properties. One may wonder how our predictions for $\Delta N_{\rm eff}$ are sensitive to the details of such an interpolating method. We take the KSVZ axion case for concreteness, and we modify the production rate as shown in the left panel of Fig.~\ref{fig:uncertainty}. The solid black line corresponds to the rate used in our analysis. We consider two extreme cases where around the temperature scale $500 \, {\rm MeV}$ the actual rate is a factor of two larger (red line) or smaller (green line), and we make sure to match these lines with our results below $\Lambda_{\rm ChPT}$ and above the mass of QCD resonances. The resulting predictions for $\Delta N_{\rm eff}$ are shown in the right panel of Fig.~\ref{fig:uncertainty}. For values of the axion decay constant not excluded experimentally, our predictions are quite insensitive to the detail of the interpolation and therefore utterly solid.

\bibliographystyle{JHEP}
\bibliography{AxionDR}

\end{document}